\documentclass[preprint2]{emulateapj}

\newcommand{\chn}{{\it Chandra}}
\newcommand{\swf}{{\it Swift}}
\newcommand{\xmm}{{\it XMM-Newton}}

\usepackage{graphicx}
\usepackage{longtable}
\usepackage{mdwlist}

\usepackage{natbib}
\usepackage{multirow}
\usepackage{upgreek}
\usepackage[colorlinks=true, pdfstartview=FitV, linkcolor=blue,citecolor=blue, urlcolor=blue]{hyperref}
\usepackage{wasysym}
\usepackage{txfonts}
\usepackage{gensymb}
\usepackage{morefloats}
\usepackage{threeparttable}
\usepackage{rotating}
\usepackage{graphicx}

\slugcomment{version \today: fm} 

\shorttitle{The \chn\ 3CR extragalactic survey with 0.5$<z<$1.0}
\shortauthors{F. Massaro et al.  2017}

\begin{document}
\title{The 3CR Chandra snapshot survey: extragalactic radio sources with 0.5$<z<$1.0} 

\author{
F. Massaro\altaffilmark{1,2,3}, 
V. Missaglia\altaffilmark{4},
C. Stuardi\altaffilmark{1,3}, 
D. E. Harris\altaffilmark{5,+},
R. P. Kraft\altaffilmark{5},
A. Paggi\altaffilmark{5},
E. Liuzzo\altaffilmark{6}, \\
G. R. Tremblay\altaffilmark{5,7},
S. A. Baum\altaffilmark{8,9},
C. P. O'Dea\altaffilmark{8,10},
B. J. Wilkes\altaffilmark{5},
J. Kuraszkiewicz\altaffilmark{5}
\&
W. R. Forman\altaffilmark{5}
}

\altaffiltext{1}{Dipartimento di Fisica, Universit\`a degli Studi di Torino, via Pietro Giuria 1, I-10125 Torino, Italy.}
\altaffiltext{2}{Istituto Nazionale di Fisica Nucleare, Sezione di Torino, I-10125 Torino, Italy.}
\altaffiltext{3}{INAF-Osservatorio Astrofisico di Torino, via Osservatorio 20, 10025 Pino Torinese, Italy}
\altaffiltext{4}{Department of Physical Sciences, University of Napoli Federico II, via Cinthia 9, 80126 Napoli, Italy.}
\altaffiltext{5}{Smithsonian Astrophysical Observatory, 60 Garden Street, Cambridge, MA 02138, USA.}
\altaffiltext{6}{Istituto di Radioastronomia, INAF, via Gobetti 101, 40129, Bologna, Italy.}
\altaffiltext{7}{Yale Center for Astronomy and Astrophysics, Physics Department, Yale University, PO Box 208120, New Haven, CT 06520-8120, USA.}
\altaffiltext{8}{University of Manitoba,  Dept. of Physics and Astronomy, Winnipeg, MB R3T 2N2, Canada.}
\altaffiltext{9}{Center for Imaging Science, Rochester Institute of Technology, 84 Lomb Memorial Dr., Rochester, NY 14623, USA.}
\altaffiltext{10}{School of Physics \& Astronomy, Rochester Institute of Technology, 84 Lomb Memorial Dr., Rochester, NY 14623, USA.}
\altaffiltext{+}{Dan Harris passed away on December 6th, 2015. His career spanned much of the history of radio and X-ray astronomy. His passion, insight, and contributions will always be remembered. A significant fraction of this work is one of his last efforts.}

\begin{abstract} 
This paper presents the analysis of \chn\ X-ray snapshot observations of a subsample of the extragalactic sources listed in the revised Third Cambridge radio catalog (3CR), previously lacking X-ray observations and thus observed during \chn\ Cycle 15. This data set extends the current \chn\ coverage of the 3CR extragalactic catalog up to redshift $z$=1.0. Our sample includes 22 sources consisting of one compact steep spectrum (CSS) source, three quasars (QSOs), and 18 FR\,II radio galaxies. {As in our previous analyses, here we report the X-ray detections of radio cores and extended structures (i.e., knots, hotspots and lobes) for all sources in the selected sample}. We measured their X-ray intensities in three energy ranges: soft (0.5--1 keV), medium (1--2 keV) and hard (2-7 keV) and we also performed standard X-ray spectral analysis for brighter nuclei. {All radio nuclei in our sample have an X-ray counterpart. We also discovered X-ray emission associated with the eastern knot of 3CR\,154, with radio hotspots in 3CR\,41, 3CR\,54 and 3CR\,225B and with the southern lobe of 3CR\,107. Extended X-ray radiation around the nuclei 3CR\,293.1 and 3CR\,323 on a scale of few tens kpc was also found.} X-ray extended emission, potentially arising from the hot gas in the intergalactic medium {and/or due to the high energy counterpart of lobes, is } detected for 3CR\,93, 3CR\,154, 3CR\,292 and 3CR\, 323 over a few hundreds kpc-scale. {Finally, this work also presents an update on the state-of-the-art of \chn\ and \xmm\ observations for the entire 3CR sample.}
\end{abstract}

\keywords{galaxies: active --- X-rays: general --- radio continuum: galaxies}

\section{Introduction}
\label{sec:intro}
 Since the early 60's, the ensemble of extragalactic sources listed in the Third Cambridge radio catalog (3C) represents one of the most attractive samples to study the physics of radio-loud active galactic nuclei (AGNs). Originally created using radio observations {performed} at 159 MHz \citep{edge59}, and subsequently at 178 MHz \citep{bennett62}, the 3C catalog went through two main revisions later in the 80's \citep[see e.g.][]{laing83,spinrad85}. 

{Since then, a vast suite of observations became available from the radio to optical wavelengths thus enriching the multifrequency database necessary to carry out broad band analyses. Radio images with arcsecond resolution for almost all 3CR extragalactic sources are already present in the NRAO Very Large Array (VLA) Archive Survey (NVAS)\footnote{\underline{http://archive.nrao.edu/nvas/}} and in the MERLIN\footnote{\underline{http://www.jb.man.ac.uk/cgi-bin/merlin\_retrieve.pl}} archives \citep[see e.g.,][]{giovannini05}.} At higher frequencies, Spitzer \citep[see e.g.,][]{werner12,dicken14}, in particular for high redshift sources \citep[see also][]{haas08,leipski10}, and Hubble Space Telescope observations cover more than 90\% of the 3CR extragalactic catalog \citep[see e.g.][]{madrid06,privon08,tremblay09,hilbert16}. {In addition near infrared observations are also available for a significant fraction of the 3CR objects} \citep[see e.g.][]{baldi10}. Recently, the Herschel Space Observatory also observed several 3CR sources, mostly focusing on the higher redshift ones, \citep[see e.g.][]{podigachoski15,westhues16}. {Moreover, a dedicated spectroscopic campaign was carried out with the Telescopio Nazionale Galileo to provide a detailed optical classification and to study their nuclear emission \citep{buttiglione09,buttiglione11}}. All these observations make the extragalactic 3CR catalog an ideal sample to investigate AGN {nuclear properties, extended radio structures, such as jet knots, hotspots and lobes, and/or their large-scale environments} \citep[see e.g.][]{ineson13,chiaberge15}.

However, although most of the 3CR extragalactic radio sources were observed thanks to extensive X-ray campaigns carried out with \chn, \xmm\ and \swf\ \citep[see e.g.,][and references therein]{hardcastle00,harvanek01,hardcastle06,evans06,balmaverde12,wilkes13,maselli16}, until {Cycle 9 the \chn\ archive covered only up to $\sim$60\% of the 3CR extragalactic sample \citep[see e.g.][for a recent review]{massaro15}, while the others, such as \xmm, covered less than 1/3 of the entire catalog. Thus we started our \chn\ snapshot survey to ensure that all 3CR extragalactic sources have at least an exploratory X-ray observation, with an angular resolution similar to those at lower energies, available to the astronomical community.} Adopting a step-wise strategy, we {requested} observations in narrow redshift, $z$, ranges {resulting in} modest proposals each cycle to minimize the impact on the \chn\ schedule. To date all the 3CR sources with $z<$1 have at least a snapshot observation (i.e., less than 20ksec total exposure time) available in the \chn\ archive \citep{massaro10,massaro12,massaro13} {and several of them inspired follow up X-ray observations on interesting objects \citep[see e.g.,][to name a few examples]{3c171,3c305fol,3c89}.}

Here, we present the analysis of the \chn\ snapshot observations {approved} during Cycle 15 including all 3CR radio sources lying between $z$=0.5 and $z$=1 that were previously unobserved by \chn. 

The paper is organized as follows. The update of the ongoing \chn\ campaign of the 3CR sources is described in \S~\ref{sec:history} together with some details on the current sample. A brief overview of the data reduction procedures are given in \S~\ref{sec:obs} while results are described in \S~\ref{sec:results}.  Then, in \S~\ref{sec:summary} we present our summary and conclusions. Finally, in the Appendix, we show X-ray images with radio contours overlaid for all the sources in the current sample (\S~\ref{sec:images}) together with the updated summary of the \chn\ observations for the entire of 3CR extragalactic catalog (\S~\ref{sec:state}).

Unless {otherwise stated we adopt cgs units for numerical results and we also assume} a flat cosmology with $H_0=69.6$ km s$^{-1}$ Mpc$^{-1}$, $\Omega_{M}=0.286$ and $\Omega_{\Lambda}=0.714$ \citep{bennett14}. Spectral indices, $\alpha$, are defined by flux density, S$_{\nu}\propto\nu^{-\alpha}$.

\section{State-of-the-art of the 3CR extragalactic Chandra snapshot survey}
\label{sec:history}
The revised 3C extragalactic catalog includes 298 sources \citep{spinrad85}. We have already {analyzed} and published all the data collected to date for the \chn\ observations carried out in Cycles 9, 12 and 13 for a total of 75 sources \citep{massaro10,massaro12,massaro13} and an additional 140 objects, {listed} in the \chn\ archive, were also presented adopting the same data reduction procedures \citep{massaro11,massaro15}. Several subsets of the 3CR sample have been also observed by other groups \citep[e.g.,][]{wilkes13,kuraszkiewicz17}. Table~\ref{tab:summary} summarizes the references for the {\chn\ observations of the 3CR extragalactic sources} already analyzed and published as part of this compilation.

In our previous archival analyses we excluded 7 sources, namely 3CR 66A \citep[e.g., ][]{abdo11}, 3CR 71 \citep[alias NGC 1068; e.g., ][]{brinkman02}, 3CR 84 \citep[alias NGC1275 or Perseus A; e.g., ][]{fabian03}, 3CR 186 \citep{siemiginowska10}, 3CR 231 \citep[alias M82; e.g., ][]{griffiths00}, 3CR 317 \citep[alias Abell 2052; e.g., ][]{blanton09} and 3CR 348 \citep[alias Hercules A; e.g., ][]{nulsen05}, since {each of these source has accumulated an exposure time of greater than 80 ks, and has been discussed extensively in the literature.} In addition, we also did not re-analyze: 3CR 236, 3CR 326, 3CR 386 since the PI of these observations is currently carrying out the analysis (M. Birkinshaw, priv. comm.).

The \chn\ archive now includes all the 3CR {extragalactic} sources up to $z$=0.5 (i.e., 150 sources), with the only exceptions {being} those objects for which spectroscopic observations, performed after the last revision \citep{spinrad85}, reported different redshift estimates, namely: 3CR\,27, at $z$=0.184, 3CR\,69 at $z$=0.458 \citep{hiltner91} and 3CR\,93 at $z$=0.357, as confirmed by Ho \& MinJin (2009). The present analysis extends {the \chn\ } database up to $z$=1.0 including 22 more targets. During these Cycle 15 3CR snapshot observations, we also observed 3CR\,142.1 and 3CR\,277 for which the redshift reported in the literature \citep{hewitt91} updates the earlier {estimates} from Spinrad et al. (1985). These two sources, together with 3CR\,93, belong to the sample of 22 targets analyzed in the present work. \chn\ snapshot observations of 3CR\,27 and 3CR\,69 {were proposed} and obtained in subsequent observing cycles (Stuardi et al. 2017 in prep.).

Twenty-five of the 298 3CR extragalactic sources are still unidentified, lacking an optical counterpart and/or an optical spectroscopic observation necessary to unveil their nature. We recently observed 21 of these 25 targets with \swf\ snapshot observations {discovering} X-ray counterparts for eleven of them, but even using optical data {available thanks to the instruments on board of the \swf\ satellite}, we could not discern and/or confirm their extragalactic nature \citep[see][for all details]{maselli16}.

A summary table on the state-of-the-art of the \chn\ observations for all the 3CR {extragalactic} sources, including detections of extended components: jet knots, hotspots, lobes and X-ray emission from the hot intergalactic medium {present} in galaxy groups/clusters, is reported in Appendix \S~\ref{sec:state}. 

\begin{table*}
\begin{center} 
\caption{Summary of the 3CR extragalactic sources analyzed in our previous \chn\ investigations}
\label{tab:summary}
\tiny
\begin{tabular}{|lcccll|}
\hline
Program  & Cycle & Proposal Number & Number of sources   & Redshift range & Reference \\
\hline 
\noalign{\smallskip}
3CR snapshot survey & 9 & 09700745 & 30$^*$ & $z<$0.3 & Massaro et al. (2010) \\
XJET$^+$ & --- & --- & 47 & --- & Massaro et al. (2011) \\
3CR snapshot survey & 12 & 12700211 & 26 & $z<$0.3 & Massaro et al. (2012) \\
3CR snapshot survey & 13 & 13700190 & 19 & 0.3$<z<$0.5 & Massaro et al. (2013) \\
Archival project$^+$ & --- & --- & 93 & --- & Massaro et al. (2015) \\
3CR snapshot survey & 15 & 15700111 & 22$^\circ$ & 0.5$<z<$1.0 & This work\\
\noalign{\smallskip}
\hline
\end{tabular}\\
\end{center}
$^*$ The AO9 sample includes 3CR 346 that was re-observed in Cycle
12 because during Cycle 9 its \chn\ observation was affected by high
background \\ \citep[see][for details]{massaro10}.\\
$^+$ The redshift ranges for both the archival and the XJET samples are
unbounded w.r.t. selection \\ (see also {\underline{http://hea-www.cfa.harvard.edu/XJET/} for more details on the database}). \\
$^\circ$ The field of view of the 3C\,255B \chn\ observation also covers the region where 3C\,255A lies but no X-ray counterpart for its nucleus is detected.
\end{table*}

\section{Data reduction and data analysis}
\label{sec:obs}
Data reduction and analysis procedures adopted for all the \chn\ observations presented here were extensively described in our previous papers, thus only a brief overview is reported in the following. We adopted the same procedures since our final aim is to create a uniform database for the entire 3CR extragalactic survey once all the sources listed therein will be observed by \chn. 

We followed the standard procedure described in the \chn\ Interactive Analysis of Observations (CIAO) threads\footnote{\underline{http://cxc.harvard.edu/ciao/guides/index.html}},  to perform our data reduction and we used CIAO v4.7 with the \chn\ Calibration Database (CALDB) version 4.6.2.

\subsection{X-ray photometry}
We initially generated level 2 event files using the $acis\_process\_events$ task and filtering for grades 0,2,3,4,6.  We checked the absence of high background intervals inspecting lightcurves extracted for each data set, but this never occurred. We aligned the X-ray position of each core with that of the radio to perform the astrometric registration \citep[see e.g.,][for details]{massaro11}. {The source 3CR\,292 has been observed twice with \chn, obsID 16065 and obsID 17488, with exposure times of $\sim$4ksec and $\sim$8ksec, respectively. In this case a merged event file was created using the CIAO routine \textsc{reproject\_obs}, thus reprojecting event files to the reference coordinates of the deeper observation (i.e., obsID 17488).} 

Then we created flux maps in the X-ray energy ranges: 0.5 -- 1 keV (soft), 1 -- 2 keV (medium), 2 -- 7 keV (hard), taking into account exposure time and effective area. In our procedure, as previously done, we used monochromatic exposure maps set to the nominal energies of 0.75, 1.4, and 4 keV for the soft, medium and hard band, respectively. All flux maps were converted from units of counts/sec/cm$^2$ to cgs units by multiplying each event by the nominal energy of each band. However, we made the necessary correction to recover the observed erg/cm$^2$/s, when performing X-ray photometry \citep[see e.g.,][for details]{3c17,3c305}. 

We measured observed fluxes for all the X-ray detected nuclei and extended components. This was done choosing a region of size and shape appropriate to the observed X-ray emission and matching the radio structure. We also chose two background regions, having the same shape and size so as to avoid X-ray emission from other parts of the source.  The flux for each energy band and region was measured using funtools\footnote{\underline{http://www.cfa.harvard.edu/$\sim$john/funtools}} as in our previous analyses. Uncertainties are computed assuming Poisson statistics (i.e., square root of the number-of-counts) in the source and background regions. X-ray fluxes, not corrected for the Galactic absorption, measured for the cores are reported in Table~\ref{tab:cores} while those for detected jet knots, hotspots and lobes are given in Table~\ref{tab:features}. The name of each component (e.g., knot or hotspot) is a combination of one letter indicating the orientation of the radio structure and one number indicating distance from the core in arcseconds.

Since the \chn\ native pixel size for the ACIS instrument is 0\arcsec.492, the data are undersampled, thus to recover the resolution we regridded our images to 1/2, 1/4, or 1/8 of the native ACIS pixel size. This was dictated by the angular size of each radio source and by the number of counts in each source components. For sources of large angular extent 1/2 or no regridding was adopted \citep[see also][for more details]{massaro12,massaro13}.

Finally, we performed a comparison between radio and X-ray images at similar angular resolution to verify if extended structures in radio sources, such as jet knots, hotspots and lobes, have an X-ray counterpart. To achieve this goal, we used radio images retrieved from publicly available websites as that of the National Radio Astronomy Observatory VLA Archive Survey (NVAS) \footnote{\underline{http://archive.nrao.edu/nvas/}}, NASA Extragalactic Database (NED) \footnote{\underline{http://ned.ipac.caltech.edu/}}, the (DRAGN)\footnote{\underline{http://www.jb.man.ac.uk/atlas/}} website as well as personal websites of our colleagues\footnote{\underline{http://3crr.extragalactic.info}}\footnote{\underline{http://www.slac.stanford.edu/~teddy/vla3cr/vla3cr.html}}. Image parameters for each radio observation used are given in the {figure captions} of \S~\ref{sec:images}.

\subsection{X-ray spectral analysis}
We performed spectral analysis for the X-ray counterparts of radio cores of those sources having more than 400 counts, to determine their X-ray spectral indices \(\alpha_X\), the presence or absence of significant intrinsic absorption, and the role played by mild pileup in artificially hardening the spectrum. 

The spectral data were extracted from a 2\arcsec\ aperture, as for photometric measurements, using the CIAO routine \textsc{specextract}, thereby automating the creation of count-weighted response matrices. Background spectra were extracted in nearby circular regions of radius 10\arcsec\ not containing obvious sources. The source spectra were then filtered in energy between 0.5-7 keV and binned to allow a minimum number of 30 counts per bin to ensure the use of the Gaussian statistic. We used the \textsc{Sherpa}\footnote{\href{http://cxc.harvard.edu/sherpa}{http://cxc.harvard.edu/sherpa}} modeling and fitting package to fit our spectra. For each source we adopted two models: (1) a redshifted power-law with Galactic and intrinsic photoelectric absorption components, and (2) the same model with an additional pileup component using the \textsc{jdpileup} \textit{Chandra} CCD pileup model developed by Davis (2001). 

Prior to fitting, the Galactic hydrogen column density and the source redshift were fixed at the measured values for each source. When considering the first (1) fitting model, the two main variable parameters - namely the intrinsic absorption (N$_H$(z)) and X-ray photon index \(\Gamma\) - were allowed to vary in a first pass fit, but subsequently stepped through a range of possible physical values to explore the parameter space, determine 68\% confidence intervals, and quantify the degree to which \(N_H(z)\) and \(\Gamma\) are degenerate. 

We have also explored the possible effect of pileup on our sources adding a \textsc{jdpileup}\footnote{\href{http://cxc.harvard.edu/sherpa/ahelp/jdpileup.html}{http://cxc.harvard.edu/sherpa/ahelp/jdpileup.html}} component to our baseline model (2). We left the parameters of the \textsc{jdpileup} model fixed to their default values, with the exception of \textsc{alpha} (the probability of a good grade when two photons pile together) and \textsc{f} (the fraction of flux falling into the pileup region). In no case were we able to constrain the value of \textsc{alpha} since it was usually degenerate with \textsc{f} and/or the intrinsic absorption, so we decided to freeze it to its default value of 1. The value of \textsc{f} was left free to vary between 0.85 and 1, and was constrained in two cases. The results of spectral fitting are given in Table~\ref{tab:spectra}. 


\section{Results}
\label{sec:results}

\subsection{General}
We detected {the X-ray counterpart of all radio cores} in our sample and we measured {their fluxes} adopting a circular region of 2$\arcsec$ centered on the radio position used for the astrometric registration. All the results of our X-ray photometry, i.e., nuclear X-ray fluxes in the three different bands (see \S~\ref{sec:obs}) together with their X-ray luminosities, are reported in Table~\ref{tab:cores} while X-ray images for the whole selected sample are presented in \S~\ref{sec:images}. In Table~\ref{tab:cores}, we {also report the values of the 'extended emission' parameter,} computed as the ratio of the total number of counts in a circular region of radius r\,=\,2\arcsec\ circle to that of a circle of radius r\,=\,10\arcsec\ both centered on the radio position of each 3CR source (i.e., Ext. Ratio ``Extent Ratio'').  This ratio is close to unity for unresolved (i.e., point-like) sources since the on-axis encircled energy for 2\arcsec\ is $\simeq$0.97, and only a small increase is expected up to 10\arcsec. Thus {parameter values} significantly less than 0.9 indicate the presence of extended emission around the nuclear component \citep[e.g.,][]{massaro10,massaro13}. {In our sample this situation clearly occurs} for 3CR\,107, 3CR\,293.1 and 3CR\,323. 

Four of the 22 sources{./, namely: 3CR\,93, 3CR\,114, 3CR\,154 and 3CR\,288.1,} show X-ray nuclei with more than 400 counts within a circular region of 2\arcsec\ radius. Thus, according to our previous works, we performed a basic X-ray spectral analysis for them. Their spectra are consistent with a mildly absorbed power-law with the absorption consistent with the Galactic value \citep{kalberla05}. 3CR\,93 and 3CR\,154 show a fraction of pileup of 10\%, however their fits do not improve significantly when {including} a jdpileup {spectral component} \footnote{\underline{http://cxc.harvard.edu/sherpa/ahelp/jdpileup.html}}. Results of the spectral analysis are reported in Table~\ref{tab:spectra}, the statistical uncertainties quoted refer to the 68\% confidence level.

We also discovered X-ray emission associated with three hotspots in three different sources; their {X-ray fluxes are reported} in Table~\ref{tab:features} together with detection significances, all above 3$\sigma$, computed assuming a Poisson distribution for the background. {The X-ray counterpart of a radio knot in the 3CR\,154 eastern lobe as well as that of the whole southern radio lobe of 3CR\,107 was also found (see Table~\ref{tab:features} for details).}

Three of 22 3CR sources in our sample {lie within optically known galaxy cluster, namely}: 3CR\,34, 3CR\,44 and 3CR\,247 \citep{spinrad85} and to search for possible X-ray emission due to the {hot gas permeating the intergalactic medium} around the radio {structure}, we adopted the following procedure for all our targets. We measured the total number of counts in a circular region with diameter equal to the total extent of the radio source and we subtracted those counts within circular regions of 2\arcsec\ corresponding to the radio nucleus and/or to background sources lying within the larger circle. Then assuming a Poisson distribution for the background events, we computed the probability of obtaining the measured value given the expected number of counts in the background. 

\begin{figure*}
\includegraphics[height=6.5cm,width=9cm,angle=0]{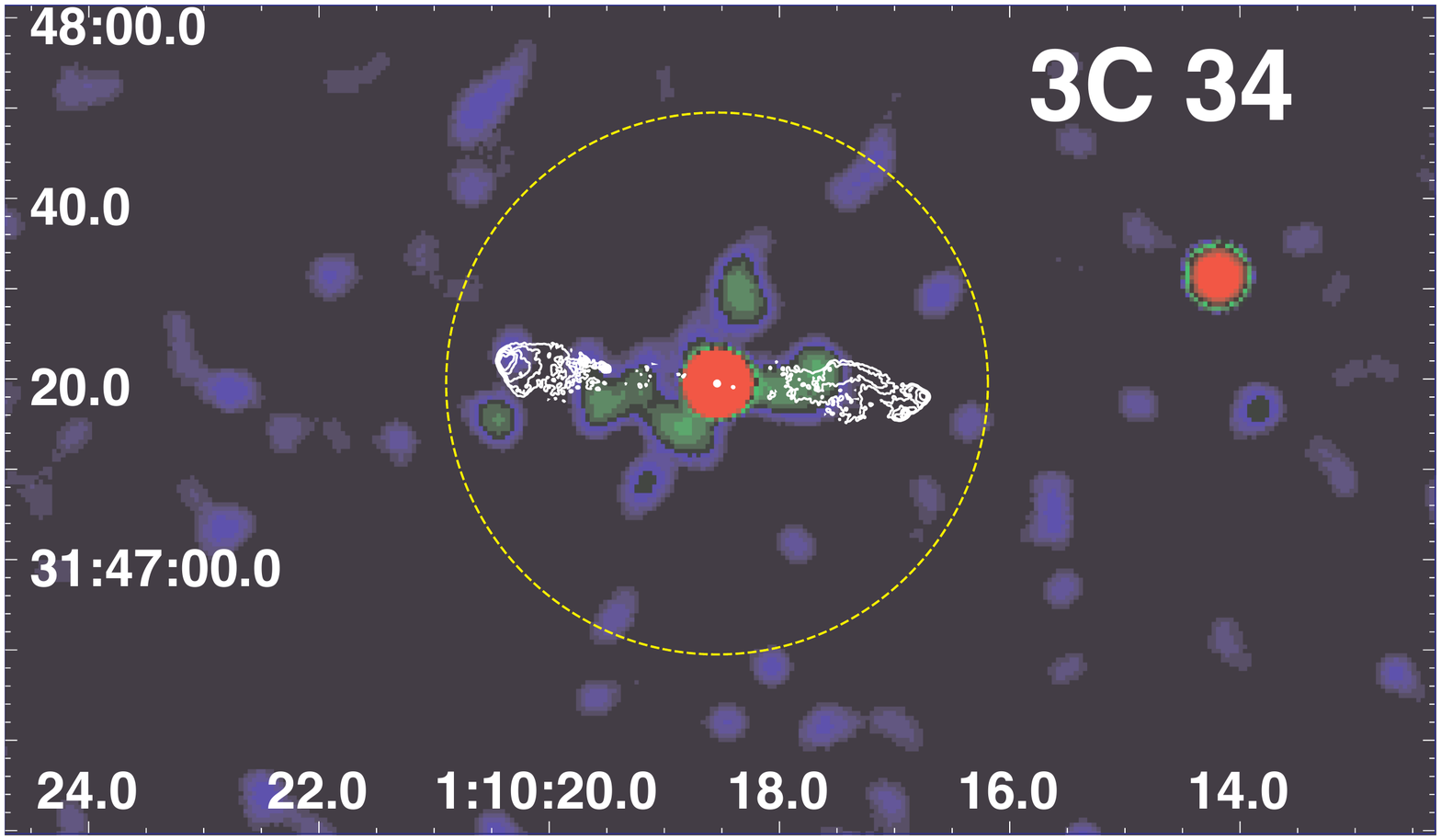}
\includegraphics[height=6.5cm,width=9cm,angle=0]{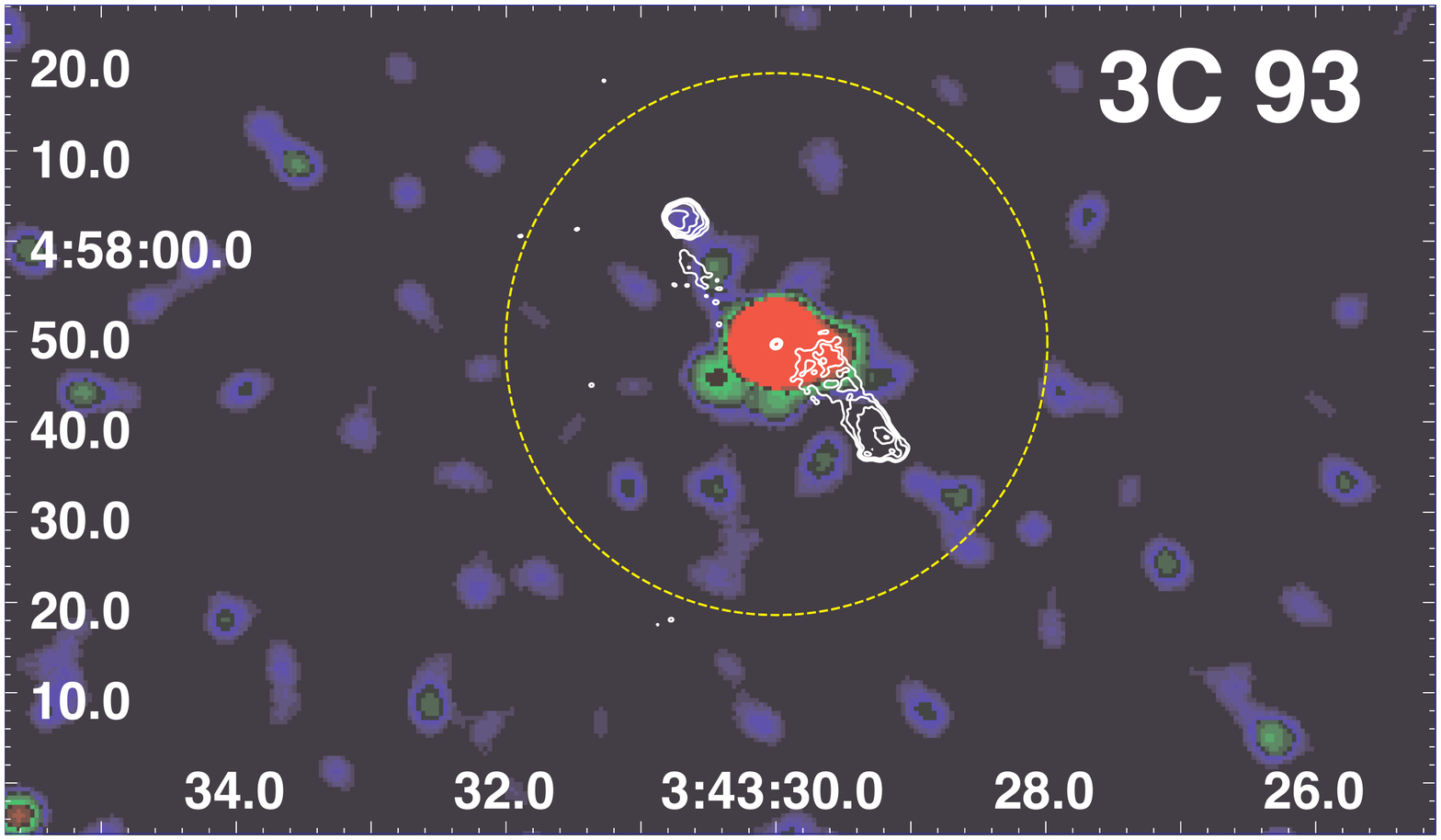}
\includegraphics[height=6.5cm,width=9cm,angle=0]{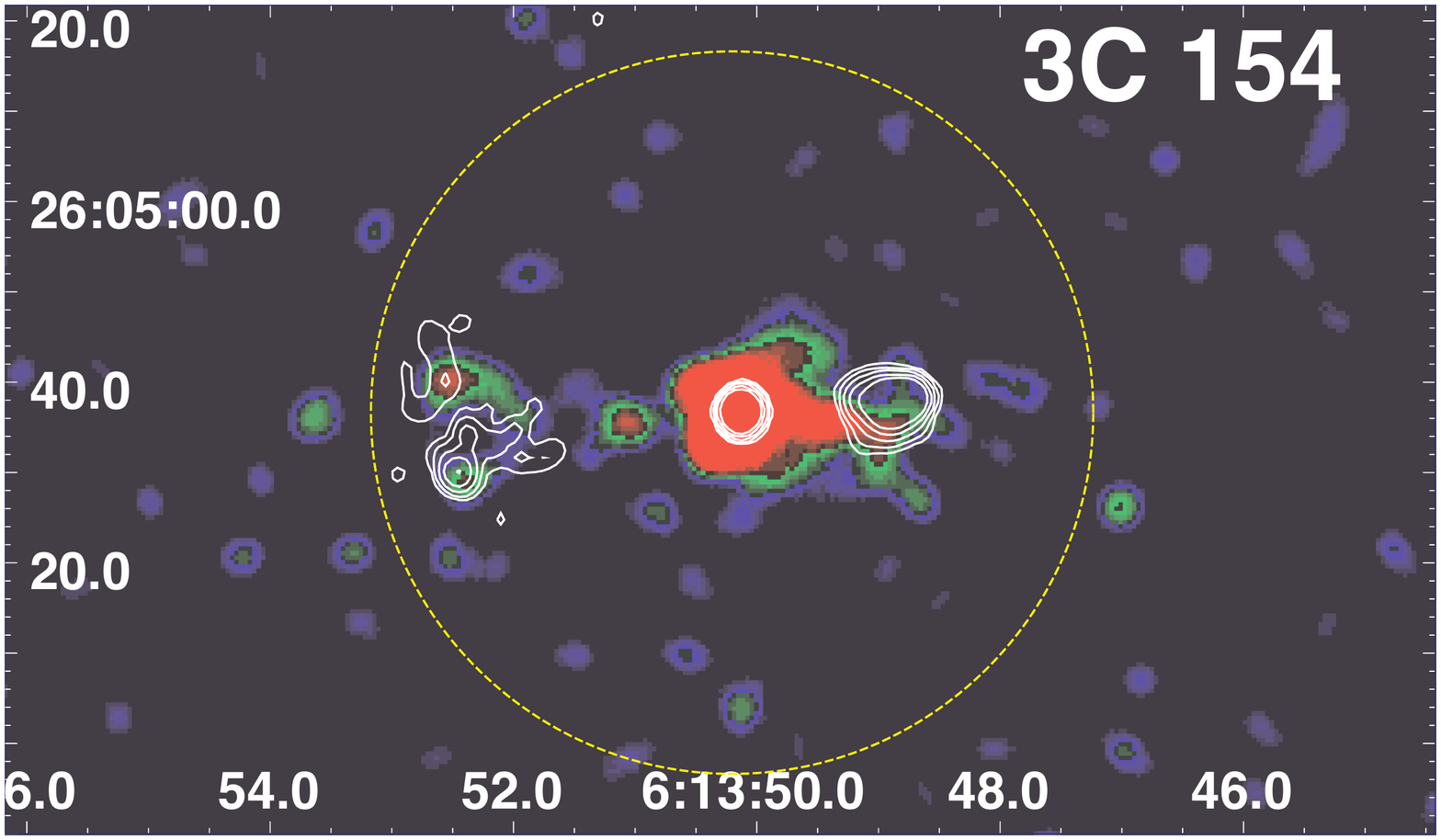}
\includegraphics[height=6.5cm,width=9cm,angle=0]{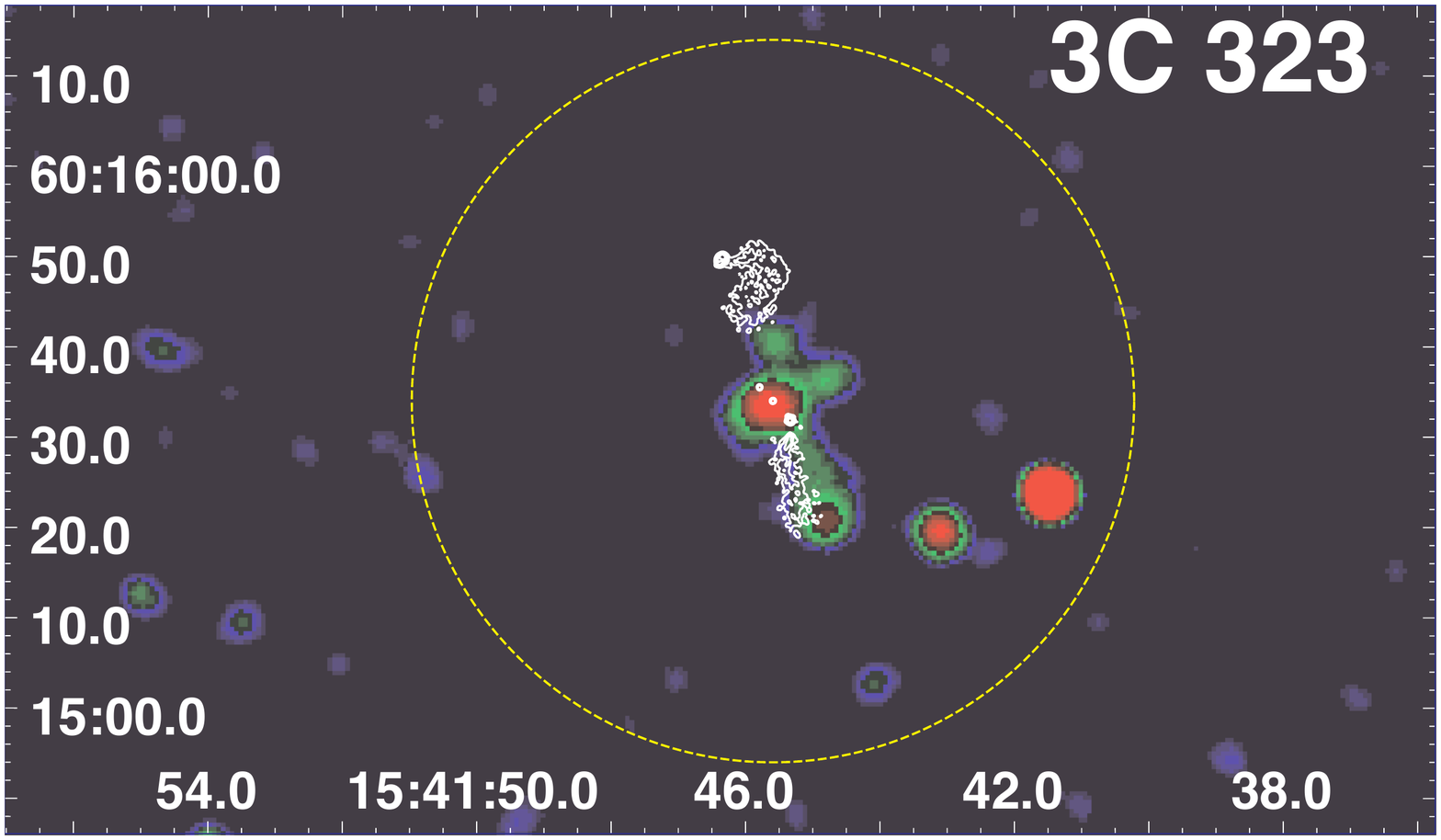}
\caption{{The \chn\ X-ray images for 3CR\,34 (top left), 3CR\,93 (top right), 3CR\,154 (bottom left) and 3CR\,323 (bottom right) in the energy range 0.5--7 keV. Event files have not been regridded. Images are all smoothed with a Gaussian function of 8 pixels kernel radius. The five radio contour levels (white) overlaid on the \chn\ image were computed starting at 0.2, 0.4, 2 and 0.3 mJy/beam, increasing by a factor of 2, respectively. Radio maps adopted to perform the comparison with the X-ray images are the same used for the registration and shown in \S~\ref{sec:images}. A circular region of similar size to the radio structure, centered at the position of the radio cores and having radii: 30\arcsec\ for 3CR\, 34 and 3CR\,93 and 40\arcsec\ for 3CR\,154 and 3CR\,323, respectively, is marked with a dashed yellow line. An excess of X-ray counts, grater than 2$\sigma$ for 3CR\,34 and grater than 3$\sigma$ for the others, is detected. This could be due to the presence of hot gas filling their large-scale X-ray environment but it could also be contaminated by the radiation arising from the lobes of these FR\,II radio galaxies being aligned with their large-scale radio structure.}}
\label{fig:clusters}
\end{figure*}

\begin{figure}
\includegraphics[height=5.5cm,width=8.5cm,angle=0]{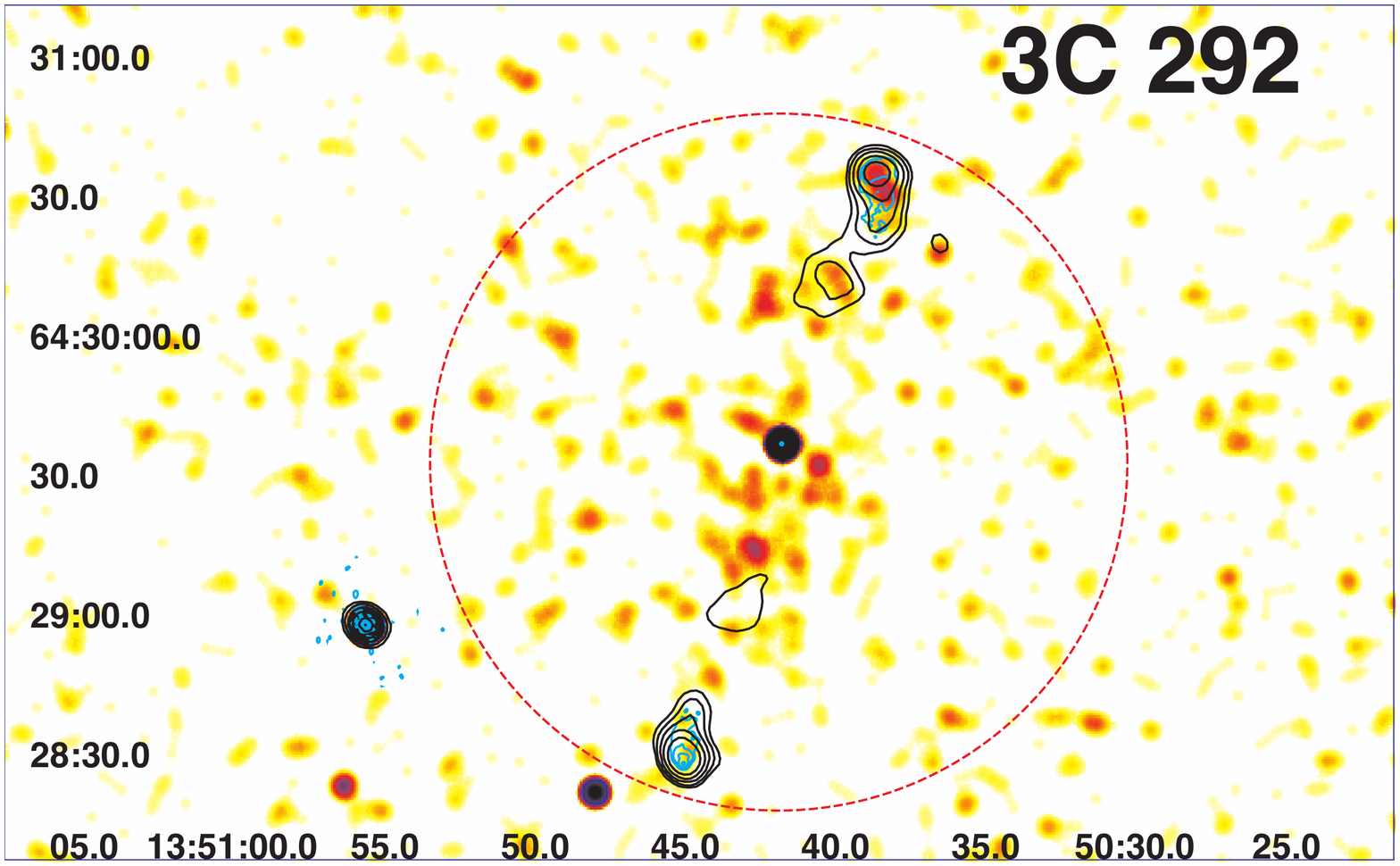}
\caption{{The \chn\ X-ray image for 3CR\,292 in the energy range 0.5--7 keV. The merged event file has been created combining both observations available in the archive (obsID 16065 and obsID 17488, $\sim$4ksec and $\sim$8ksec exposure time, respectively). X-ray image has been smoothed with a Gaussian function of 8 pixels kernel radius. The five radio contour levels, black at 1.4 GHz and cyan at 8.5 GHz, overlaid on the \chn\ image were computed starting at 8 mJy/beam and 0.4, increasing by a factor of 2 and 4, respectively. The radio map at 1.4 GHz is not registered because the radio core is undetected.}}
\label{fig:3c292cluster}
\end{figure}

{We detected an excess of X-ray photons, above 3$\sigma$ significance, around the radio structures of 3CR\,93, 3CR\,154 and 3CR\,323, on a scale of few hundreds kpc as shown in Fig.~\ref{fig:clusters}. Such X-ray extended emission could be due to the presence of hot gas filling their large-scale X-ray environment (see Fig.~\ref{fig:clusters}) but it could also be contaminated by the radiation arising from the lobes of these FR\,II radio galaxies which would be aligned with their large-scale radio structure. These scenarios are indistinguishable due to the low number of counts available, however we tend to favor the former (i.e. presence of hot gas in the intergalactic medium) since the peaks of the X-ray surface brightness do not appear coincident with those of the radio intensity (i.e., hotspots in the lobes). We also found a marginal detection (i.e. 2$\sigma$ significance) of X-ray emission from the group/cluster around 3CR\,34 (see also Fig.~\ref{fig:clusters}). For this source we do not claim that the extended X-ray emission could be associated with the radio lobes because it lies in a galaxy-rich large scale environment \citep{mccarthy95}. Then we discovered extended X-ray emission also for 3CR\,292 on a hundreds of kpc scale (see Fig.~\ref{fig:3c292cluster}) above 3$\sigma$ significance. Here an excess of X-ray photons is also found associated with the northern hotspot/lobe but it is consistent with a 3$\sigma$ fluctuation of the X-ray diffuse background (more details are given in the following section). A specific analysis to search for galaxy groups or cluster signatures surrounding the 3CR sources observed during the \chn\ snapshot survey is out of the scope of the paper and it will be presented in a forthcoming work.}

Finally, we remark that X-ray images with radio contours overlaid for all the sources in the current sample are shown in the Appendix (\S~\ref{sec:images}).

\subsection{Source details}

\textit{3CR\,34}. This is an FR\,II radio galaxy at $z$=0.69 and optically classified as a high-excitation line radio galaxy \citep[HERG;][]{mullin06}. It lies near the center of a compact cluster of galaxies \citep{mccarthy95} appearing as one of the reddest members and being surrounded by fainter companions. There is a double hotspot in the western lobe. Best et al. (1997) reported the detection of a strong jet-cloud interaction at 120 kpc distance from the core and HST images also show a narrow region of blue emission orientated along the radio axis and directed towards a radio hotspot. In the \chn\ observation analyzed here we only detected the X-ray core. We tested the presence of diffuse X-ray emission due to the intergalactic medium of the known galaxy cluster where 3C 34 lies {according to the procedure previously described}. We found an X-ray excess of about 3$\sigma$ significance (see Fig.~\ref{fig:clusters}) where the number of counts in an annular region of radii 2\arcsec and 30\arcsec, centered on the radio core position, is 90, {compared with 45.7 expected background counts in that region}. {We also estimated the luminosity of the X-ray extended emission, $L_X$, adopting an annular region of inner radius 2\arcsec and outer radius 30\arcsec, centered on the position of the 3CR\,34 radio nucleus, to exclude the nuclear contribution of the central radio galaxy. Our estimate is $L_X=$(5.2$\pm$2.2)$\cdot$10$^{43}$erg/s. We note that this $L_X$ estimate, as in the following cases of 3CR\,93, 3CR\,154, 3CR\,292 and 3CR\,323, is an upper limit on the X-ray luminosity of the hot gas in the intergalactic medium since there could be a possible contamination due to the lobe X-ray emission.}

\textit{3CR\,41}. This is an FR\,II radio galaxy with $z=$0.79, classified as a narrow emission line radio galaxy. In the radio image at 8.45 GHz there is no detection of a jet in the northern lobe while extended radio emission associated with the jet pointing towards the south-eastern hotspot is detected \citep[see e.g.]{mullin08}. Both radio lobes {are also visible even} at 8.45 GHz. At higher energies, in our \chn\ snapshot observations, the southern hotspot is detected {as well as the core of the radio galaxy}.

\textit{3CR\,44}. This quasar, with a radio morphology {similar to that of an FR\,II radio galaxy}, is at $z$=0.66 {being} associated with a galaxy cluster visible in the optical image\citep{odea09}. Kharb et al. (2008) reported a hint of a jet-like structure extending towards the southern hotspot in the 5GHz {radio map} while the HST image shows this radio galaxy to be either composed of two {structures} oriented north-south or, more likely, bisected by a dust lane running east-west \citep{mccarthy97}. We clearly detected the nuclear emission in the X-rays and we also found an excess of X-ray counts associated with the northern lobe but at $<$3$\sigma$ significance.

\textit{3CR\,54}. This is an FR\,II radio galaxy at redshift $z$=0.8274. The lobe morphology on the southern side of the source is more extended than the northern one \citep{kharb08}. In the HST image, the galaxy has close companions and the source appears extended towards the southwest side but it is unclear if it is a bridge, a tidal tail, or a jet feature \citep{mccarthy97}. In the X-ray image the radio core is clearly detected together with the southern hotspot.

\textit{3CR\,55}. This is an FR\,II radio galaxy at $z$=0.735 and {optically classified} as a narrow emission line radio galaxy. {The radio core is detected in the X-rays while the two hotspots are not.}

\textit{3CR\,93}. This {is a $z$=0.358 quasar with a lobe-dominated radio} morphology \citep[see e.g.][]{bogers94}. In the optical image, {available in the HST archive,} 3CR\,93 has a host galaxy with $\sim$3\arcsec\ diameter \citep{lehnert99}. In the X-ray image the core is clearly detected with more than 1000 counts but there is no detection of hotspots. X-ray spectral analysis of the nuclear emission shows a power law spectrum with absorption consistent with the Galactic column density. Significant extended X-ray emission on hundreds of kpc scale was found around 3CR\,93 with a significance above 3$\sigma$ (see Fig.~\ref{fig:clusters}). The number of counts in an annular region of radii 2\arcsec and 30\arcsec, centered on the radio core position, is 114 while those expected for the same region in the background is 45.4. {The X-ray luminosity of the extended emission, $L_X$, estimated adopting an annular region of inner radius 2\arcsec and outer radius 30\arcsec, centered on the location of 3CR\,93, as previously done for 3CR\,34, is $L_X=$(2.3$\pm$0.5)$\cdot$10$^{43}$erg/s.}

\begin{figure}
\includegraphics[height=6.5cm,width=9.cm,angle=0]{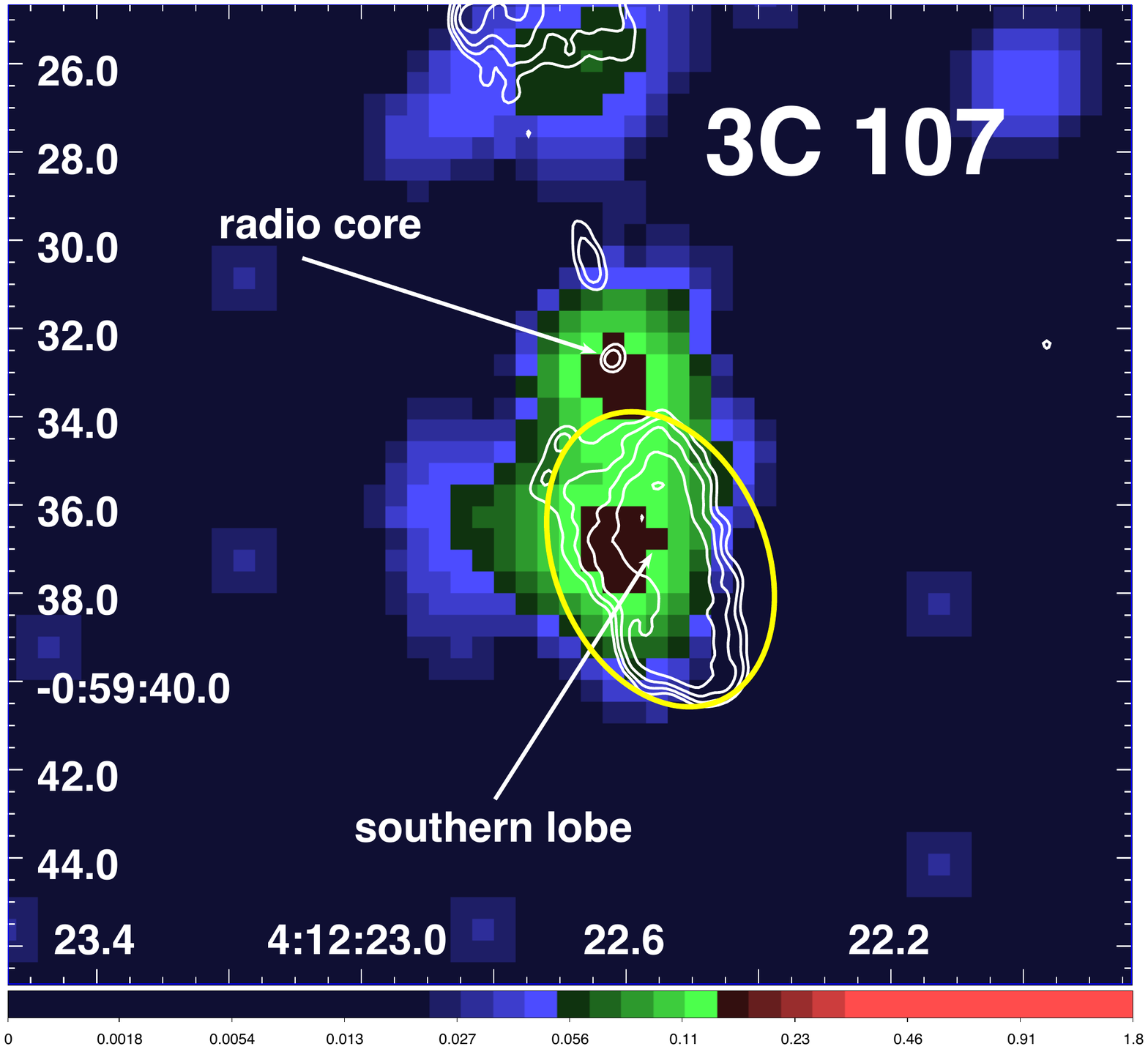}
\caption{{The \chn\ X-ray image for 3CR\,107, centered on its southern lobe, in the energy range 0.5--7 keV. The image has been smoothed with a Gaussian function of 5 pixels kernel radius. Level radio contours (white) overlaid on the \chn\ image were computed starting at 0.2 mJy/beam, increasing by a factor of 2 and drawn using the same 4.9 GHz radio map adopted for the registration. The X ray flux of the southern lobe was measured using the yellow elliptical region drawn in the figure. Here there is spatial coincidence between the excess of X-ray photons and the lobe radio structure.}}
\label{fig:3c107lobe}
\end{figure}

\textit{3CR\,107}. {This FR\,II - HERG radio galaxy lies at $z$=0.785 \citep{mccarthy97}. In our \chn\ snapshot observation we clearly detected the nuclear emission as well as X-ray extended emission associated with the southern lobe (see Fig.~\ref{fig:3c107lobe} and Table~\ref{tab:features}), as also indicated by the low value of the ``ext ratio'' in Table~\ref{tab:cores}.}

\textit{3CR\,114}. This FR\,II radio galaxy is at $z$=0.815 showing fairly weak emission lines in its optical spectrum that led to a LERG classification \citep{strom90}. A faint compact nucleus with several clumps within the few arcseconds is observed at radio frequencies and a jet-like feature appears on the northern side in the 1.4GHz image \citep{kharb08} while single hotspots are detected both in the southern and in the northern lobes. Storm et al. (1990) claimed that this radio galaxy could lie in the core of a rich, distant cluster with some signatures of a merger on the basis of their optical observations. In the X-ray image the nucleus is detected but there is no evidence of hotspots and no signatures of X-ray emission from intracluster medium was found.

\textit{3CR\,142.1}. This double radio source (i.e., FR\,II morphology) at $z=$0.4061. In the radio there are two clear extended radio lobes with a radio bridge showing a constant spectral index \cite{kharb08,odea09}. In the X-ray image only the radio nucleus is detected.

\begin{figure}
\includegraphics[height=5.5cm,width=9cm,angle=0]{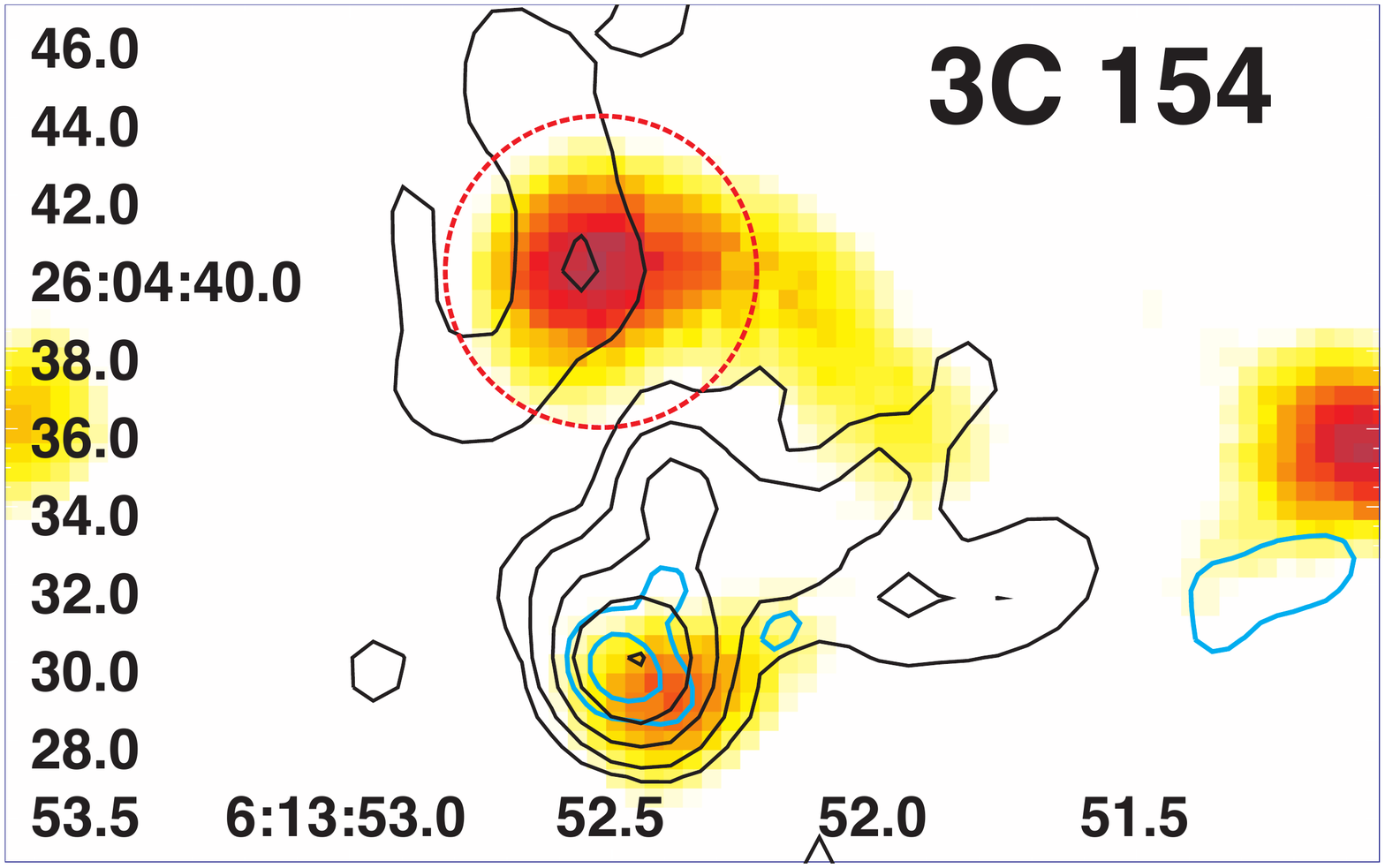}
\caption{{The \chn\ X-ray image for 3CR\,154, centered on its eastern lobe, in the energy range 0.5--7 keV. The image has been smoothed with a Gaussian function of 8 pixels kernel radius. Radio contour levels, overlaid on the \chn\ image, were computed using radio maps at 1.4 and at 8.46 GHz, starting at 1 and 2 mJy/beam and increasing by a factor of 2 and 4, respectively. A circular region of 4\arcsec\ radius, marked with a red dashed line, was used to measure the X-ray flux of the eastern knot. The southern radio knot, in the eastern lobe, has an offset of 1.7\arcsec\ between the peak of the radio surface brightness and the X-ray one. This second excess of X-ray photons could be due to diffuse hot gas in the large scale environment.}}
\label{fig:3c154knot}
\end{figure}

\textit{3CR\,154}. This is a nearby {lobe-dominated quasar} \citep{bogers94} at $z$=0.58 \citep{sokolovsky11} {appearing as a point-like} source in the HST optical image. In {the \chn\ snapshot observation only} the relatively {bright radio core} is detected, for which {X-ray spectral analysis was performed}. As in 3CR\,34 and 3CR\,93, 3CR\,154 shows extended X-ray emission on kpc scale detected above 5$\sigma$ level of confidence, {measured adopting the same method previously described (see Fig.~\ref{fig:clusters})}. We measured 243 X-ray photons in a annular region of radii 2\arcsec and 40\arcsec, centered on the location of the radio core, while 70.1 are expected in the background region. {This X-ray extended emission has a luminosity $L_X=$(2.2$\pm$0.3)$\cdot$10$^{44}$erg/s, measured using an annular region of inner radius 2\arcsec and outer radius 40\arcsec, centered on the location of radio nucleus of 3CR\,154. In the eastern lobe we found an association between the peak of the radio and the X-ray intensities for a radio knot lying at 33\arcsec\ from the nucleus (see marked region in Fig.~\ref{fig:3c154knot}). Given this spatial coincidence, this excess of X-ray photons is probably due to the high energy counterpart of the radio extended structure. On the other hand, the southern knot in the same eastern lobe does not show a correspondence between the radio and the X-ray emissions ($\sim$1.5\arcsec\ offset), indicating that this high energy emission could be linked to the hot gas in the intergalactic medium.}

\textit{3CR\,169.1}. This classical FR\,II radio galaxy is at $z$=0.633. There is no detection of the jet {and/or of the hotspots in the radio map at} 8 GHz \cite{kharb08}. Harvanek et al. (1998) classified this source as a narrow emission line radio galaxy according to its optical spectrum. In the X-ray snapshot image the  nucleus is {detected but there is no detection of the extended radio structure}.

\textit{3CR\,217}. This is a FR\,II radio galaxy with narrow emission lines at $z$=0.898. {In the radio image at} 8 GHz there are no jet signatures, in either the eastern or the western lobe \citep{mullin08} and only the western lobe is detected. In the X-ray band the radio core is  detected but the \chn\ image is not registered because {we were not able to locate precisely its position}. Only a single photon is associated with the western hotspot {being undetected in the X-rays.}

\textit{3CR\,225B}. This is a FR\,II radio galaxy at $z$=0.582, optically classified as a narrow emission line galaxy. It does show only hotspots within its radio lobes \cite{mullin08}. In the X-ray there is a hint of diffuse emission near the nucleus but no detection for the two hotspots. We note that the nearby radio source 3CR\,225A is also in the field of view of the \chn\ snapshot observation on a nearby CCD, but we did not detect any signature of X-ray emission arising from this radio object.

\textit{3CR\,237}.This is the only compact steep spectrum (CSS) radio source in our sample. It shows an FR\,II radio morphology and is at $z$=0.877. The radio source size is less than 2\arcsec\ {(i.e., less than $\sim$15 kpc at $z=$0.877)} and it has a clear X-ray counterpart.

\textit{3CR\,247}. This is an FR\,II radio galaxy at $z$=0.75, {showing two hotspots on both the eastern and the western side}, optically classified as a narrow emission line radio galaxy. The host galaxy associated with the radio source 3C247 lies in a very crowded field. {In the HST optical image it appears as a symmetrical central galaxy} with a close companion lying about 0.8 arcsec to the south. Two nearby galaxies also lie within the envelope of the infrared emission \citep[see e.g.][]{best98,best00}. Radio spectral ageing analysis of this source has yielded an age of 3-5 million years, corresponding to a hotspot advance velocity of about 0.1c \citep{liu92}. In the \chn\ X-ray image the radio core and the northern hotspot are detected.

\textit{3CR\,272}. A classical double FR\,II radio source at $z$=0.944, showing a relatively faint nuclear component with respect to the two hotspots at radio frequencies. {Its optical spectrum shows a typical HERG spectrum} with high ionization emission lines \citep{strom90}. There is nothing to report in the \chn\ snapshot image other than the core detection.

\textit{3CR\,277}. This is a giant FR\,II radio galaxy at $z$=0.414 \citep{strom90}, optically classified as a low excitation type radio galaxy (i.e., LERG). At 1.4 GHz the core and a jet-like structure are detected in the eastern lobe. Both radio lobes show double hotspots in high resolution radio images \citep[see e.g.,][]{harvanek98}. According to McCarthy et al. (1997) the source reveals a basic double morphology with the nuclear component slightly offset from the geometrical center. In the X-ray image the {radio core} is detected while {while neither double hotspot has an X-ray counterpart}.

\textit{3C\,277.2}. This FR\,II radio galaxy at $z$=0.767 is optically classified as a narrow line emission galaxy. It has a clear jet-like component in the southern lobe that is closer to the radio core but no signatures in the X-rays.

\textit{3CR\,288.1}. This is a lobe-dominated QSO lying at $z$=0.961 having a clear point-like optical counterpart \citep{gendre08}. In the X-ray image the nucleus is clearly detected and given the high number of photons with respect to the majority of the other targets, we also performed its spectral analysis finding its X-ray spectrum consistent with an absorbed power-law (see Table~\ref{tab:spectra}). Extended emission on kpc scale seems to be present around the core (see Table~\ref{tab:cores}) but no evidence of a galaxy cluster or a group of galaxies is present in the optical images.

\begin{figure}
\includegraphics[height=6.5cm,width=9.cm,angle=0]{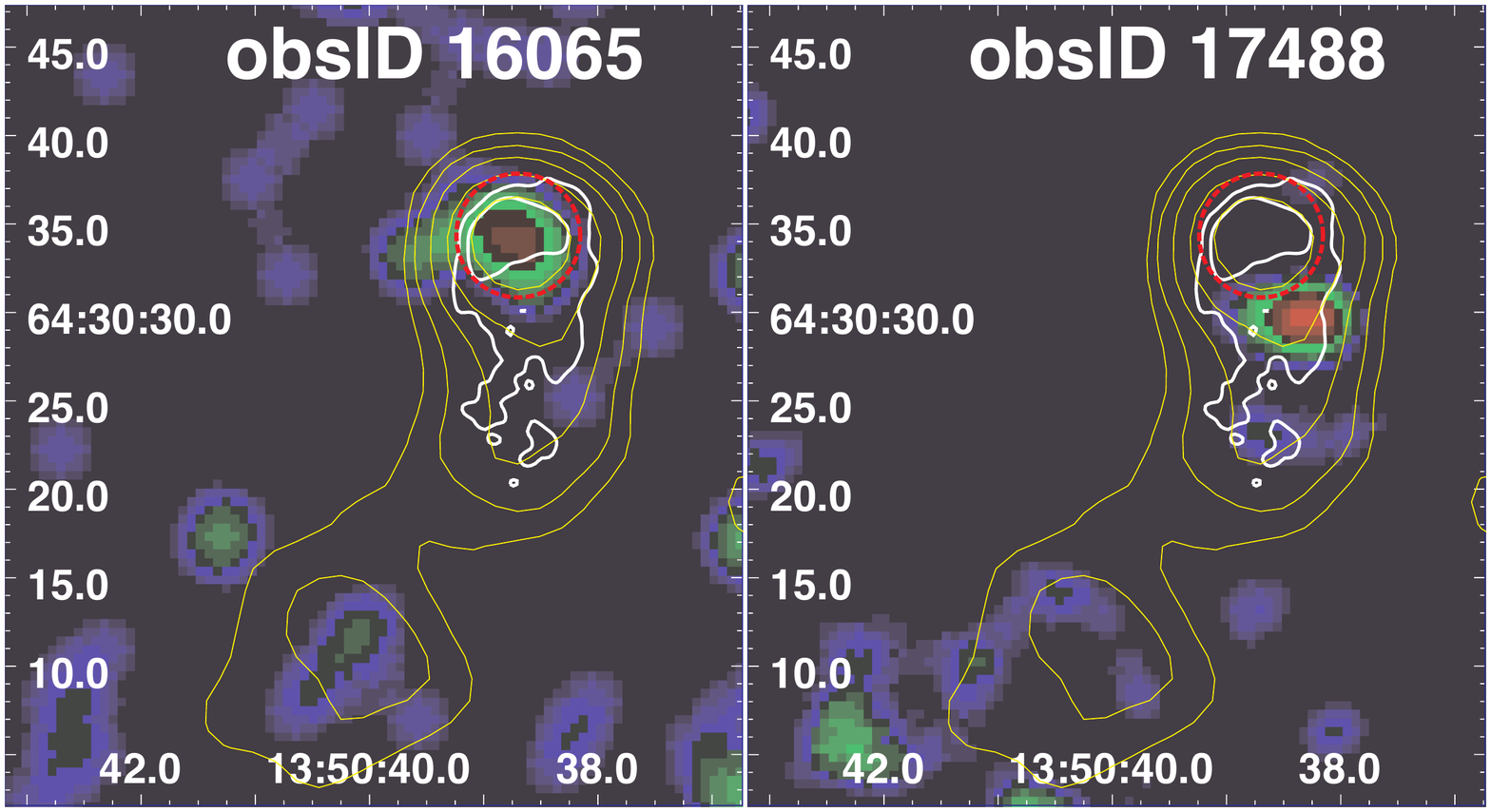}
\caption{{The \chn\ X-ray images for 3CR\,292, centered on its northern lobe, in the energy range 0.5--7 keV. The left image corresponds to the \chn\ observation with shorter exposure time (obsID 16065, $\sim$4ksec) while the right panel to the deeper one $(obsID 17488 \sim$8ksec of exposure time). Both images have been smoothed with a Gaussian function of 8 pixels kernel radius. The five radio contours levels, yellow at 1.4 GHz and white at 8.5 GHz, overlaid on the \chn\ image were computed starting at 8 and 0.4mJy/beam, increasing by a factor of 2 and 4, respectively. An elliptical region, marked with a red dashed line, was used to compute the X-ray flux of the northern lobe and it is also shown.}}
\label{fig:3c292lobe}
\end{figure}

{\textit{3CR\,292}. A narrow line FR\,II radio galaxy at redshift 0.713. Mullin et al. (2006) reported the detection of the two hotspots and the core in the radio map at 8.45 GHz but a part from the core none of these features is detected in the X-rays. Adopting the same procedure as in previous cases we found that 3CR\,292 shows extended X-ray emission (more than 3$\sigma$ detection significance) detected on kpc scale and potentially due to the presence of hot gas in the intergalactic medium (see Fig.~\ref{fig:clusters}). The number of X-ray photons measured in a annular region of radii 2\arcsec and 75\arcsec, centered on the radio core position, is 525 while those expected in the background, for the same area, is 364.3, where both X-ray counts have been measured using the merged event file. We estimated an X-ray luminosity for the extended emission of $L_X=$(2.5$\pm$0.7)$\cdot$10$^{44}$erg/s, using the same annular region. Belsole et al. (2004) reported the X-ray detection of radio lobes in an \xmm\ observation and as previously stated in the merged event file we clearly detected an excess of X-ray photons correspondent to the location of the northern hotspot. However, as shown in Fig.~\ref{fig:3c292lobe}, in a circular region of 3.5\arcsec\ radius, centered on the northern hotspot, we measured 5 X-ray counts in the $\sim$4ksec \chn\ observation (i.e. obsID 16065) but only 1 in the deeper one (i.e. obsID 17488, $\sim$8ksec exposure time). Then, in the latter observation, the peak of the radio surface brightness is not coincident with the X-ray excess. Since this radio structure lies in a large scale environment permeated by X-ray emission from the hot intergalactic medium, we measured the level fo the X-ray background of the $\sim$4ksec observation within 75\arcsec\ from the radio core and excluding the nuclear X-ray emission. We found that, over the same area (i.e., circle of 3.5\arcsec\ radius), the expected number of counts is 1.5$\pm$1.2 (obsID 16065), thus this X-ray excess is consistent with a 3$\sigma$ fluctuation of the diffuse X-ray background. A deeper investigation will be necessary to distinguish between X-ray emission associated with the hotspot or that of the intergalactic medium. Variability on monthly time scale has been excluded because the hotspot size, even if being relatively compact with respect to the lobe structure, covers a region of $\sim$25 kpc.}

\textit{3CR\,293.1}. This faint FR\,II radio galaxy is at $z$=0.709. {Its radio structures}, i.e, the nucleus and both hotspots, are barely detected in the radio image available to us at 4.9 GHz and the {comparison with the X-ray image indicates} that the radio core is detected with {possible extended X-ray emission around it (see Table~\ref{tab:cores})}.

\textit{3CR\, 323}. This is a FR\,II radio galaxy at $z$=0.679. McCarthy et al. (1997) pointed out that the outer lobes are irregular and rather different in structure, the northern one undergoing a sharp bend to the east and ending in what may be a hotspot. In the X-ray we only detected the nucleus and there is also significant X-ray extended emission surrounding it both on scales of few tens and $\sim$300 kpc (above 5$\sigma$ significance), potentially due to its intergalactic medium (see Fig.~\ref{fig:clusters}). There are 329 X-ray photons in a annular region of radii 2\arcsec and 75\arcsec, centered on the radio core position, while 269.3 background photons are expected in this region. In this case the number of X-ray counts within a circular region of 2\arcsec\ positioned on two point-like sources lying on the western side of 3CR\,323 were also subtracted from the X-ray photons measured within the annular region. This does not affect the significance of the detection of the X-ray extended emission since only 8 and 20 photons were measured for the two nearby objects.  {We estimated an X-ray luminosity of $L_X=$(4.7$\pm$1.6)$\cdot$10$^{43}$erg/s for the extended emission surrounding 3CR\,323, this was measured over an annular region of inner radius 2\arcsec and outer radius 21\arcsec, centered on the location of its radio core. A lower value of the outer radius for the annular region has been chosen to avoid contamination due to the presence of nearby foreground/background X-ray point-like sources.}

\section{Summary and Conclusions}
\label{sec:summary}
This paper presents the analysis of \chn\ snapshot observations of a subsample of the extragalactic sources listed in the revised Third Cambridge radio catalog (3CR), previously lacking \chn\ observations and observed during Cycle 15. This data set extends the current \chn\ coverage of the 3CR extragalactic catalog up to redshift $z$=1.0. {The 3CR extragalactic sample} includes 22 sources listing one compact steep spectrum (CSS) source and three quasars (QSOs) while all the remaining sources are FR\,II radio galaxies. Nineteen targets lie at $z$ in the range 0.5--1.0 plus 3C\,93, 3CR\,142.1 and 3CR\,277. One additional target, 3CR\,255A, lies in the \chn\ field of view of a nearby source (i.e., 3CR\,255B) observed during these Cycle 15 observations, but its radio core is not detected in the X-rays.
 
The main aims of the 3CR \chn\ snapshot survey are: (i) to search for X-ray emission from jet knots, hotspots and lobes, (ii) to study the nuclear emission of their host galaxies and (iii) to {investigate} their environments at all scales, aiming to discover new galaxy {groups/clusters via the X-ray emission of the intergalatic medium}. 

In the present work, the basic source parameters for the newly acquired {\chn\ data are presented. We created fluxmaps for all the X-ray snapshot} observations and compared them with radio images to search for the high energy counterparts of extended radio structures (i.e., jet knots, hotspots, lobes). We measured their X-ray intensities in three energy ranges, {namely soft, medium and hard band, for all radio cores and hotspots detected in the X-rays}. Then, for the nuclei brighter than 400 X-ray photons, measured in a circular region of 2\arcsec\ radius in the 0.5--7 keV energy range, we also performed X-ray spectral analysis {showing nuclear spectra all consistent with simple power-law model, with an eventual mild intrinsic absorption in a single case.}

We found X-ray emission arising from three hotspots in 3CR\,41, 3CR\,54 and 3CR\,225B. We also report the discovery of extended X-ray emission, on tens of kpc scale, around the radio nuclei of 3CR\,107 and 3CR\,293.1, {for the former source due to the X-ray counterpart of the southern radio lobe}. Three sources in our sample are members of, optically-known galaxy groups/clusters: 3CR\,41, 3CR\,44 and 3CR\,247. In the last two cases we did not detect X-ray emission arising from the intergalactic medium while a marginal detection (i.e., 2$\sigma$) was found for 3CR\,34. Moreover, we discovered extended X-ray emission on a scale of few hundreds of kpc around the radio structures of 3CR\,93, 3CR\,154, 3CR\,292 and 3CR\, 323, all above 3$\sigma$ significance. {Then, for 3CR\,154, we also detected the X-ray counterpart of a knot in the eastern radio lobe at 33\arcsec\ distance from the nucleus.}

{Finally, we highlight that a table summarizing the state-of-the-art of the X-ray (i.e., \chn\ and \xmm) observations carried out to date is reported at the end of the present manuscript (see \S~\ref{sec:state}). \chn\ detections are all based on both our current and previous analysis and represent an update with respect to previous works while those regarding \xmm, shown here for the first time, are only based on literature search.}

\acknowledgments 
We thank the anonymous referee for useful comments that led to improvements in the paper.
We are grateful to M. Hardcastle and C. C. Cheung for providing
several radio images of the 3CR sources while the remaining ones were
downloaded from the NVAS\footnote{http://archive.nrao.edu/nvas/} (NRAO
VLA Archive Survey), NED\footnote{http://ned.ipac.caltech.edu/} (Nasa
Extragalactic Database) and from the DRAGN
webpage\footnote{http://www.jb.man.ac.uk/atlas/}.
This investigation is supported by the NASA grants GO1-12125A,
GO2-13115X, and GO4-15097X.  G.R.T. acknowledges support from the National Aeronautics and Space Administration (NASA) through Einstein Postdoctoral Fellowship Award Number PF-150128, issued by the Chandra X-ray Observatory Center, which is operated by the Smithsonian Astrophysical Observatory for and on behalf of NASA under contract NAS8-03060. This work was also supported by
contributions of European Union, Valle D'Aosta Region and the
Italian Minister for Work and Welfare. 
The National Radio Astronomy Observatory is operated by Associated Universities, Inc.,
under contract with the National Science Foundation.
This research has made use of data obtained from the High-Energy Astrophysics Science Archive
Research Center (HEASARC) provided by NASA's Goddard Space Flight Center; 
the SIMBAD database operated at CDS,
Strasbourg, France; the NASA/IPAC Extragalactic Database
(NED) operated by the Jet Propulsion Laboratory, California
Institute of Technology, under contract with the National Aeronautics and Space Administration.
TOPCAT\footnote{\underline{http://www.star.bris.ac.uk/$\sim$mbt/topcat/}} 
\citep{taylor05} for the preparation and manipulation of the tabular data and the images.
SAOImage DS9 were used extensively in this work
for the preparation and manipulation of the
images.  SAOImage DS9 was developed by the Smithsonian Astrophysical
Observatory. 

{Facilities:} \facility{VLA}, \facility{MERLIN}, \facility{CXO (ACIS)}

\clearpage
\begin{table*} 
\caption{Source List of the \chn\ AO15 Snapshot survey}
\label{tab:log}
\tiny
\begin{center}
\begin{tabular}{|llllllllllll|}
\hline
3CR  & Class & R.A. (J2000) & Dec. (J2000) & z & kpc scale    & D$_L$ & N$_H$ & m$_v$ & S$_{178}$ &\chn\   & Obs. Date  \\
name &          & (hh mm ss)   & (dd mm ss)    &   & (kpc/arcsec) & (Mpc)  & cm -2   &            & Jy            & Obs. ID & yyyy-mm-dd \\ 
\hline 
\noalign{\smallskip}
  34 & FR\,II - HERG & 01 10 18.542 & +31 47 19.51 & 0.69 & 7.191 & 4236.3 & 5.50e20 & 21 & 11.9 & 16046 & 2014-09-25 \\
  41 & FR\,II - HERG & 01 26 44.325 & +33 13 10.96 & 0.794 & 7.586 & 5035.7 & 5.09e20 & 21 & 10.6 & 16047 & 2014-09-03 \\
  44 & QSO & 01 31 21.647 & +06 23 43.14 & 0.66 & 7.058 & 4011.6 & 3.18e20 & 22 & 7.9 & 16048 & 2014-06-14 \\
  54 & FR\,II - HERG & 01 55 30.258 & +43 45 59.06 & 0.8274 & 7.693 & 5298.7 & 7.80e20 & 22 & 8.8 & 16049 & 2014-06-15 \\
  55 & FR\,II - HERG & 01 57 10.539 & +28 51 39.70 & 0.735 & 7.374 & 4578.4 & 5.38e20 & 22 & 21.5 & 16050 & 2014-06-15 \\
  93 & QSO & 03 43 29.996 & +04 57 48.60 & 0.358 & 5.059 & 1924.5 & 1.15e21 & 18.1 & 9.9 & 16051 & 2014-10-10 \\
  107 & FR\,II - HERG & 04 12 22.620 & -00 59 32.69 & 0.785 & 7.555 & 4965.3 & 8.42e20 & 22 & 10.8 & 16052 & 2014-09-02 \\
  114 & FR\,II - LERG & 04 20 22.243 & +17 53 56.97 & 0.815 & 7.654 & 5200.7 & 1.61e21 & 22 & 6.5 & 16053 & 2014-09-02 \\
  142.1 & FR\,II & 05 31 29.334 & +06 30 26.90 & 0.4061 & 5.476 & 2233 & 1.79e21 & 21 & 19.4 & 16054 & 2014-08-16 \\
  154 & QSO & 06 13 50.139 & +26 04 36.64 & 0.580 & 6.654 & 3426.3 & 3.47e21 & 18.0 & 23.1 & 16055 & 2014-08-13 \\
  169.1 & FR\,II - HERG & 06 51 14.816 & +45 09 28.56 & 0.633 & 6.930 & 3811.7 & 9.30e20 & 20.5 & 7.3 & 16056 & 2014-08-16 \\
  217 & FR\,II - HERG & 09 08 50.6 & +37 48 19 & 0.898 & 7.892 & 5863.9 & 1.68e20 & 22 & 11.3 & 16057 & 2014-06-13 \\
  225B & FR\,II -HERG & 09 42 15.396 & +13 45 50.49 & 0.582 & 6.665 & 3440.7 &  3.45e20 & 19 & 21.3 & 16058 & 2014-10-18 \\
  237 & FR\,II-CSS & 10 08 00.0 & +07 30 16 & 0.877 & 7.836 & 5694.5 & 1.89e20 & 21 & 20.9 & 16059 & 2014-10-31 \\
  247 & FR\,II - HERG & 10 58 58.973 & +43 01 24.66 & 0.750 & 7.430 & 4693.7 & 8.79e19 & 21.5 & 10.6 & 16060 & 2014-09-26 \\
  272 & FR\,II & 12 24 28.5 & +42 06 36 & 0.944 & 8.003 & 6238.6 & 2.23e20 & 22 & 8 & 16061 & 2015-03-01 \\
  277 & FR\,II & 12 51 43.6 & +50 34 25 & 0.414 & 5.540 & 2284.7 & 1.04e20 & 20 & 7.5 & 16062 & 2015-03-03 \\
  277.2 & FR\,II - HERG & 12 53 33.330 & +15 42 31.18 & 0.767 & 7.492 & 4825.2 & 1.70e20 & 21.5 & 12 & 16063 & 2015-05-07 \\
  288.1 & QSO & 13 42 13.267 & +60 21 42.79 & 0.9610 & 8.041 & 6378.3 & 1.75e20 & 18.1 & 9 & 16064 & 2014-06-08\\
  292 & FR\,II - HERG & 13 50 41.852 & +64 29 35.86 & 0.713 & 7.287 & 4410.4 & 1.66e20 & 20.7 & 10.1 & 16065$^+$ & 2014-09-12\\
  293.1 & FR\,II & 13 54 40.519 & +16 14 43.14 & 0.709 & 7.271 & 4380.0 & 1.82e20 & 21 & 9.2 & 16066 & 2014-06-05 \\
  323 & FR\,II & 15 41 45.594 & +60 15 34.03 & 0.6790 & 7.143 & 4153.6 & 1.49e20 & 21 & 8.4 & 16067 & 2014-04-30 \\
\noalign{\smallskip}
\hline
\end{tabular}\\
\end{center}
Col. (1): The 3CR name.
Col. (2): The `class' column contains both a radio descriptor (Fanaroff-Riley class I or II), Compact Steep Spectrum (CSS) and 
the optical spectroscopic designation, LERG, ``Low Excitation Radio Galaxy'', HERG, ``High Excitation Radio Galaxy'' and QSO for quasars  \citep[see also][for more details]{perryman84,hes96,grimes04}.
Col. (3-4): The celestial positions listed are those of the radio nuclei which we used to register the X-ray images: Right ascension and Declination (equinox J2000, see \S~\ref{sec:obs} for details). We reported here the original 3CR position \citep{spinrad85} of the sources for which the radio core was not clearly detected.
Col. (5): Redshift $z$. We also verified in the literature (e.g., NED and/or SIMBAD databases) if new $z$ values were reported after the release of the 3CR catalog.
Col. (6): The angular to linear scale factor in arcseconds. Cosmological parameters used to compute it are reported in \S~\ref{sec:intro}.
Col. (7): Luminosity Distance in Mpc. Cosmological parameters used to compute it are reported in \S~\ref{sec:intro}. 
Col. (8): Galactic Neutral hydrogen column densities N$_{H,Gal}$ along the line of sight \citep{kalberla05}.
Col. (8): The optical magnitude in the V band taken from the 3CR catalog \citep{spinrad85}.
Col. (9): S$_{178}$ is the flux density at 178 MHz, taken from Spinrad et al. (1985).
Col. (10): The \chn\ observation ID.
Col. (11): The date when the \chn\ observation was performed.\\
{$^+$ For 3CR\,292 two Chandra observations are available with obsID: 16065 and 17488, the latter performed in 2014-11-21, (see \S~\ref{sec:results} for details).}
\end{table*}

\begin{table*} 
\caption{X-ray emission from radio cores.}
\label{tab:cores}
\tiny
\begin{center}
\begin{tabular}{|rrrrrrr|}
\hline
3CR  & Ext. Ratio & F$_{0.5-1~keV}^*$ & F$_{1-2~keV}^*$ & F$_{2-7~keV}^*$ & F$_{0.5-7~keV}^*$ & L$_X$ \\
name &         & (cgs)                 & (cgs)           & (cgs)           & (cgs)             & (10$^{44}$erg~s$^{-1}$) \\
\hline 
\noalign{\smallskip}
34 & 0.72 (0.04) & 0.25 (0.57) & 4.69 (1.43) & 77.53 (9.95) & 82.47 (10.06) & 1.76 (0.22)\\
41 & 0.59 (0.06) & 0.32 (0.32) & 0.94 (0.68) & 41.33 (7.42) & 42.6 (7.46) & 1.29 (0.23)\\
44 & 0.45 (0.09) & 0.82 (0.59) & 0.68 (0.64) & 9.71 (3.67) & 11.22 (3.77) & 0.22 (0.07)\\
54 & 0.83 (0.05) & 1.06 (0.75) & 4.76 (1.33) & 41.04 (7.25) & 46.85 (7.41) & 1.57 (0.25)\\
55 & 0.48 (0.09) & 0.82 (0.58) & 2.41 (0.91) & 8.39 (3.47) & 11.61 (3.64) & 0.29 (0.09)\\
93 & 0.948 (0.006) & 77.48 (6.26) & 157.41 (7.41) & 429.53 (22.58) & 664.42 (24.58) & 2.94 (0.11)\\
107 & 0.15 (0.06) & 0.38 (0.38) & 0.49 (0.49) & 2.95 (2.12) & 3.82 (2.21) & 0.11 (0.07)\\
114 & 0.91 (0.02) & 1.16 (0.82) & 43.23 (4.05) & 255.54 (17.3) & 299.94 (17.78) & 9.71 (0.58)\\
142.1 & 0.61 (0.09) & 0.57 (0.57) & 2.24 (0.91) & 11.19 (3.77) & 13.99 (3.92) & 0.08 (0.02)\\
154 & 0.947 (0.006) & 41.35 (4.59) & 241.5 (9.31) & 923.47 (33.73) & 1206.32 (35.29) & 16.95 (0.5)\\
169.1 & 0.64 (0.08) & 2.1 (0.94) & 0.23 (0.51) & 18.65 (4.98) & 20.98 (5.1) & 0.36 (0.09)\\
217 & 0.91 (0.20) & 1.42 (1.0) & 15.01 (2.29) & 100.01 (10.98) & 116.44 (11.26) & 4.78 (0.46)\\
225B & 0.48 (0.11) & 0.68 (0.48) & 0.12 (0.26) & 5.18 (2.59) & 5.98 (2.65) & 0.08 (0.04)\\
237 & 0.84 (0.05) & 4.11 (1.45) & 7.0 (1.53) & 13.76 (4.01) & 24.87 (4.53) & 0.97 (0.18)\\
247 & 0.7 (0.06) & 0.99 (0.7) & 3.71 (1.25) & 36.22 (6.61) & 40.92 (6.77) & 1.07 (0.18)\\
272 & 0.41 (0.12) & 0.0 (0.0) & 0.46 (0.46) & 4.85 (2.48) & 5.31 (2.52) & 0.25 (0.12)\\
277 & 0.84 (0.06) & 0.33 (0.33) & 2.2 (1.0) & 26.81 (6.0) & 29.34 (6.09) & 0.18 (0.04)\\
277.2 & 0.43 (0.10) & 3.15 (1.29) & 0.21 (0.47) & 2.26 (2.26) & 5.63 (2.65) & 0.16 (0.07)\\
288.1 & 0.946 (0.08) & 76.35 (5.65) & 114.05 (6.1) & 236.49 (16.45) & 426.88 (18.43) & 20.78 (0.9)\\
292 & 0.86 (0.04) & 0.9 (0.9) & 1.8 (1.04) & 103.58 (14.13) & 106.29 (14.19) & 2.47 (0.33)\\
293.1 & 0.26 (0.10) & 0.0 (0.0) & 0.84 (0.6) & 1.76 (1.24) & 2.59 (1.38) & 0.06 (0.03)\\
323 & 0.18 (0.06) & 0.0 (0.0) & 0.9 (0.65) & 5.03 (2.25) & 5.93 (2.34) & 0.12 (0.05)\\
\noalign{\smallskip}
\hline
\end{tabular}\\
\end{center}
Col. (1): The 3CR name.
Col. (2): The Ext. Ratio defined as the ratio of the net counts in the r\,=\,2\arcsec\ circle to the net counts in the
r\,=\,10\arcsec\ circular region surrounding the core of each 3CR source. The 1$\sigma$ uncertainties are given in parenthesis.
Col. (3): Measured X-ray flux between 0.5 and 1 keV.
Col. (4): Measured X-ray flux between 1 and 2 keV.
Col. (5): Measured X-ray flux between 2 and 7 keV.
Col. (6): Measured X-ray flux between 0.5 and 7 keV.
Col. (7): X-ray luminosity in the range 0.5 to 7 keV with the 1$\sigma$ uncertainties given in parenthesis.\\
Note:\\
($^*$) Fluxes are given in units of 10$^{-15}$erg~cm$^{-2}$s$^{-1}$ and 1$\sigma$ uncertainties are given in parenthesis. 
The uncertainties on the flux measurements are computed as described in \S~\ref{sec:obs}\\
($^+$) Sources having count rates above the threshold of 0.1 counts per frame 
for which the X-ray flux measurement is affected by pileup \citep[see][and references therein for additional details]{massaro13}.
\end{table*}

\begin{table*} 
\caption{Results of the X-ray spectral analysis for the brighter nuclei.}
\label{tab:spectra}
\tiny
\begin{center}
\begin{tabular}{|lllll|}
\hline
3CR & $\Gamma_X$ & $N_{H,int}$ & \textsc{f} & $\chi^2/dof$ \\
\hline 
\noalign{\smallskip}
93 & 1.78(-0.08,0.15) & $<$0.06 & $<$0.91 & 34.91/25 \\
114 & 2.06(-0.34,0.36) & 7.21(-2.4,2.54) & -& 3.91/7 \\
154 & 1.85(-0.14,0.18) & 0.61(-0.26,0.28) & $<$0.92 & 41.52/39 \\
288.1 & 1.77(-0.08,0.09) & $<$0.1 & - &16.99/19 \\
\noalign{\smallskip}
\hline
\end{tabular}\\
\end{center}
Col. (1): The 3CR name.
Col. (2): The X-ray photon index $\Gamma_X$.
Col. (3): The intrinsic absorption at the source redshift.
Col. (4): The fraction of flux falling into the pileup region.
Col. (5): The $\chi^2$ value divided by the degrees of freedom.
Statistical uncertainties quoted refer to the 68\% confidence level.
\end{table*}

\begin{table*} 
\caption{X-ray emission from radio extended structures (i.e., knots and hotspots).}
\label{tab:features}
\tiny
\begin{center}
\begin{tabular}{|rrrrrrrrr|}
\hline
3CR  & Component & class & Counts & F$_{0.5-1~keV}^*$ & F$_{1-2~keV}^*$ & F$_{2-7~keV}^*$ & F$_{0.5-7~keV}^*$ & L$_X$ \\
name &        & & & (cgs)                 & (cgs)           & (cgs)           & (cgs)             & (10$^{42}$erg~s$^{-1}$) \\
\hline 
\noalign{\smallskip}
41 & s11.3 & h & 6 (0.1) & 0.56 (0.56) & 0.55 (0.55) & 4.7 (2.7) & 5.8 (2.8) & 17.5 (8.5)\\
54 & s9.3 & h & 3 (0.4) & 0.0 (0.0) & 0.91 (0.52) & 0.0 (0.0) & 0.91 (0.52) & 3.05 (1.74)\\
107 & s5.0 & l & 11 (1.5) & 0.62 (0.62) & 1.3 (0.8) & 3.9 (2.3) & 5.9 (2.5) & 17.3 (7.2) \\
154 & e33.0 & k & 8 (0.9) & 0.0 (0.0) & <0.24 & 6.3 (2.6) & 6.6 (2.6) & 9.2 (3.7) \\
225B & w2.0 & h & 4 (0.1) & 0.0 (0.0) & 0.79 (0.58) & 0.69 (0.69) & 1.5 (0.9) & 2.1 (1.3)\\
\noalign{\smallskip}
\hline
\end{tabular}\\
\end{center}
Col. (1): The 3CR name.
Col. (2): The component name (e.g., knot or hotspot) is a combination of one letter indicating the orientation of the radio structure and one number indicating distance from the core in arcseconds.
Col. (3): The component class: ``h'' = hotspot - ``k'' = knot - ``l'' = lobe.
Col. (4): The total counts in the photometric circle together with the average of the 8 background regions, in parentheses; both for the 0.5 to 7 keV band.
Col. (5): Measured X-ray flux between 0.5 and 1 keV.
Col. (6): Measured X-ray flux between 1 and 2 keV.
Col. (7): Measured X-ray flux between 2 and 7 keV.
Col. (8): Measured X-ray flux between 0.5 and 7 keV.
Col. (9): X-ray luminosity in the range 0.5 to 7 keV with the 1$\sigma$ uncertainties given in parenthesis.\\
Note:\\
($^*$) Fluxes are given in units of 10$^{-15}$erg~cm$^{-2}$s$^{-1}$ and 1$\sigma$ uncertainties are given in parenthesis.
The uncertainties on the flux measurements were computed as described in \S~\ref{sec:obs}\\
\end{table*}

\appendix

\section{A: Images of the sources}
\label{sec:images}

For all the 3CR sources in our sample, radio morphologies are shown here 
as contours superposed on the re-gridded/smoothed X-ray event files.  
The full width half maximum (FWHM) of the Gaussian
smoothing function and the binning factor are reported in the figure captions.
X-ray event files were limited to the 0.5 to 7 keV band and rebinned
to change the pixel size with a binning factor 'f' (e.g. f=1/4
produces pixels 4 times smaller than the native ACIS pixel of 0.492\arcsec).  
The labels on the color bar for each X-ray map are in units of
counts/pixel. We included in each caption also the radio
brightness of the lowest contour, the factor (usually 2 or 4) by which
each subsequent contour exceeds the previous one, the frequency of the
radio map, and the FWHM of the clean beam.
Figures appear so different from
each other mainly because of the wide range in angular size of the radio sources.

\begin{figure}
\includegraphics[keepaspectratio=true,scale=0.5,angle=-90]{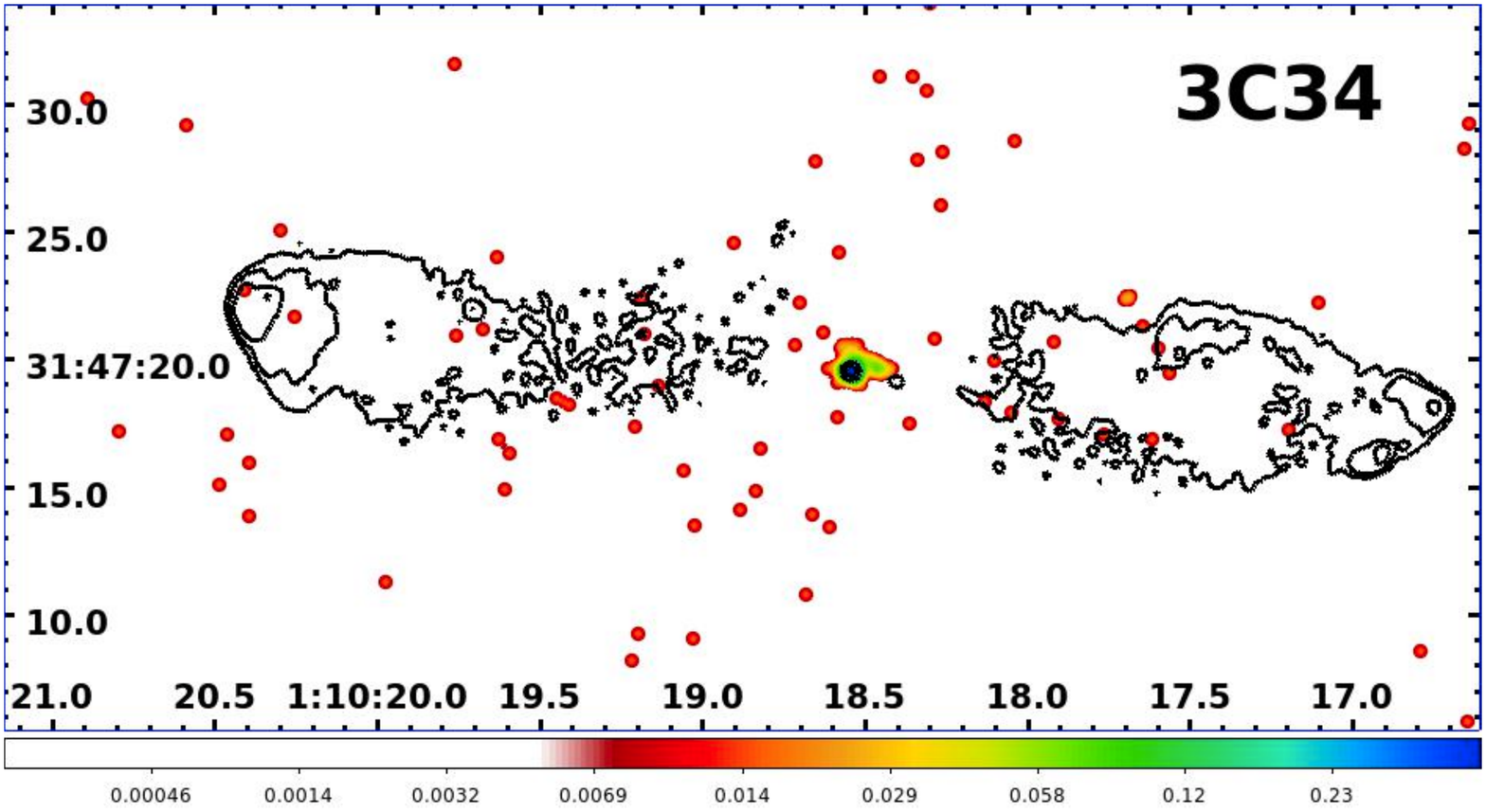}
\caption{The X-ray image of 3CR\,34 for the energy band 0.5-7 keV.  The
  event file has been regridded to 1/8 of the native pixel size (i.e., 0.492\arcsec).
  The image has been smoothed with a Gaussian of FWHM=7\arcsec.   
  The radio contours (black) were computed using a 4.9 GHz radio
  map and start at 0.125 mJy/beam, increasing by factors of four.}
\label{fig:3c34app}
\end{figure}

\begin{figure}
\includegraphics[keepaspectratio=true,scale=0.65,angle=-90]{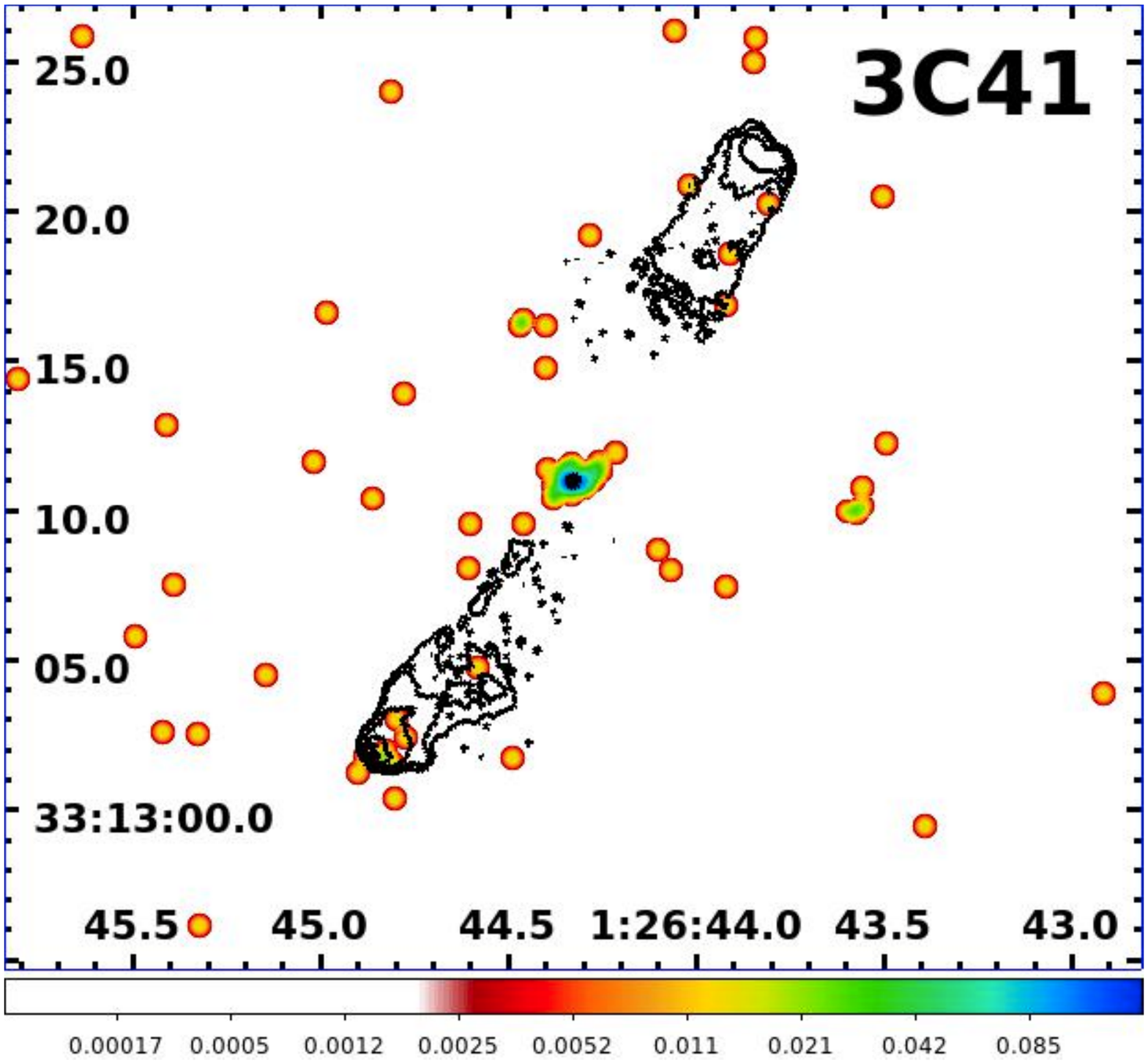}
\caption{The X-ray image of 3CR\,41 for the energy band 0.5-7 keV.  The
  event file has been regridded to 1/8 of the native pixel size (i.e., 0.492\arcsec).
  The image has been smoothed with a Gaussian of FWHM=7\arcsec.   
  The radio contours (black) were computed using a 8.5 GHz radio
  map and start at 0.125 mJy/beam, increasing by factors of four.}
\label{fig:3c41app}
\end{figure}

\begin{figure}
\includegraphics[keepaspectratio=true,scale=0.65]{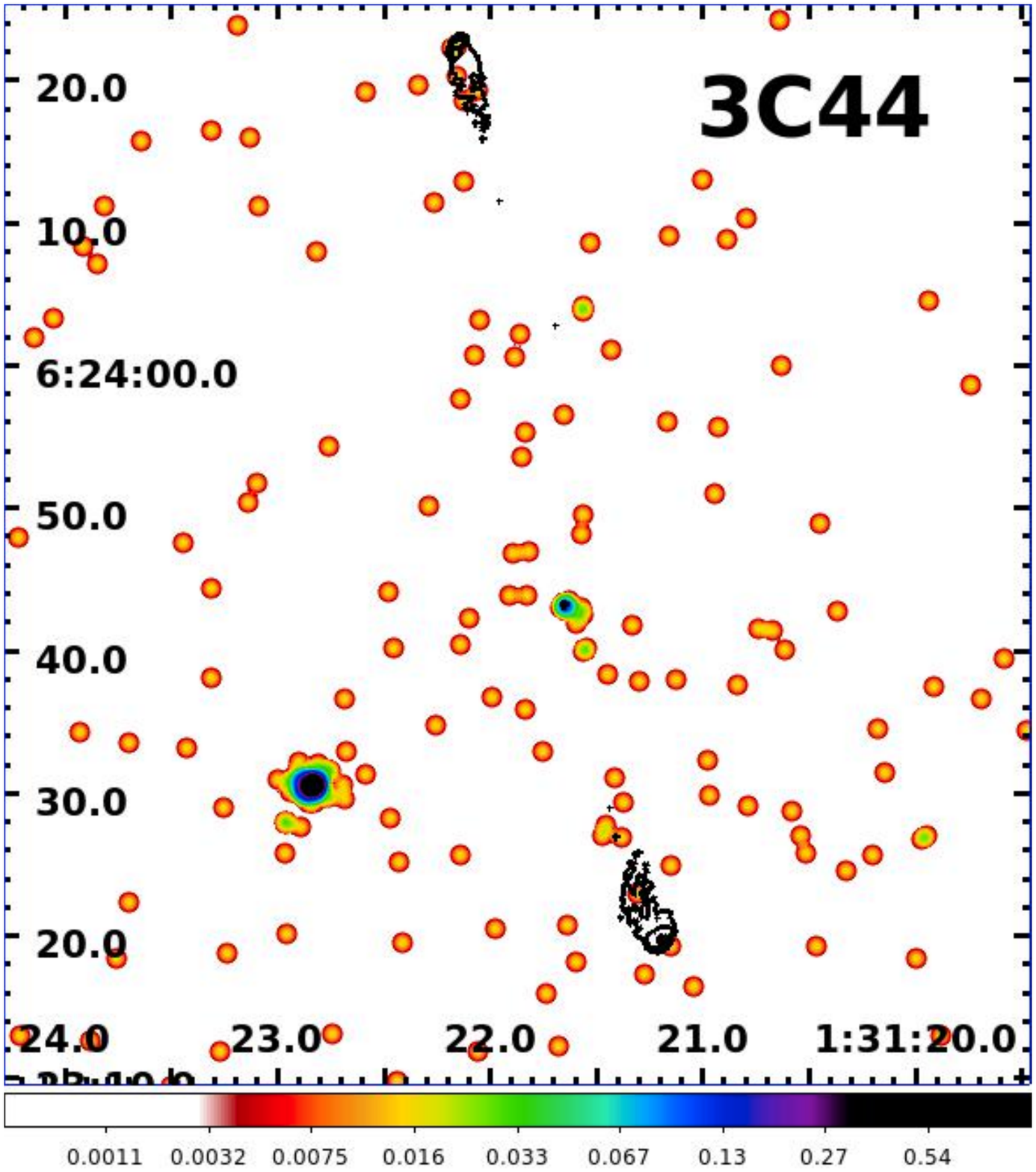}
\caption{The X-ray image of 3CR\,44 for the energy band 0.5-7 keV.  The
  event file has been regridded to 1/8 of the native pixel size (i.e., 0.492\arcsec).
  The image has been smoothed with a Gaussian of FWHM=5\arcsec.   
  The radio contours (black) were computed using a 8.4 GHz radio
  map and start at 0.25 mJy/beam, increasing by factors of four. 
  Radio core is weak but detected in our 8.4 GHz radio image; thus the \chn\ image is registered.}
\label{fig:3c44app}
\end{figure}

\begin{figure}
\includegraphics[keepaspectratio=true,scale=0.65]{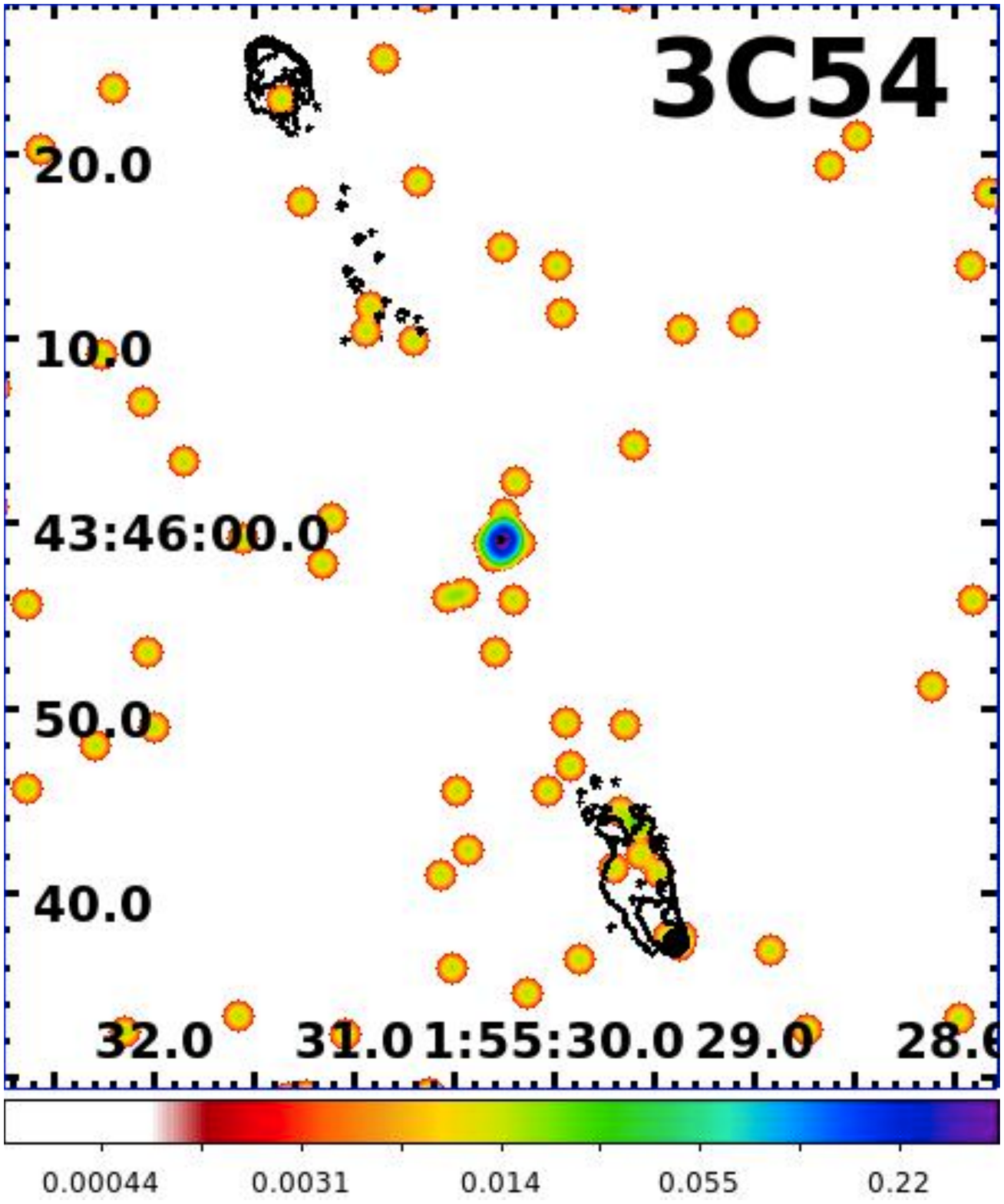}
\caption{The X-ray image of 3CR\,54 for the energy band 0.5-7 keV.  The
  event file has been regridded to 1/8 of the native pixel size (i.e., 0.492\arcsec).
  The image has been smoothed with a Gaussian of FWHM=7\arcsec.   
  The radio contours (black) were computed using a 8.4 GHz radio
  map and start at 0.25 mJy/beam, increasing by factors of four.}
\label{fig:3c54app}
\end{figure}

\begin{figure}
\includegraphics[keepaspectratio=true,scale=0.5,angle=-90]{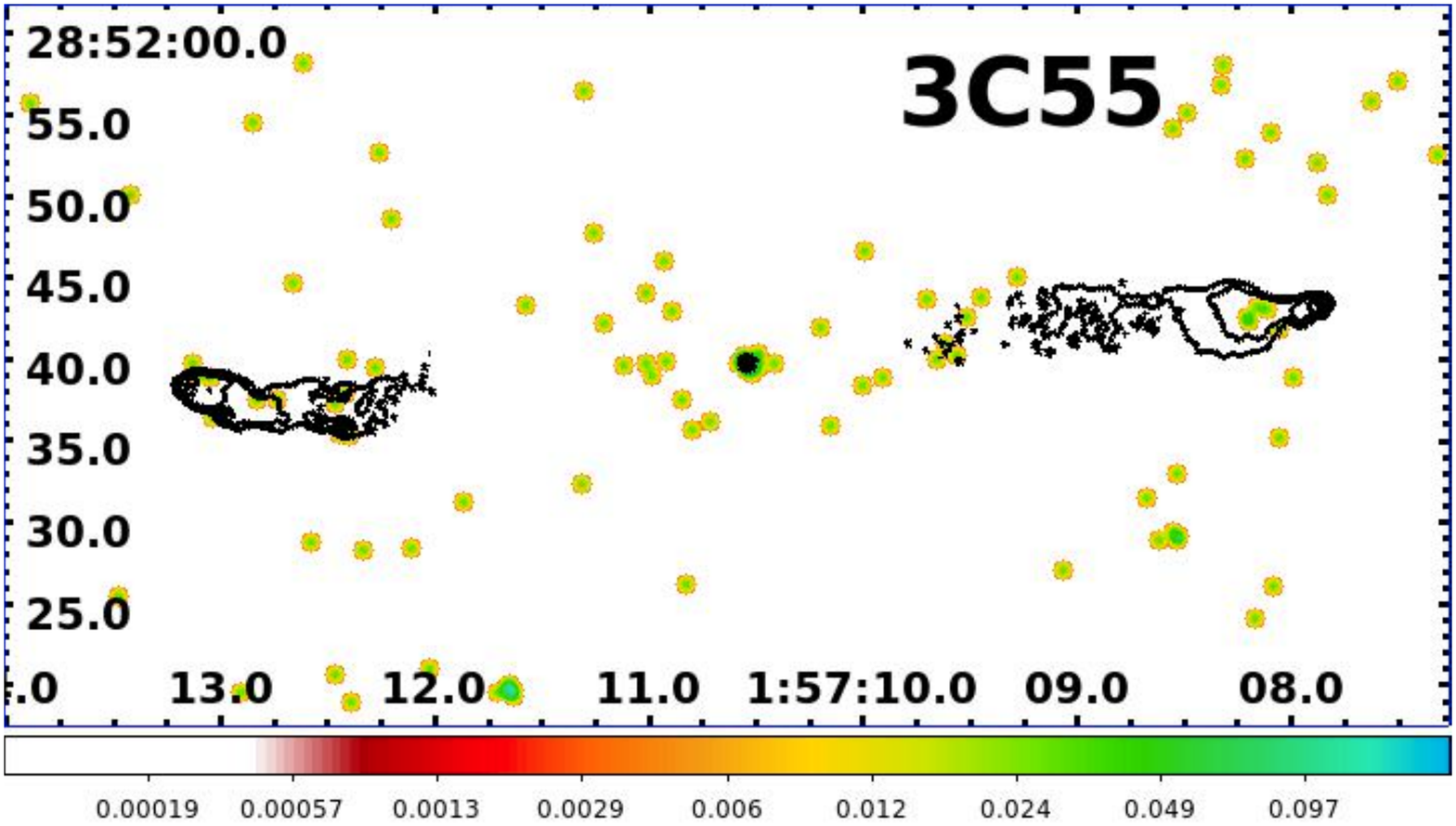}
\caption{The X-ray image of 3CR\,55 for the energy band 0.5-7 keV.  The
  event file has been regridded to 1/4 of the native pixel size (i.e., 0.492\arcsec).
  The image has been smoothed with a Gaussian of FWHM=5\arcsec.   
  The radio contours (black) were computed using a 4.8 GHz radio
  map and start at 0.125 mJy/beam, increasing by factors of four.}
\label{fig:3c55app}
\end{figure}

\begin{figure}
\includegraphics[keepaspectratio=true,scale=0.65,angle=-90]{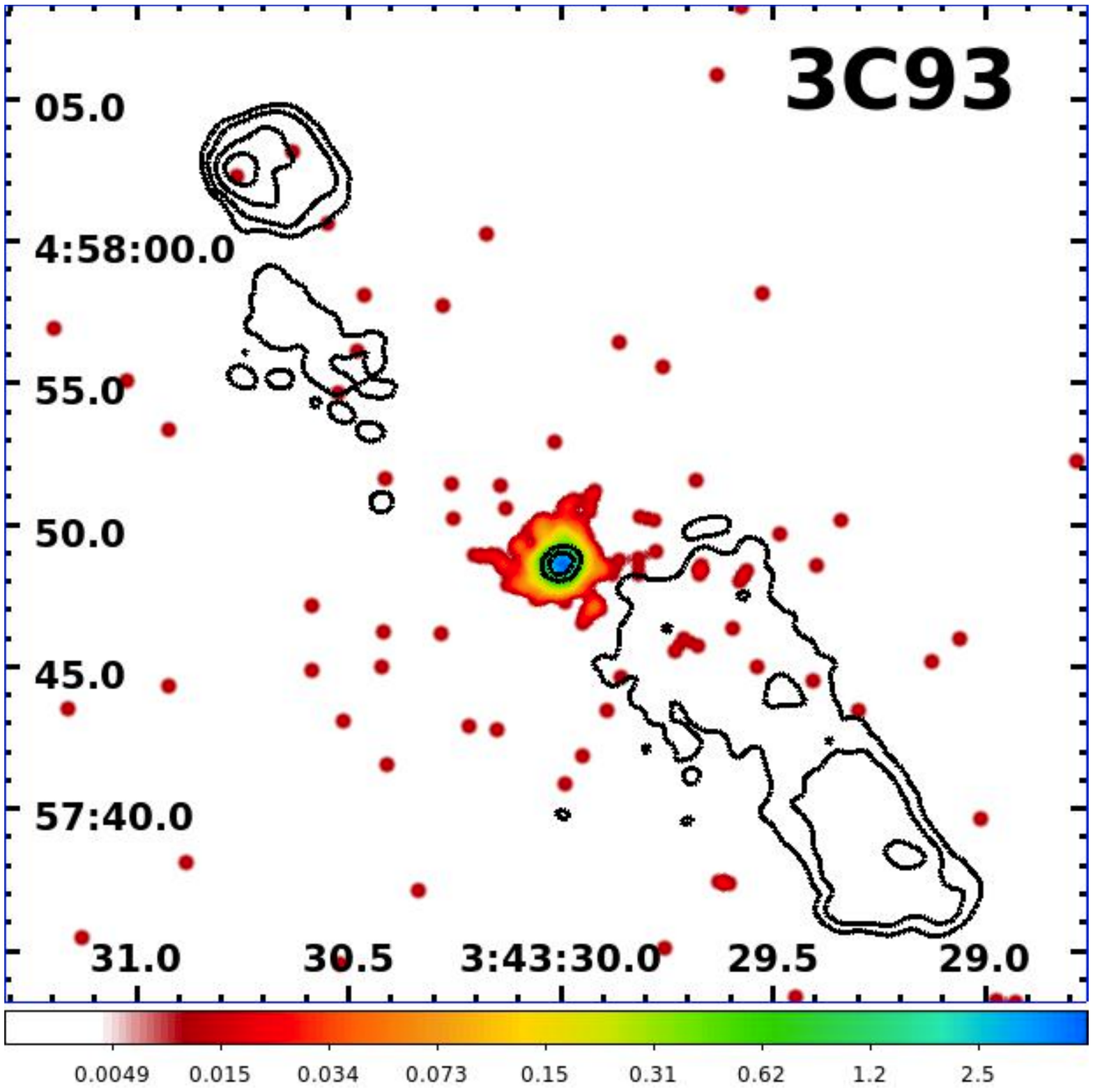}
\caption{The X-ray image of 3CR\,93 for the energy band 0.5-7 keV.  The
  event file has been regridded to 1/8 of the native pixel size (i.e., 0.492\arcsec).
  The image has been smoothed with a Gaussian of FWHM=7\arcsec.   
  The radio contours (black) were computed using a 8.5 GHz radio
  map and start at 0.3 mJy/beam, increasing by factors of four.}
\label{fig:3c93app}
\end{figure}

\begin{figure}
\includegraphics[keepaspectratio=true,scale=0.65]{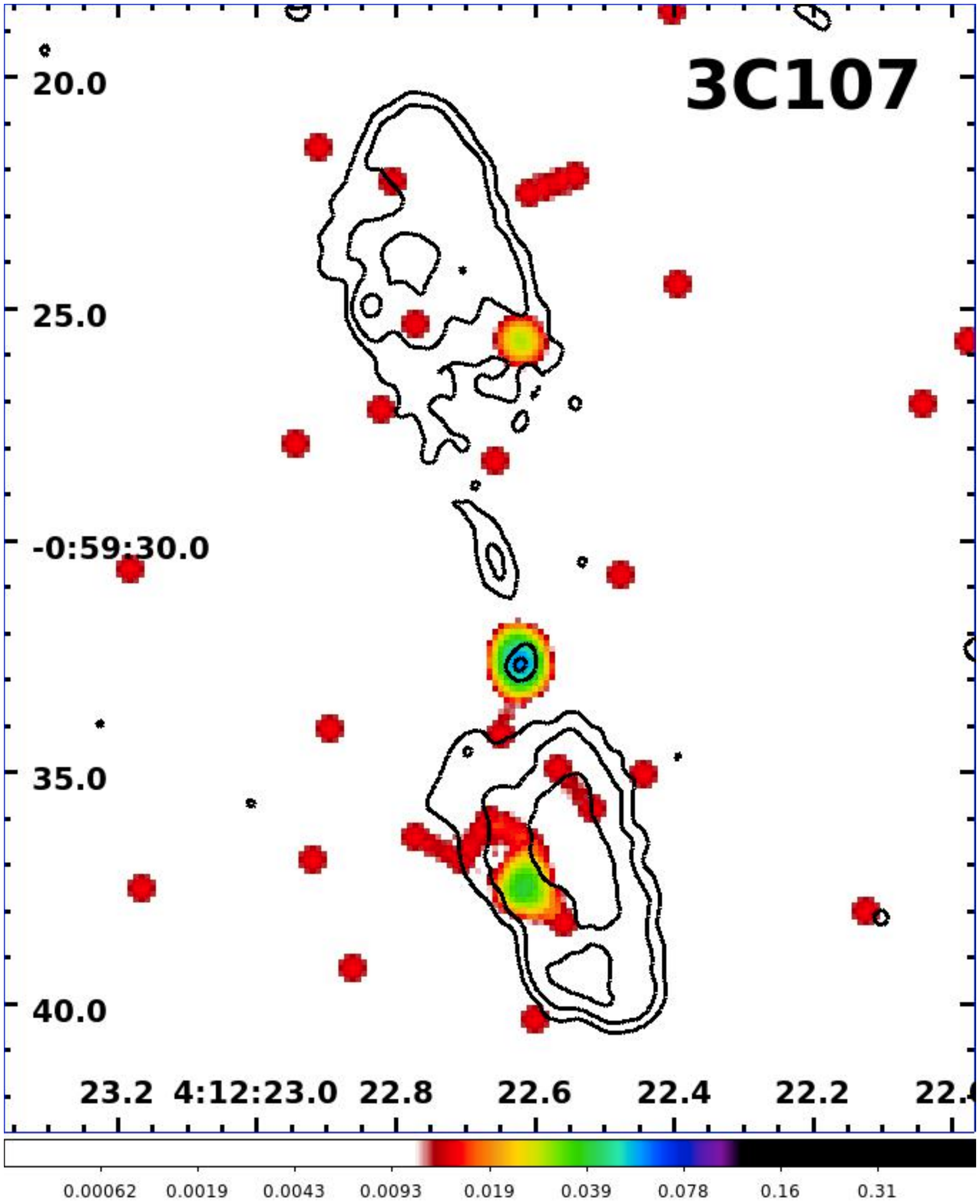}
\caption{The X-ray image of 3CR\,107 for the energy band 0.5-7 keV.  The
  event file has been regridded to 1/4 of the native pixel size (i.e., 0.492\arcsec).
  The image has been smoothed with a Gaussian of FWHM=7\arcsec.   
  The radio contours (black) were computed using a 4.9 GHz radio
  map and start at 0.125 mJy/beam, increasing by factors of four.}
\label{fig:3c107app}
\end{figure}

\begin{figure}
\includegraphics[keepaspectratio=true,scale=0.65]{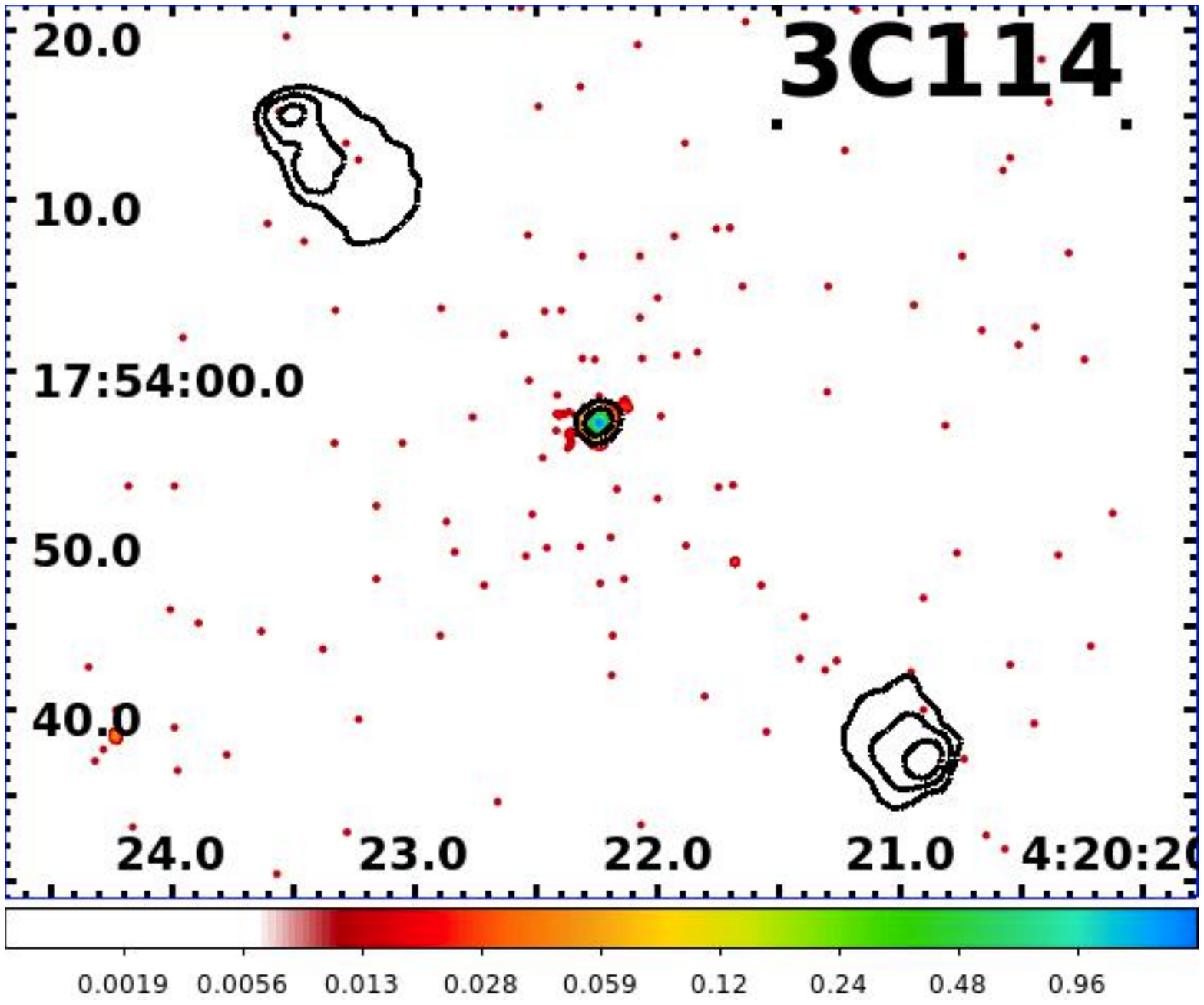}
\caption{The X-ray image of 3CR\,114 for the energy band 0.5-7 keV.  The
  event file has been regridded to 1/8 of the native pixel size (i.e., 0.492\arcsec).
  The image has been smoothed with a Gaussian of FWHM=7\arcsec.   
  The radio contours (black) were computed using a 4.9 GHz radio
  map and start at 1.0 mJy/beam, increasing by factors of four.}
\label{fig:3c114app}
\end{figure}

\begin{figure}
\includegraphics[keepaspectratio=true,scale=0.65,angle=-90]{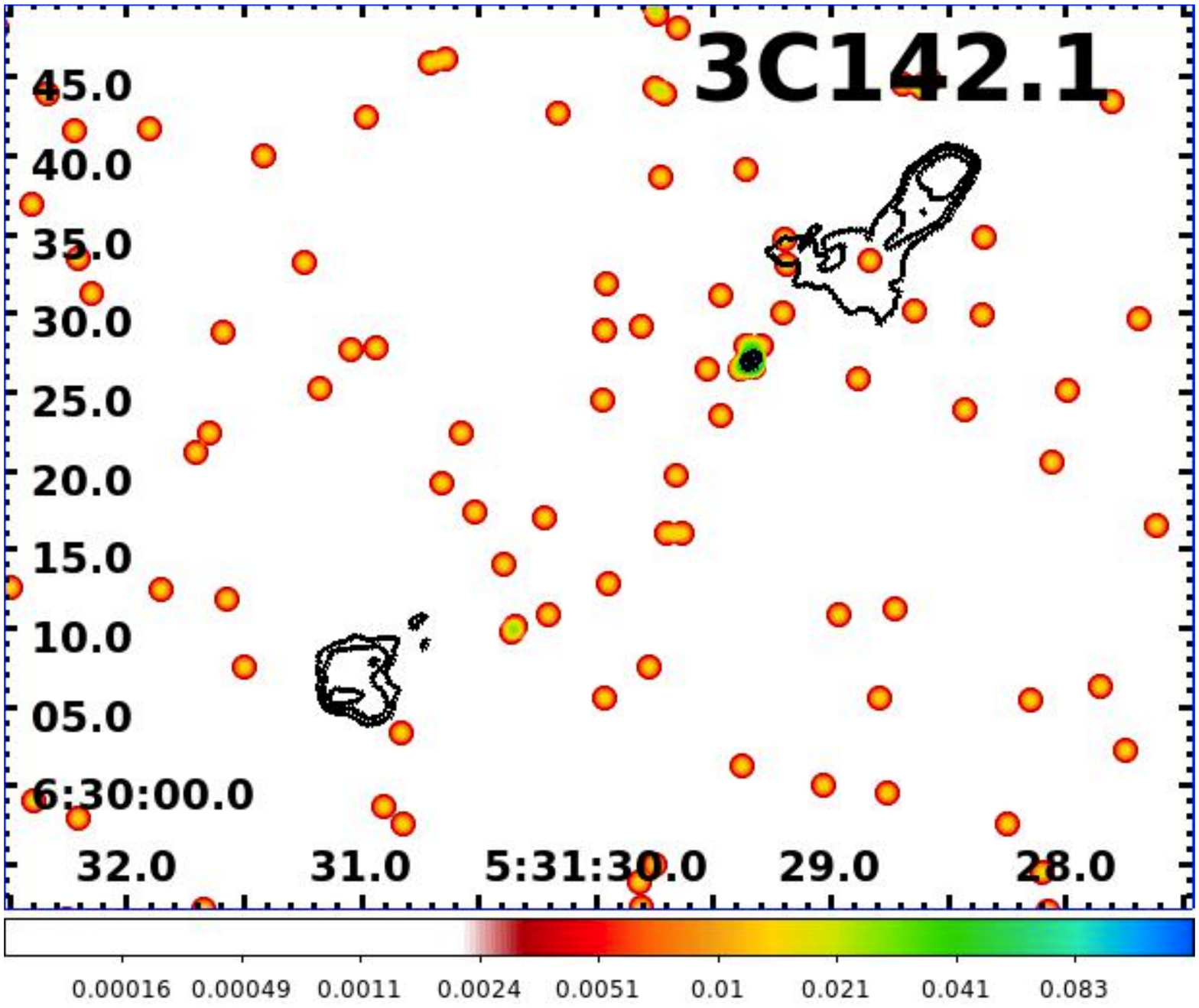}
\caption{The X-ray image of 3CR\,142.1 for the energy band 0.5-7 keV.  The
  event file has been regridded to 1/4 of the native pixel size (i.e., 0.492\arcsec).
  The image has been smoothed with a Gaussian of FWHM=7\arcsec.   
  The radio contours (black) were computed using a 8.5 GHz radio
  map and start at 0.4 mJy/beam, increasing by factors of four.}
\label{fig:3c142.1app}
\end{figure}

\begin{figure}
\includegraphics[keepaspectratio=true,scale=0.5,angle=-90]{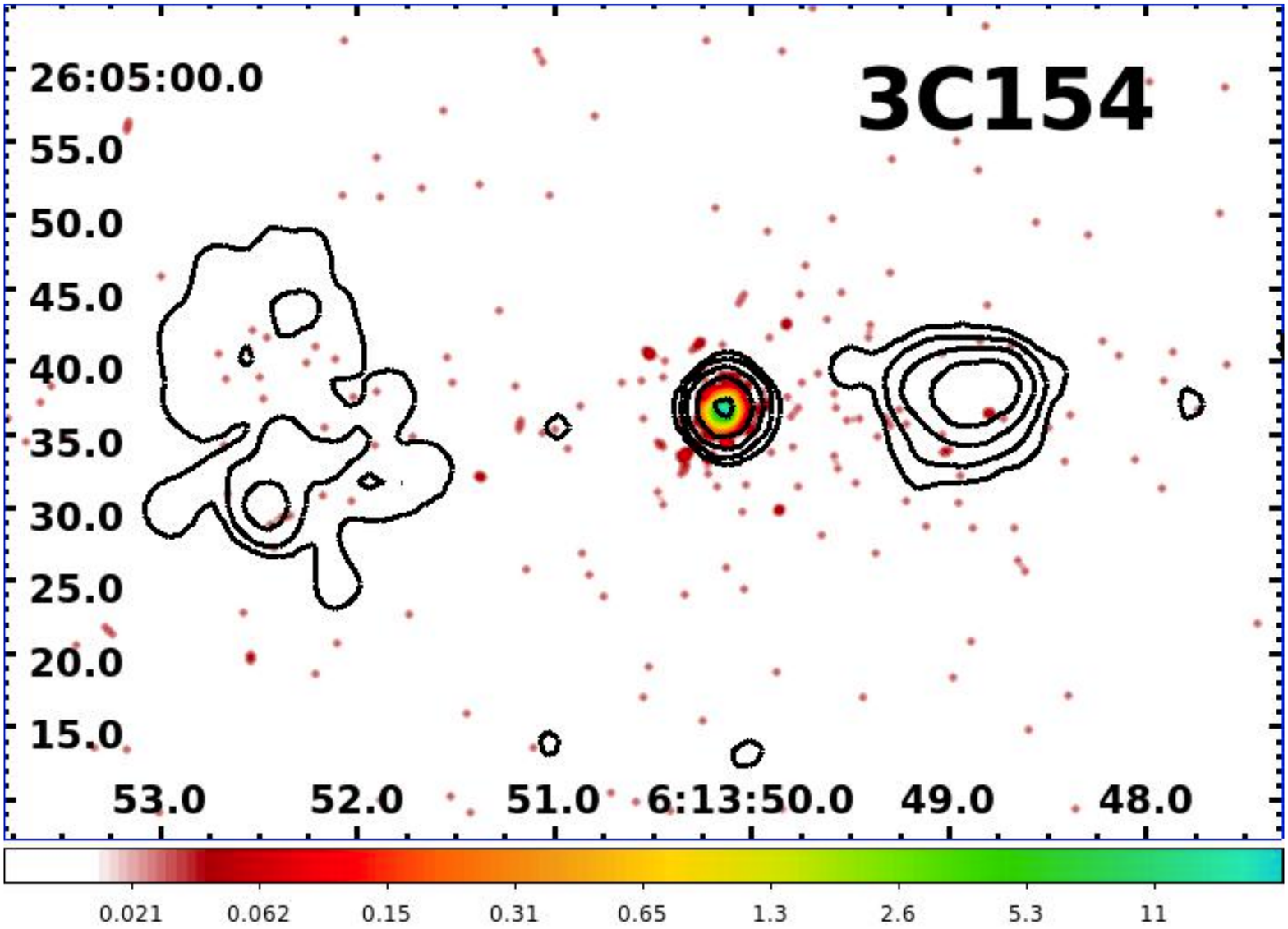}
\caption{The X-ray image of 3CR\,154 for the energy band 0.5-7 keV.  The
  event file has been regridded to 1/4 of the native pixel size (i.e., 0.492\arcsec).
  The image has been smoothed with a Gaussian of FWHM=5\arcsec.   
  The radio contours (black) were computed using a 8.5 GHz radio
  map and start at 1 mJy/beam, increasing by factors of four.}
\label{fig:3c154app}
\end{figure}

\begin{figure}
\includegraphics[keepaspectratio=true,scale=0.65]{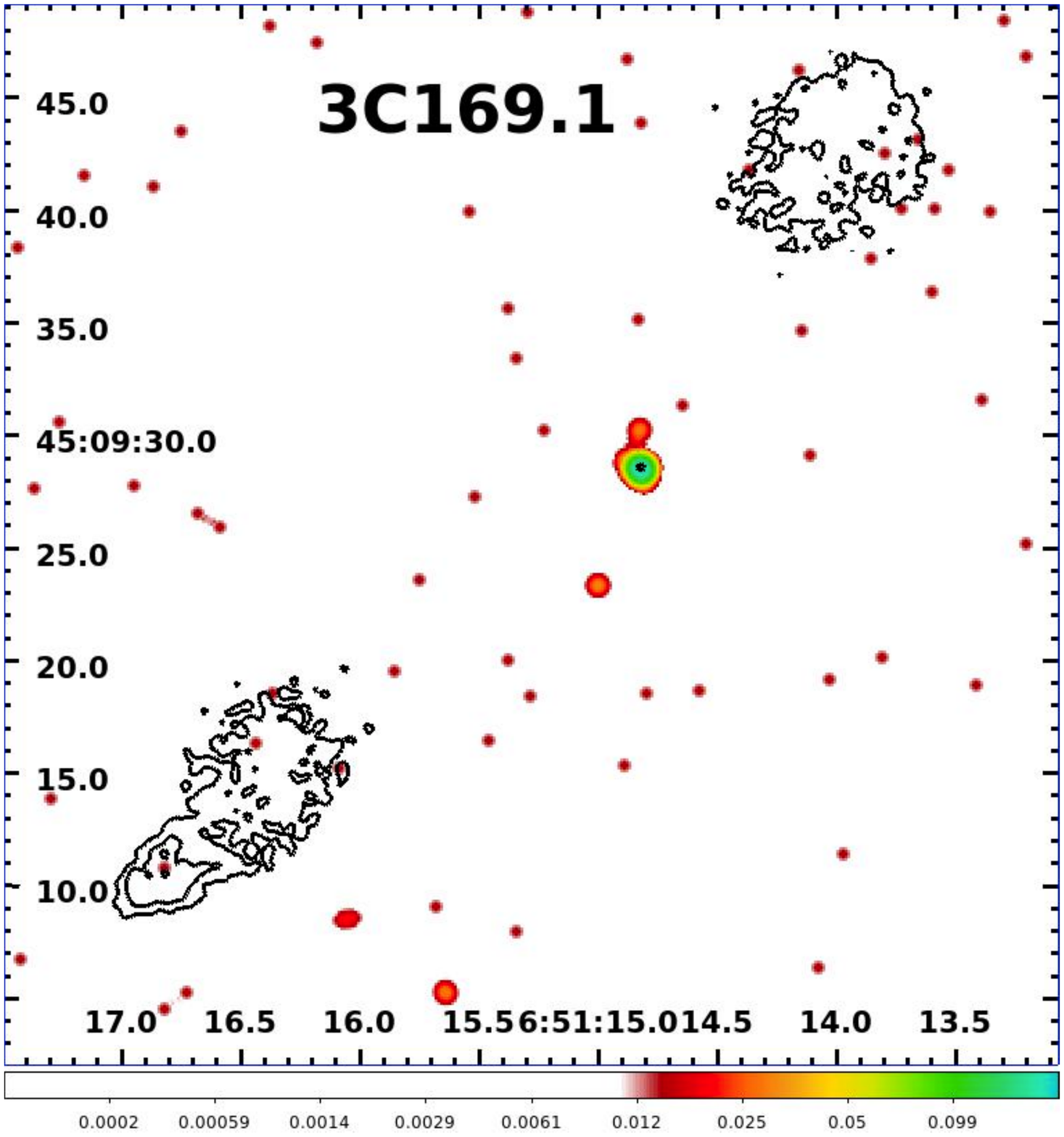}
\caption{The X-ray image of 3CR\,169.1 for the energy band 0.5-7 keV.  The
  event file has been regridded to 1/4 of the native pixel size (i.e., 0.492\arcsec).
  The image has been smoothed with a Gaussian of FWHM=7\arcsec.   
  The radio contours (black) were computed using a 8.4 GHz radio
  map and start at 0.25 mJy/beam, increasing by factors of four.}
\label{fig:3c169.1app}
\end{figure}

\begin{figure}
\includegraphics[keepaspectratio=true,scale=0.5,angle=-90]{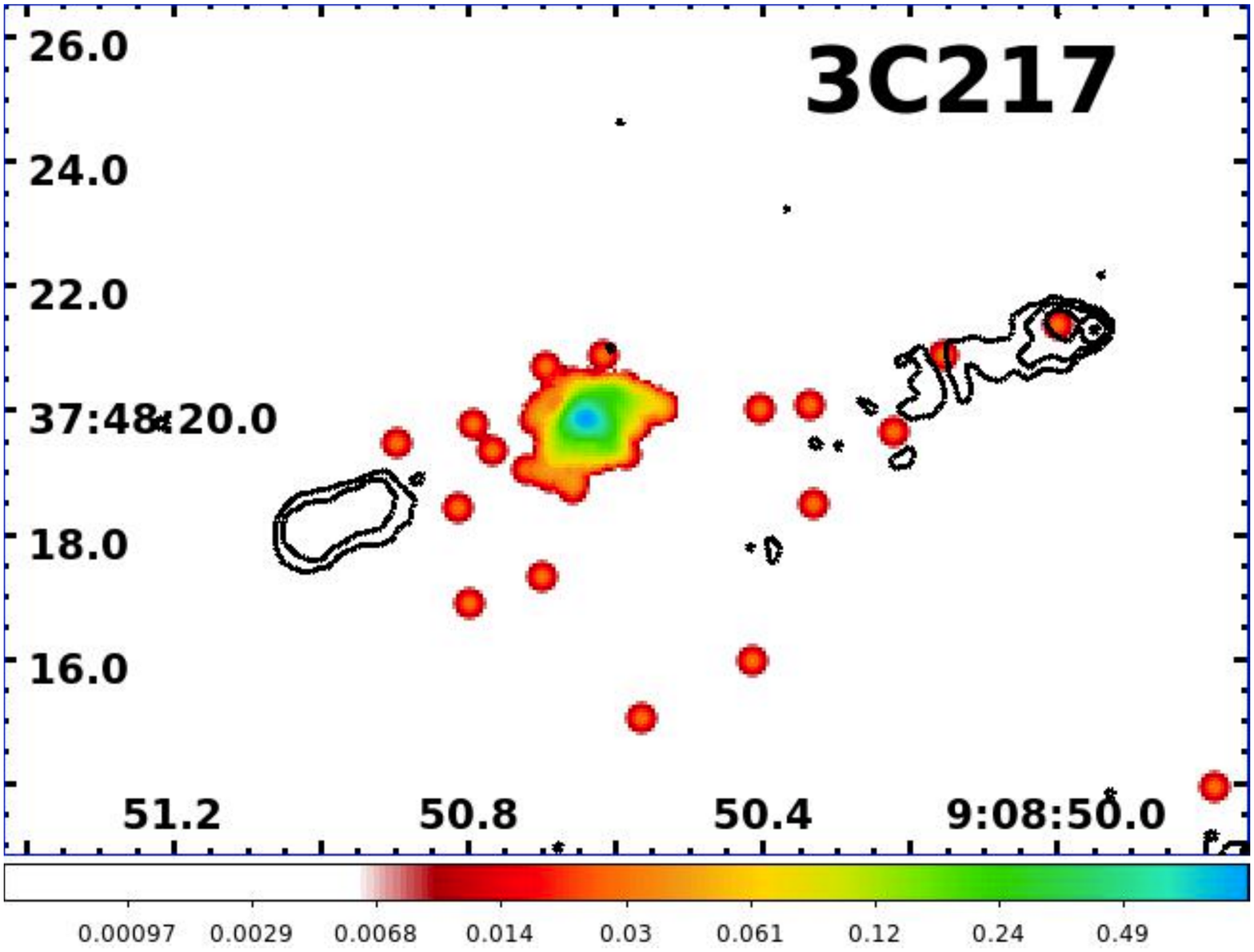}
\caption{The X-ray image of 3CR\,217 for the energy band 0.5-7 keV.  The
  event file has been regridded to 1/8 of the native pixel size (i.e., 0.492\arcsec).
  The image has been smoothed with a Gaussian of FWHM=5\arcsec.   
  The radio contours (black) were computed using a 8.5 GHz radio
  map and start at 0.3 mJy/beam, increasing by factors of four. Since
  there is only a marginal detection of the radio nucleus, the X-ray image was not been
  registered.}
\label{fig:3c217app}
\end{figure}

\begin{figure}
\includegraphics[keepaspectratio=true,scale=0.5,angle=-90]{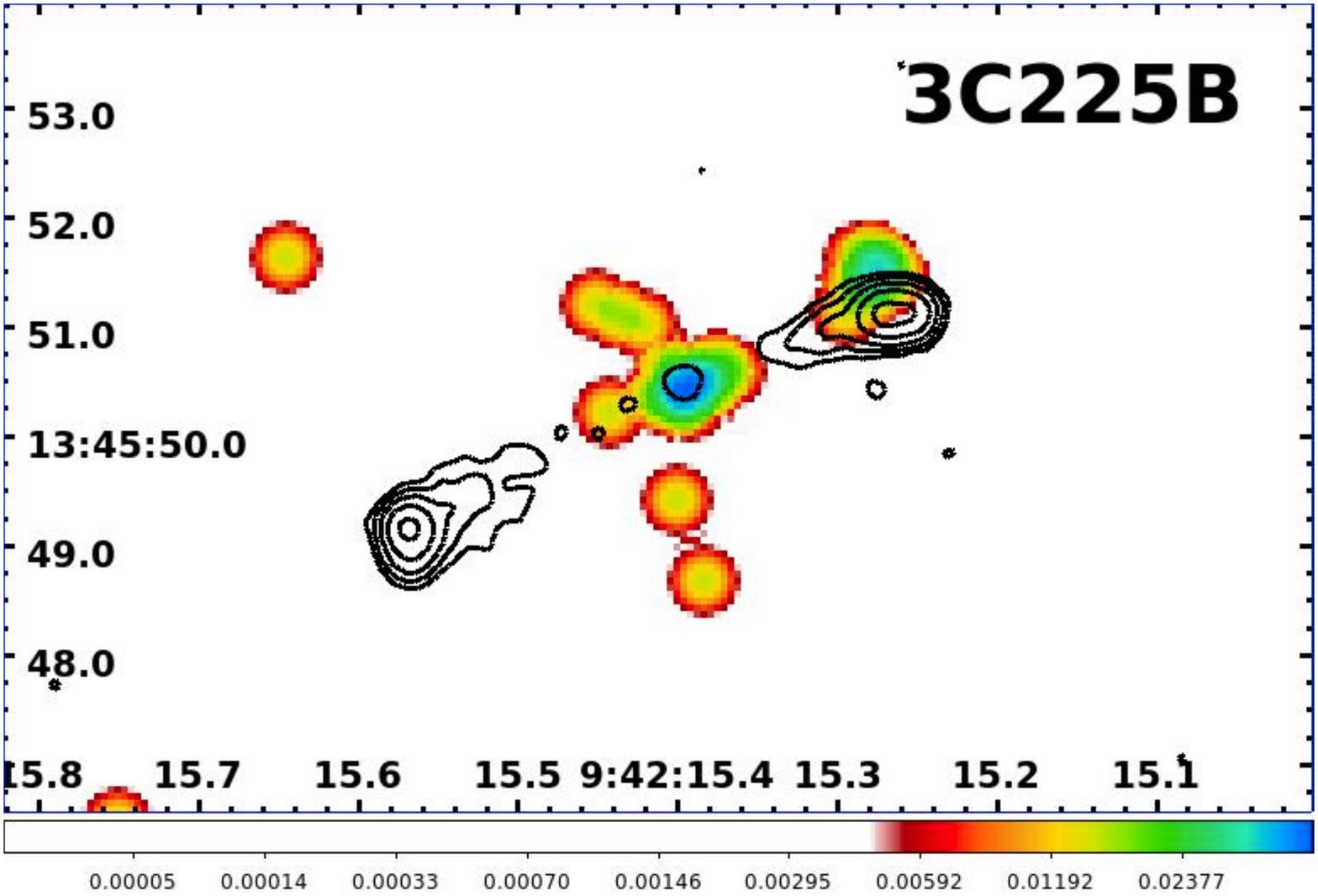}
\caption{The X-ray image of 3CR\,255B for the energy band 0.5-7 keV.  The
  event file has been regridded to 1/8 of the native pixel size (i.e., 0.492\arcsec).
  The image has been smoothed with a Gaussian of FWHM=7\arcsec.   
  The radio contours (black) were computed using a 8.4 GHz radio
  map and start at 0.3 mJy/beam, increasing by factors of four.}
\label{fig:3c225Bapp}
\end{figure}

\begin{figure}
\includegraphics[keepaspectratio=true,scale=0.5,angle=-90]{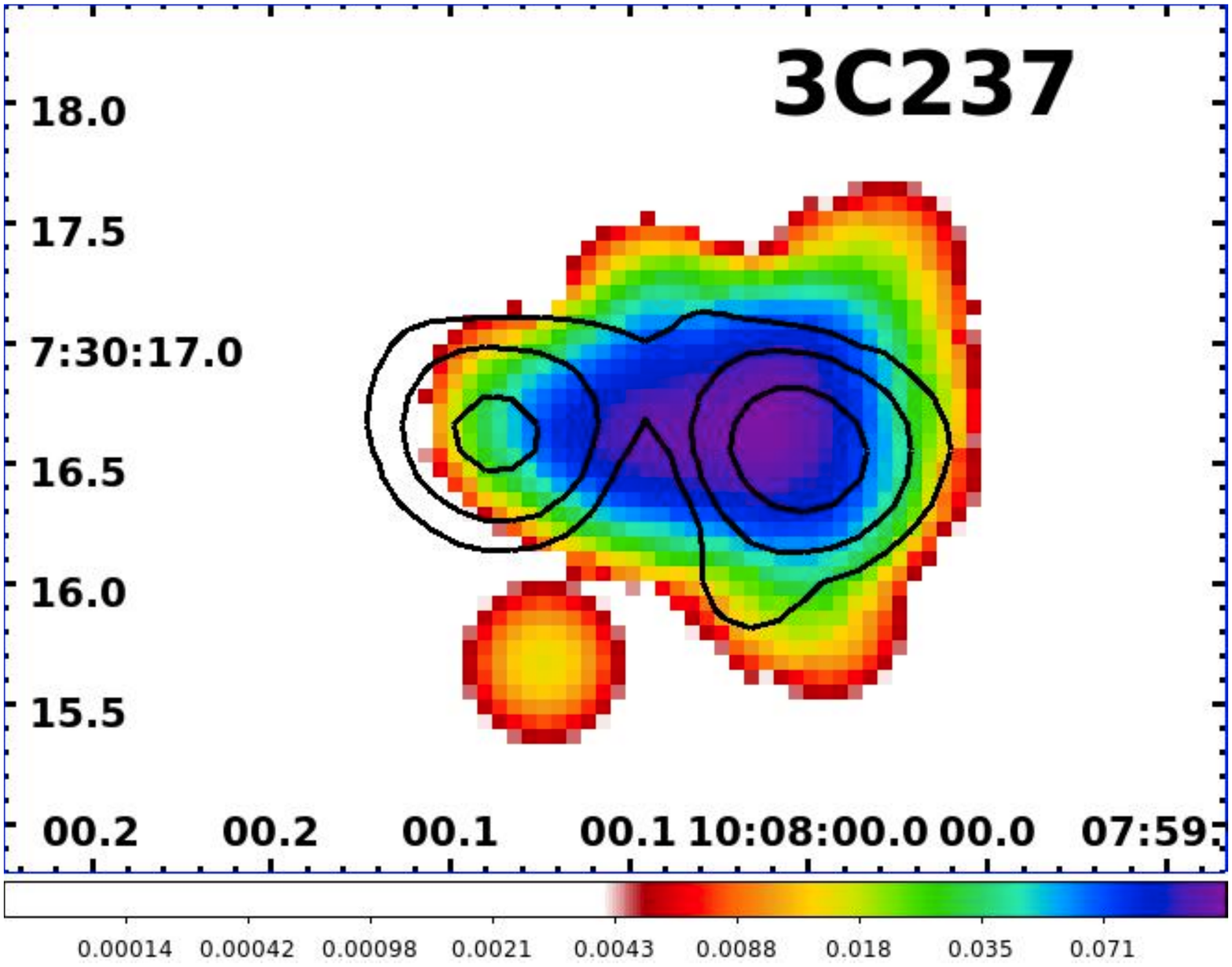}
\caption{The X-ray image of 3CR\,44 for the energy band 0.5-7 keV.  The
  event file has been regridded to 1/8 of the native pixel size (i.e., 0.492\arcsec).
  The image has been smoothed with a Gaussian of FWHM=7\arcsec.   
  The radio contours (black) were computed using a 14.9 GHz radio
  map and start at 4.8 mJy/beam, increasing by factors of four. Given the small source size
  the X-ray image was not been registered. }
\label{fig:3c237app}
\end{figure}

\begin{figure}
\includegraphics[keepaspectratio=true,scale=0.5,angle=-90]{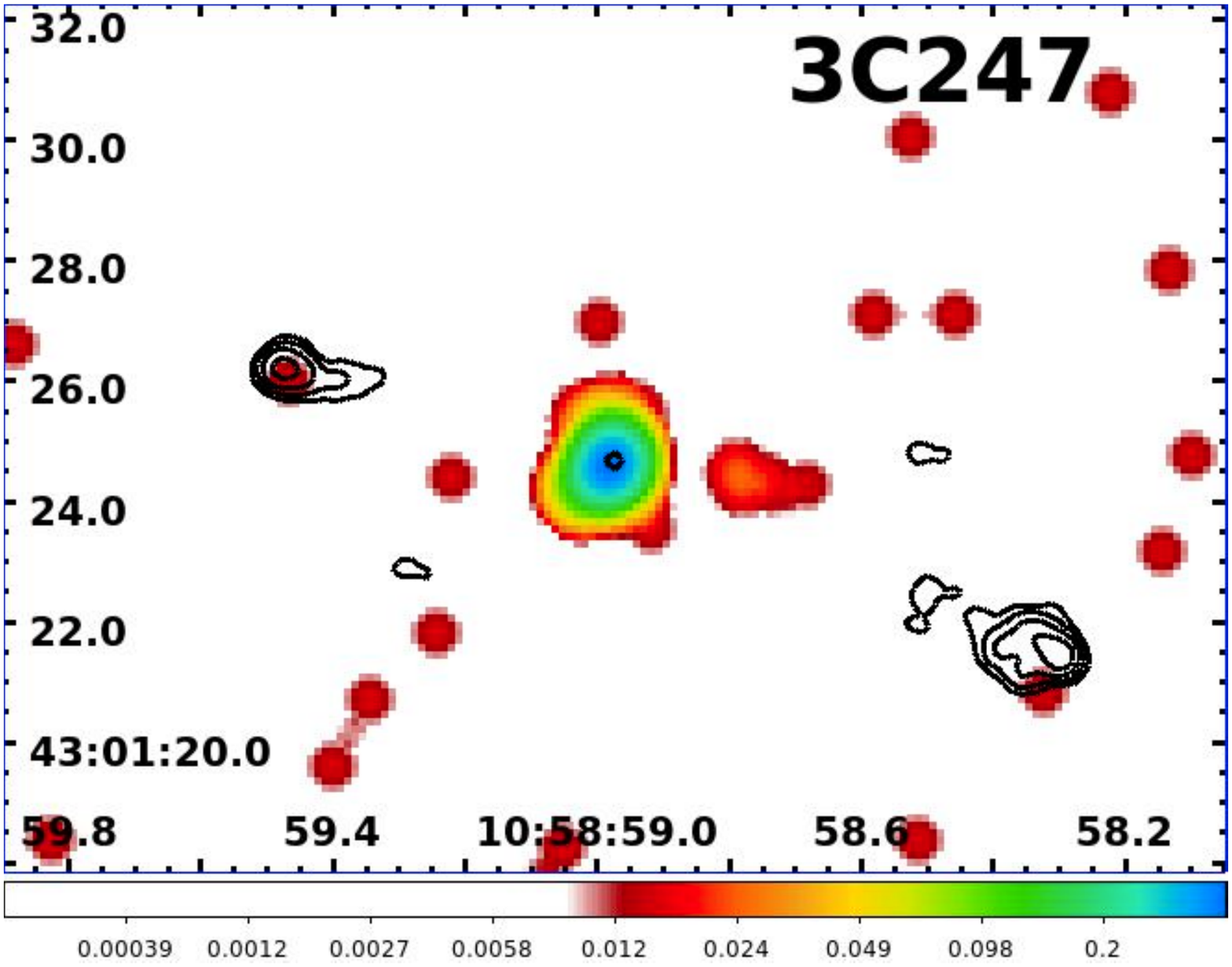}
\caption{The X-ray image of 3CR\,247 for the energy band 0.5-7 keV.  The
  event file has been regridded to 1/4 of the native pixel size (i.e., 0.492\arcsec).
  The image has been smoothed with a Gaussian of FWHM=7\arcsec.   
  The radio contours (black) were computed using a 4.9 GHz radio
  map and start at 1.2 mJy/beam, increasing by factors of four.}
\label{fig:3c247app}
\end{figure}

\begin{figure}
\includegraphics[keepaspectratio=true,scale=0.65,angle=-90]{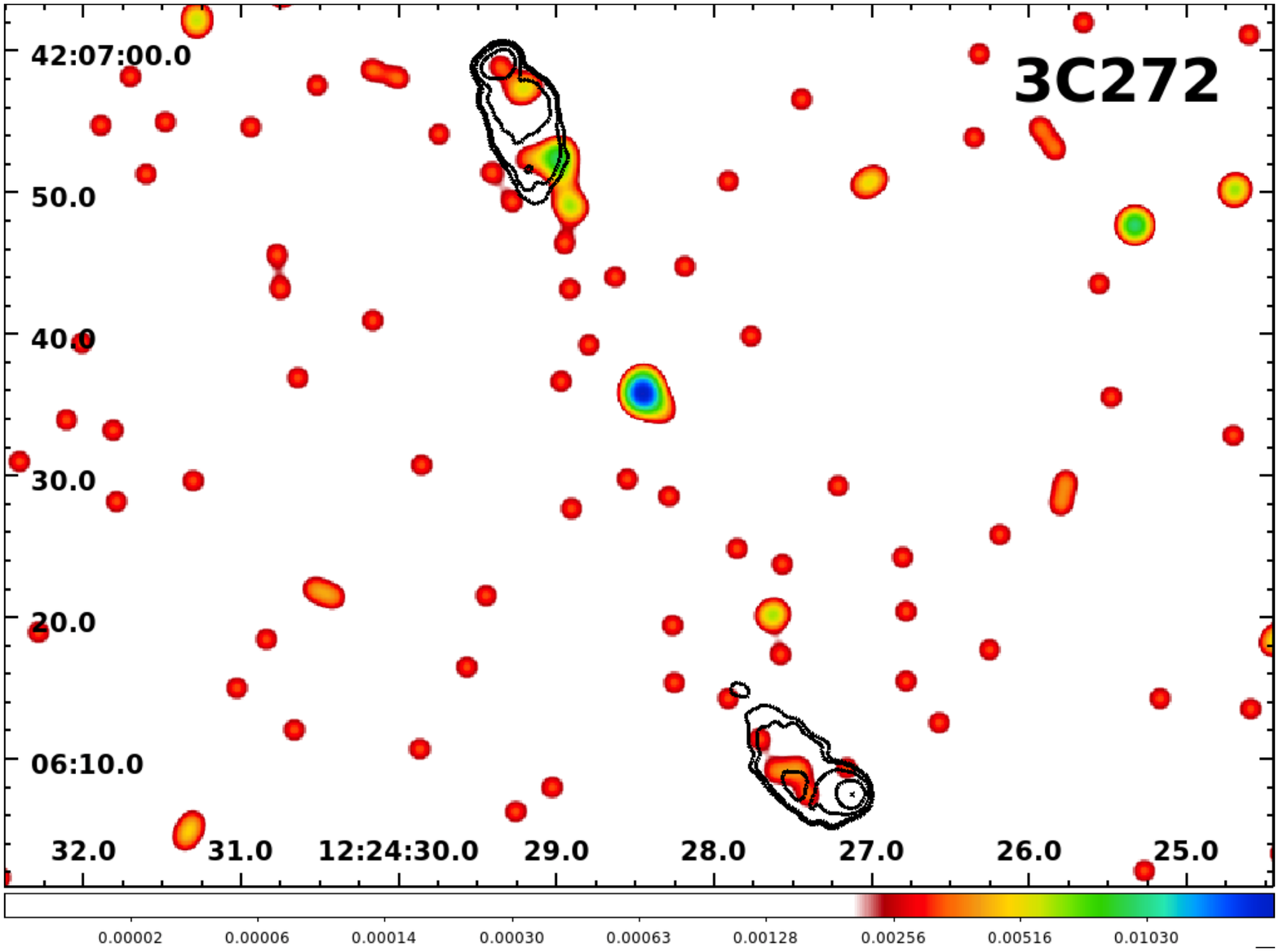}
\caption{The X-ray image of 3CR\,272 for the energy band 0.5-7 keV.  The
  event file has been regridded to 1/4 of the native pixel size (i.e., 0.492\arcsec).
  The image has been smoothed with a Gaussian of FWHM=2\arcsec.   
  The radio contours (black) were computed using a 4.86 GHz radio
  map and start at 0.8 mJy/beam up to 0.08 Jy/beam. Since
  there is only a marginal detection of the radio nucleus, the X-ray image was not been
  registered. }
\label{fig:3c247app}
\end{figure}

\begin{figure}
\includegraphics[keepaspectratio=true,scale=0.65,angle=-90]{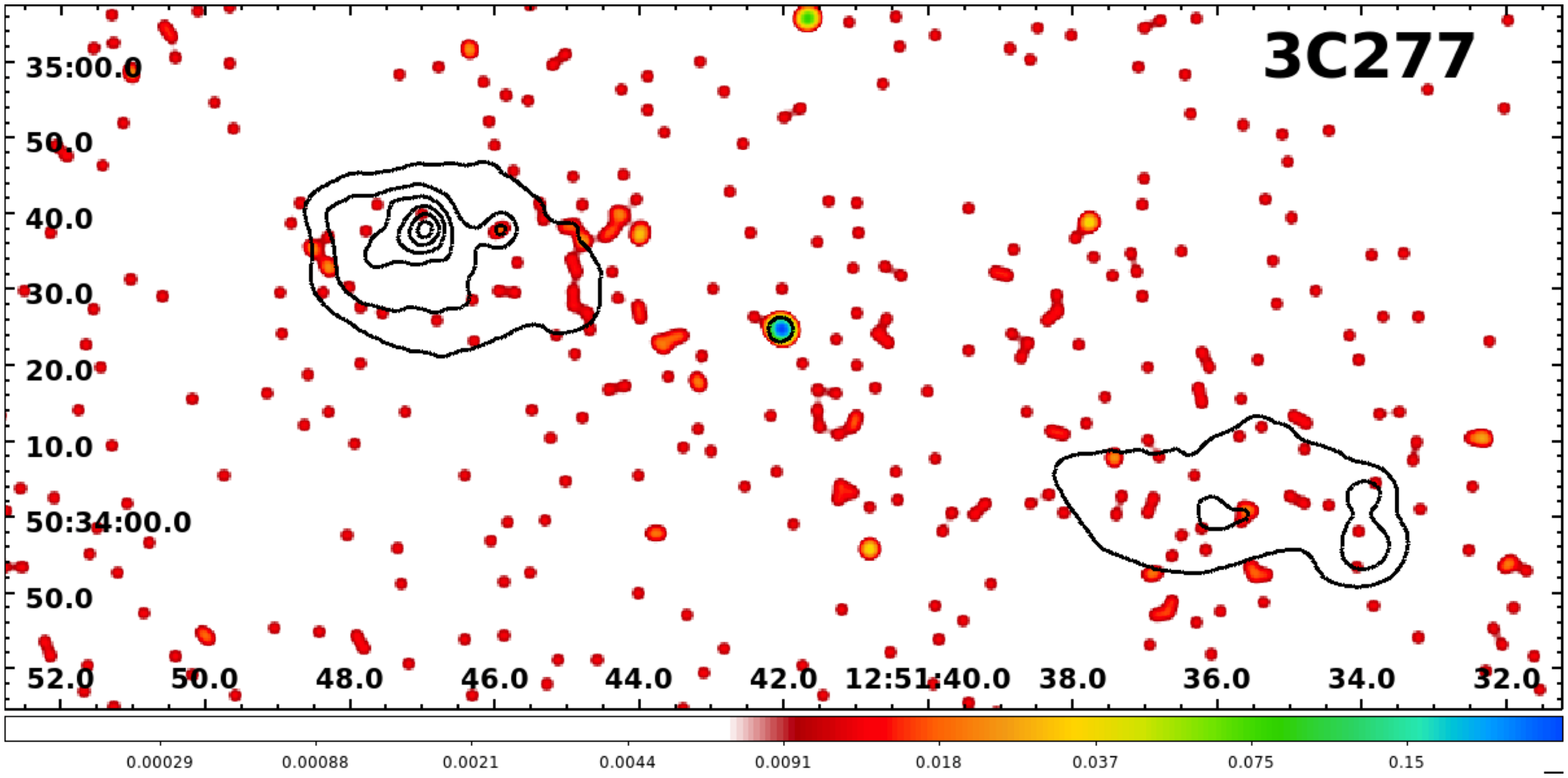}
\caption{The X-ray image of 3CR\,277 for the energy band 0.5-7 keV.  The
  event file has been regridded to 1/2 of the native pixel size (i.e., 0.492\arcsec).
  The image has been smoothed with a Gaussian of FWHM=2\arcsec.   
{ The six radio contours (black) were computed using a 1.4 GHz radio
  map and start at 2 mJy/beam up to 0.08 Jy/beam (linear scale).}}
\label{fig:3c247app}
\end{figure}

\begin{figure}
\includegraphics[keepaspectratio=true,scale=0.55,angle=-90]{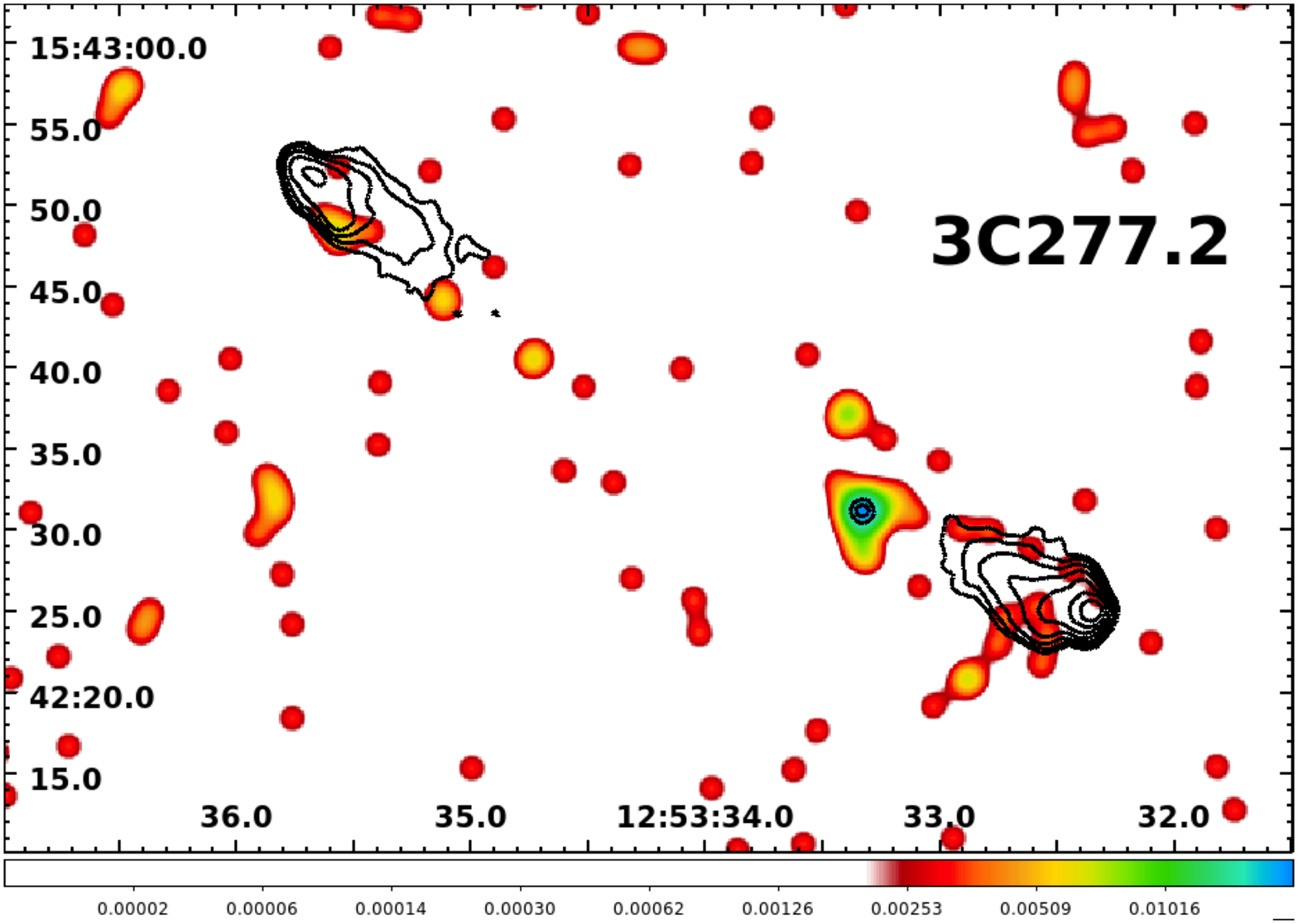}
\caption{The X-ray image of 3CR\,277.2 for the energy band 0.5-7 keV.  The
  event file has been regridded to 1/4 of the native pixel size (i.e., 0.492\arcsec).
  The image has been smoothed with a Gaussian of FWHM=2\arcsec.   
  The eight radio contours (black) were computed using a 1.4 GHz radio
  map and start at 0.2 mJy/beam up to 0.2 Jy/beam. Since
  there is only a marginal detection of the radio nucleus, the X-ray image was not been
  registered. 
}
\label{fig:3c247app}
\end{figure}

\begin{figure}
\includegraphics[keepaspectratio=true,scale=0.5,angle=-90]{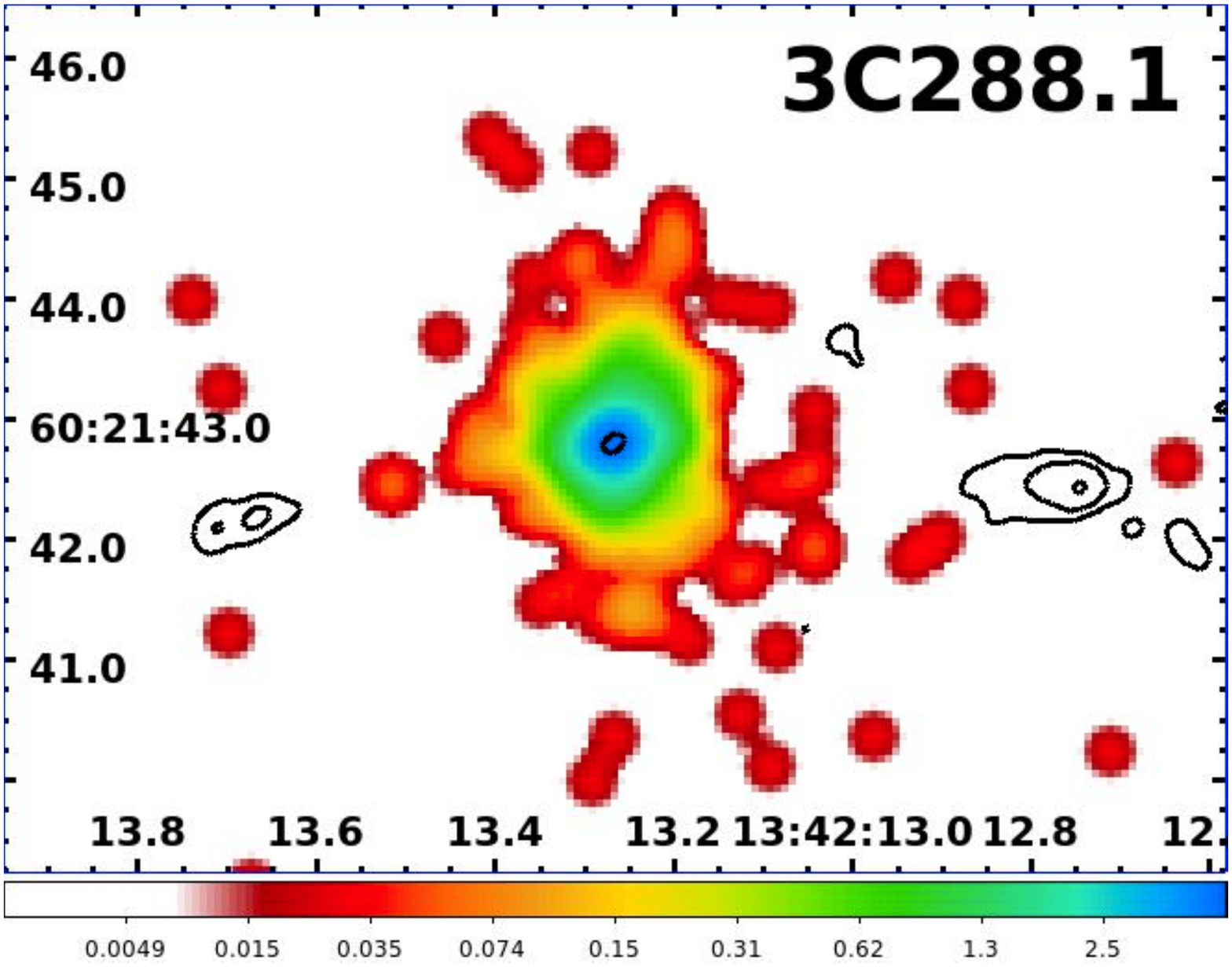}
\caption{The X-ray image of 3CR\,288.1 for the energy band 0.5-7 keV.  The
  event file has been regridded to 1/8 of the native pixel size (i.e., 0.492\arcsec).
  The image has been smoothed with a Gaussian of FWHM=5\arcsec.   
  The radio contours (black) were computed using a 8.4 GHz radio
  map and start at 2.0 mJy/beam, increasing by factors of four.}
\label{fig:3c288.1app}
\end{figure}

\begin{figure}
\includegraphics[keepaspectratio=true,scale=0.65]{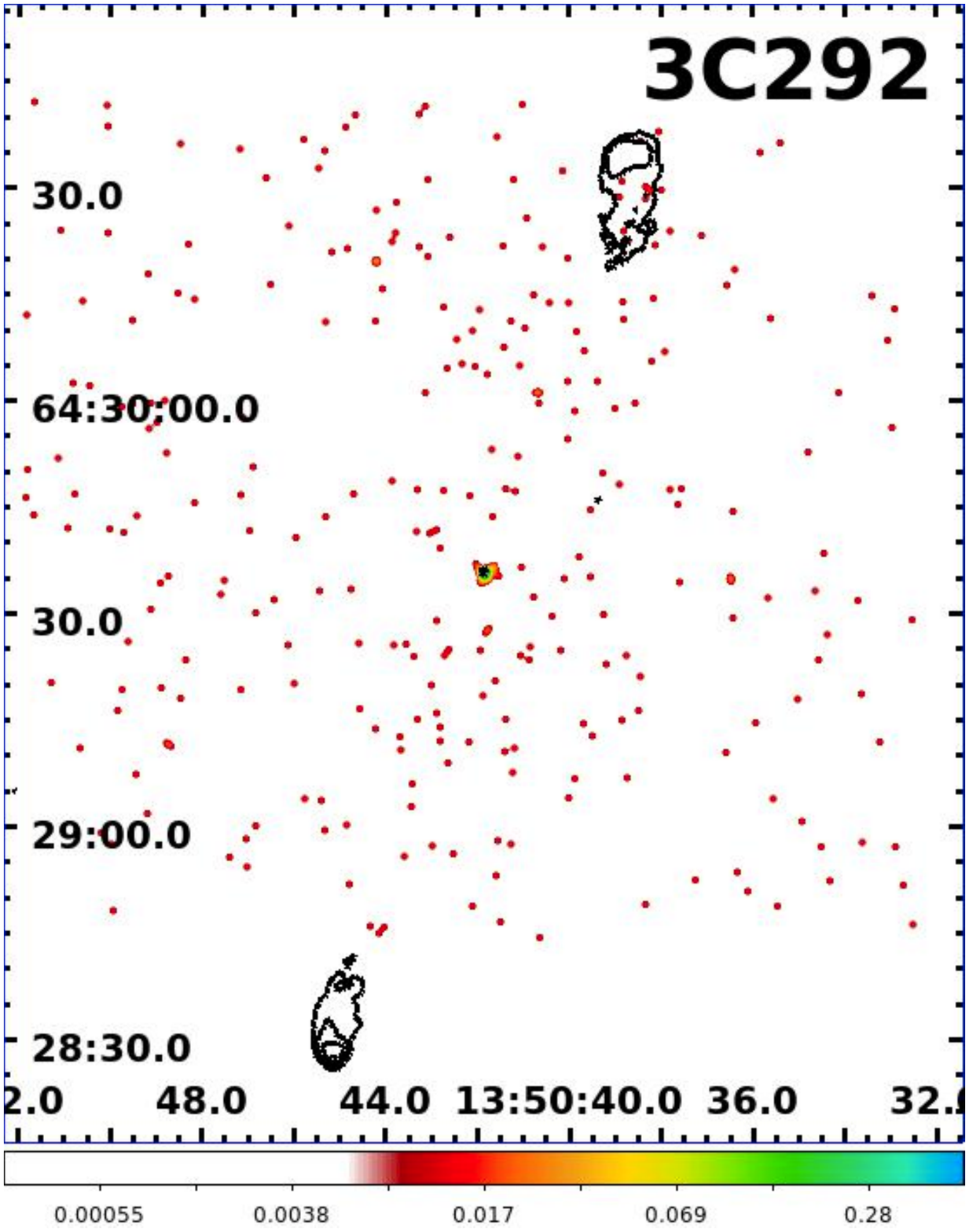}
\caption{The X-ray image of 3CR\,292 for the energy band 0.5-7 keV (obsID 17488).  The
  event file has been regridded to 1/4 of the native pixel size (i.e., 0.492\arcsec).
  The image has been smoothed with a Gaussian of FWHM=5\arcsec.   
  The radio contours (black) were computed using a 8.5 GHz radio
  map and start at 0.3 mJy/beam, increasing by factors of four.}
\label{fig:3c292app}
\end{figure}

\begin{figure}
\includegraphics[keepaspectratio=true,scale=0.5,angle=-90]{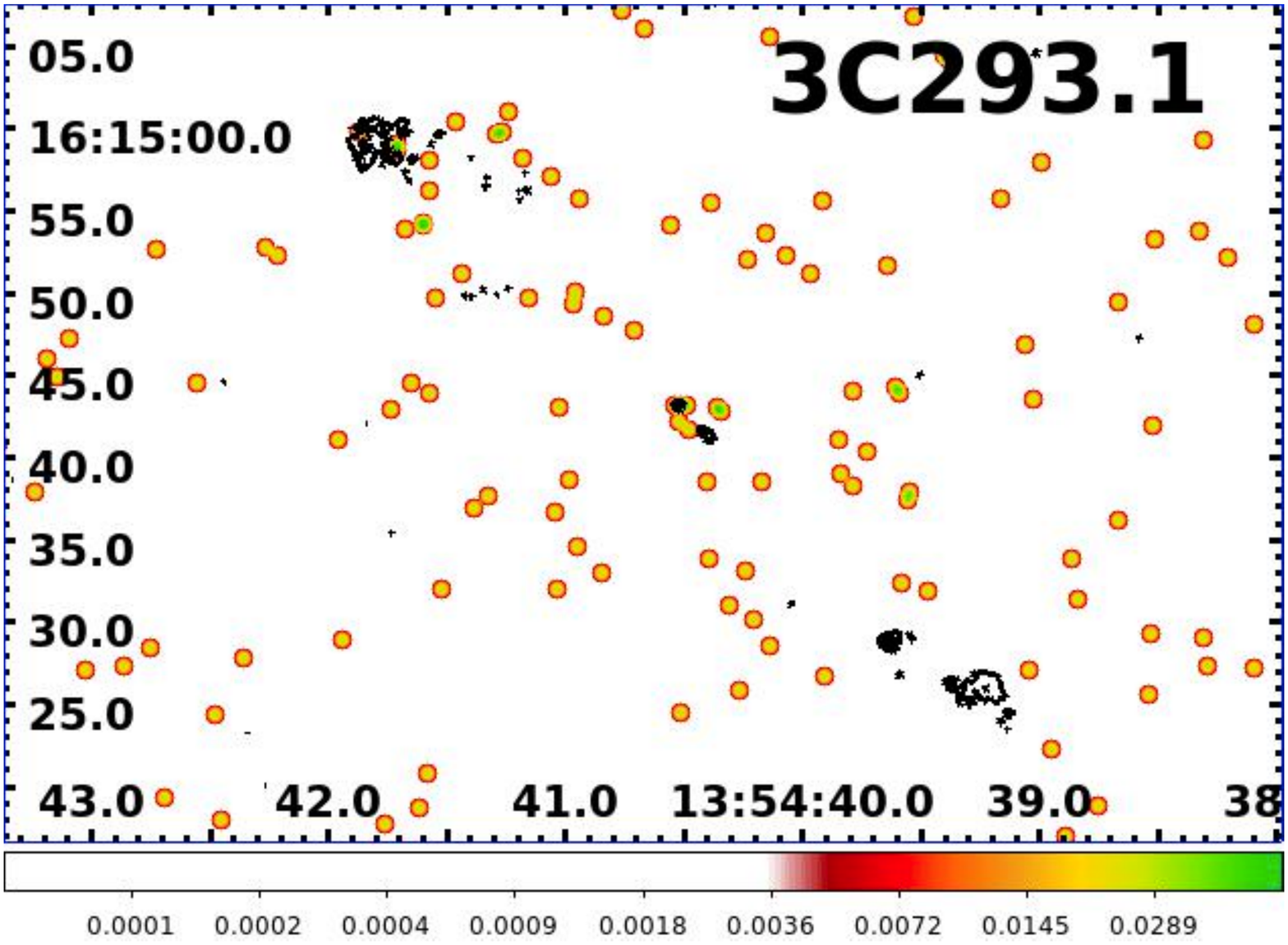}
\caption{The X-ray image of 3CR\,292 for the energy band 0.5-7 keV.  The
  event file has been regridded to 1/4 of the native pixel size (i.e., 0.492\arcsec).
  The image has been smoothed with a Gaussian of FWHM=5\arcsec.   
  The radio contours (black) were computed using a 4.9 GHz radio
  map and start at 0.15 mJy/beam, increasing by factors of wo.}
\label{fig:3c293.1app}
\end{figure}

\begin{figure}
\includegraphics[keepaspectratio=true,scale=0.65]{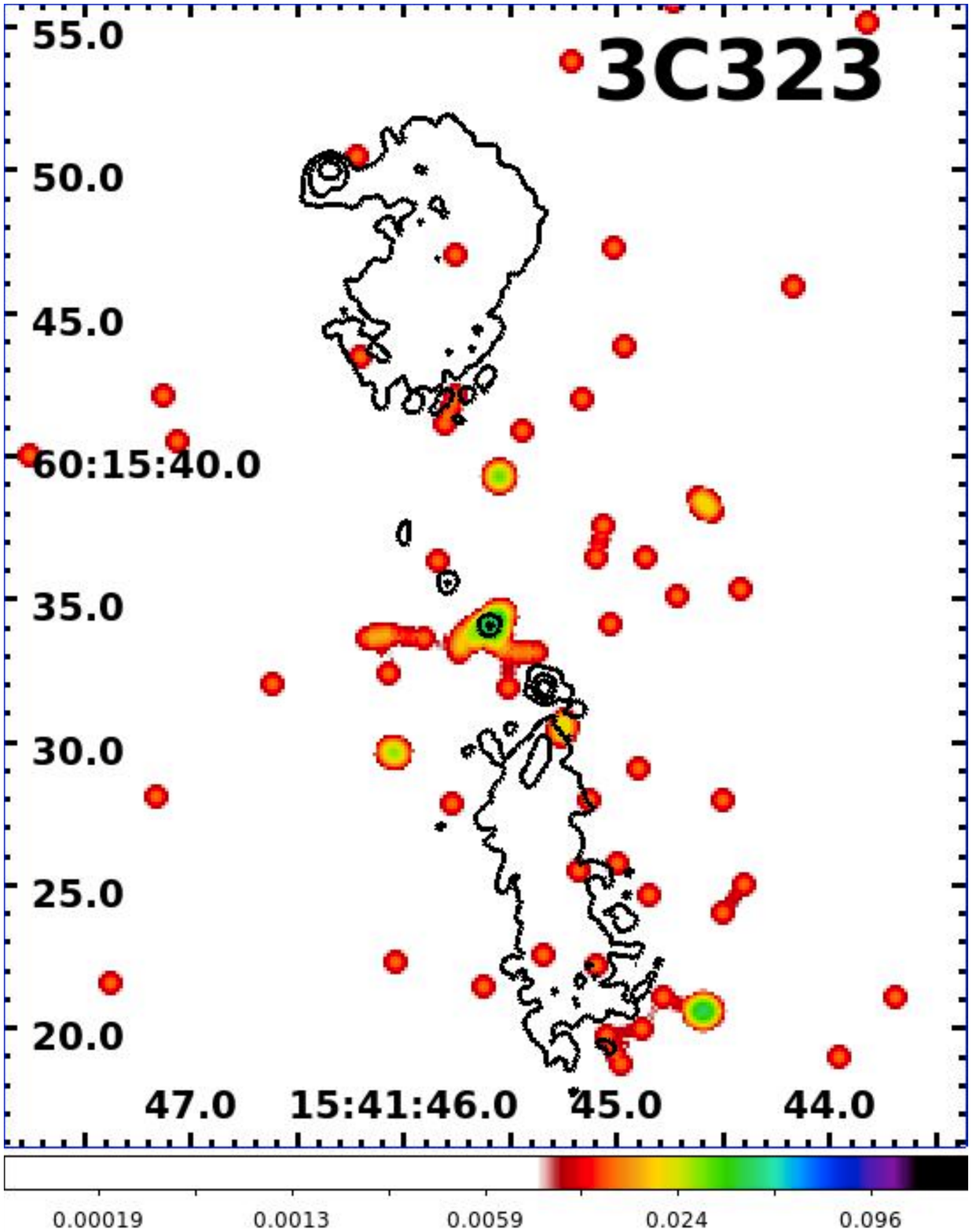}
\caption{The X-ray image of 3CR\,292 for the energy band 0.5-7 keV.  The
  event file has been regridded to 1/4 of the native pixel size (i.e., 0.492\arcsec).
  The image has been smoothed with a Gaussian of FWHM=7\arcsec.   
  The radio contours (black) were computed using a 8.4 GHz radio
  map and start at 0.25 mJy/beam, increasing by factors of four.}
\label{fig:3c323app}
\end{figure}

\newpage
\section{B: The Status of the \chn\ X-ray 3CR observations}
\label{sec:state}

{Here we present the current status of the \chn\ and \xmm\ observations for the entire 3CR catalog. While \chn\ X-ray observations have been uniformly re-analyzed, as reported in our previous investigations, all the \xmm\ information provided here is based on a literature search \citep[see e.g.][and references therein for more details]{croston05,belsole07,croston08,laskar10,shelton11,ineson13,mannering13,ineson17}.

For all 3CR sources, we report their classification, labeling: radio galaxies (RG), according to the Fanaroff \& Riley criterion \citep{fanaroff74}; quasars (i.e., QSRs);
Seyfert galaxies (Sy) and BL Lac objects (BL). We also indicate as ``UNID'' those sources which, lacking optical spectroscopy, remain unidentified. We include a column reporting the radio morphology for the radio galaxies (FR\,I $vs$ FR\,II types) and indicating those objects that also show the radio structure of: (i) Compact Steep Spectrum (CSS) or X-shaped (XS) radio sources or (ii) have been classified in the literature as wide-angle tailed or narrow-angle tailed radio galaxies (WAT and NAT, resepctively). We also devoted a column to the optical classification of radio galaxies distinguishing them as HERG or LERG. The most updated value of the redshift $z$ is also reported and we used a ``cluster flag'' to label sources that belong to a known galaxy group/cluster. We considered sources belonging to a galaxy-rich large scale environment those for which there is a known optical group/cluster reported in the literature and/or those for which there is an archival X-ray observation confirming the presence of hot gas in the intergalactic medium.

Regarding the X-ray analysis, we report X-ray detections of radio components adopting the following symbols: $k=$ jet knot; $h=$ hotspot; $l=$ lobe and $igm$ for sources that belong to a galaxy-rich large scale environment. No distinction has been made between sources lying in group or clusters of galaxies. We also adopt the symbol $e$ for those radio objects that show extended X-ray emission of kpc scale as highlighted in our analyses using the ``extent ratio'' measurements. For \xmm\ observations we only adopted $l$ and $igm$ symbols due to the lower angular resolution with respect to \chn\ that does not allow to see counterparts of jet knots and hotspots in the large fraction fo the 3CR sources.

Finally, the ``X-ray flag'' indicates if the source was already observed by \chn\ (c) and/or \xmm\ (x). Sources marked with a $^*$ close to their 3CR name are those not re-analyzed in our previous studies (see \S~\ref{sec:history} for more details). The table present in this work updates and thus supersedes those included in previous publications.}

\begin{table} 
\caption{The current status of the 3CR \chn\ observations.}
\label{tab:main}
\tiny
\begin{center}
\begin{tabular}{|rrrrrrrrr|}
\hline
3CR  & $z$ & class & radio & optical & Cluster & \chn\ & \xmm\ & X-ray \\
name  &  &  & morph. & class & flag & detections & detections & obs. \\
\hline 
\noalign{\smallskip}
  2.0 & 1.03658 & QSO &  &  &  & e &  & c \\
  6.1 & 0.8404 & RG & FRII & HERG &  & h &  & c \\
  9.0 & 2.01976 & QSO & LDQ &  &  & k;l &  & c \\
  11.1 & ? & UNID & - & - &  &  &  & x \\
  13.0 & 1.351 & RG & FRII & HERG &  & e;h &  & c-x \\
  14.0 & 1.469 & QSO &  &  &  &  &  & c \\
  14.1 & ? & UNID & - & - &  &  &  & x \\
  15.0 & 0.07368 & RG & FRI & LERG &  & k;l &  & c \\
  16.0 & 0.405 & RG & FRII & HERG &  & e;h;l &  & c-x \\
  17.0 & 0.219685 & QSO &  &  &  & k &  & c \\
  18.0 & 0.188 & RG & FRII &  &  & e &  & c \\
  19.0 & 0.482 & RG & FRII & LERG & yes & h;igm &  & c-x \\
  20.0 & 0.174 & RG & FRII & HERG &  &  &  & c-x \\
  21.1 & ? & UNID & - & - &  &  &  & x \\
  22.0 & 0.936 & RG & FRII &  &  &  &  & c \\
  27.0 & 0.184 & RG & FRII & HERG &  &  &  & c \\
  28.0 & 0.19544 & RG & FRI & LERG & yes & igm & igm & c-x \\
  29.0 & 0.045031 & RG & FRI & LERG & yes & k;igm &  & c \\
  31.0 & 0.017005 & RG & FRI & LERG & yes & k & igm & c-x \\
  33.0 & 0.0597 & RG & FRII & HERG & yes & h:l & l;igm & c-x \\
  33.1 & 0.181 & RG & FRII & HERG &  &  &  & c \\
  33.2 & ? & UNID & - & - &  &  &  & x \\
  34.0 & 0.69 & RG & FRII & HERG & yes & igm &  & c \\
  35.0 & 0.067013 & RG & FRII & LERG & yes & e;l & l;igm & c-x \\
  36.0 & 1.301 & RG & FRII & HERG &  &  &  & c \\
  40.0 & 0.018 & RG & FRI-WAT & LERG & yes & igm & igm & c-x \\
  41.0 & 0.795 & RG & FRII & HERG &  & h &  & c \\
  42.0 & 0.39598 & RG & FRII & HERG &  &  &  & c-x \\
  43.0 & 1.459 & QSO & CSS &  &  &  &  & c \\
  44.0 & 0.66 & QSO &  &  & yes &  &  & c \\
  46.0 & 0.4373 & RG & FRII & HERG & yes &  & igm & c-x \\
  47.0 & 0.425 & QSO & LDQ &  &  & h;l &  & c \\
  48.0 & 0.367 & QSO & CSS &  &  &  &  & c \\
  49.0 & 0.23568 & RG & FRII-CSS & HERG &  &  &  & c \\
  52.0 & 0.29 & RG & FRII-XS & HERG & yes & h &  & c \\
  54.0 & 0.8274 & RG & FRII & HERG &  & h &  & c \\
  55.0 & 0.7348 & RG & FRII & HERG &  &  &  & c-x \\
  61.1 & 0.18781 & RG & FRII & HERG &  & h &  & c-x \\
  63.0 & 0.175 & RG & FRII & HERG &  &  &  & c \\
  65.0 & 1.176 & RG & FRII & HERG &  & e;h &  & c-x \\
  66.0A & ? & BLL & BL & - & yes &  &  & c-x \\
  66.0B & 0.021258 & RG & FRI-XS & LERG & yes & k & igm & c-x \\
  67.0 & 0.3102 & RG & FRII-CSS &  &  &  &  & c-x \\
  68.1 & 1.238 & QSO &  &  &  &  &  & c \\
  68.2 & 1.575 & RG & FRII & HERG &  & e;h &  & c \\
  69.0 & 0.458 & RG & FRII & HERG &  &  &  & c \\
  71.0 & 0.003793 & SEY & - & Sy2 &  &  &  & c-x \\
  75.0 & 0.023153 & RG & FRI-WAT & LERG & yes & igm & igm & c-x \\
  76.1 & 0.032489 & RG & FRI & LERG & yes & igm & igm & c-x \\
  78.0 & 0.028653 & RG & FRI & LERG &  & k &  & c \\
\noalign{\smallskip}
\hline
\end{tabular}\\
\end{center}
Col. (1): The 3CR name.
Col. (2): Redshift $z$. We also verified in the literature (e.g., NED and/or SIMBAD databases) if new $z$ values were reported after the release of the 3CR catalog.
Col. (3): The source classification of the sources: RG stands for radio galaxies, QSO for quasars; Sy for Seyfert galaxies and BLL for BL Lac objects. We used the acronym UNID for sources that are still unidentified; i.e., lacking of an optical spectroscopic observation.
Col. (4): The radio morphological classification: FR\,I and FR\,II refer to the Fanaroff and Riley classification criterion \citep{fanaroff74} while LDQ and CDQ is sometimes adopted for lobe-dominated and core-dominated quasars; we also indicated if in the literature the source is classified as CSS or if presents an X-shaped radio structure (XS) or if it is a narrow or wide angle tailed radio galaxy (NAT and WAT, respectively).
Col. (5): The optical spectroscopic designation: LERG, ``Low Excitation Radio Galaxy'', HERG, ``High Excitation Radio Galaxy''.
Col. (6): The ``cluster flag'' as discussed in \S~\ref{sec:state}.
Col. (7): In this column we report if the source has a radio component with an X-ray counterpart in a \chn\ observation. We used the following labels: k = jet knot; h = hotspot; l = lobe; e = extended X-ray emission around the nucleus on kpc scale and igm whenever this extended emission is associated with the hot gas present in the intergalactic medium.
Col. (8): The same of the previous column (i.e. Col. 7) but for the \xmm\ observations.
Col. (9): The c flag indicates that at least one \chn\ observation is present in its archive while the x flag refers to the \xmm\ archive.
\end{table}

\begin{table} 
\caption{The current status of the 3CR \chn\ observations.}
\tiny
\begin{center}
\begin{tabular}{|rrrrrrrrr|}
\hline
3CR  & $z$ & class & radio & optical & Cluster & \chn\ & \xmm\ & X-ray \\
name  &  &  & morph. & class & flag & detections & detections & obs. \\
\hline 
\noalign{\smallskip}
  79.0 & 0.2559 & RG & FRII & HERG & yes &  & igm & c-x \\
  83.1 & 0.025137 & RG & FRI-NAT & LERG & yes & k;igm & igm & c-x \\
  84.0 & 0.017559 & RG & FRI & LERG & yes & igm & igm & c-x \\
  86.0 & ? & UNID & - & - &  &  &  & \\
  88.0 & 0.030221 & RG & FRI & LERG & yes & k;igm & igm & c-x \\
  89.0 & 0.13981 & RG & FRI-WAT & LERG & yes & igm &  & c \\
  91.0 & ? & UNID & - & - &  &  &  & \\
  93.0 & 0.35712 & QSO &  &  &  & e &  & c \\
  93.1 & 0.243 & RG & FRII-CSS & HERG & yes &  &  & c \\
  98.0 & 0.030454 & RG & FRII-XS & HERG &  & l & l & c-x \\
  99.0 & 0.426 & SEY & - & Sy2 &  &  &  & c \\
  103.0 & 0.33 & RG & FRII &  &  &  &  & c \\
  105.0 & 0.089 & RG & FRII & HERG &  & k;h &  & c-x \\
  107.0 & 0.785 & RG & FRII & HERG &  & l &  & c \\
  109.0 & 0.3056 & RG & FRII & HERG &  & h;l &  & c-x \\
  111.0 & 0.0485 & RG & FRII &  &  & k;h &  & c-x \\
  114.0 & 0.815 & RG & FRII & LERG &  &  &  & c \\
  119.0 & 1.023 & QSO & CSS &  &  &  &  & c \\
  123.0 & 0.2177 & RG & FRII & LERG & yes & h;igm &  & c \\
  124.0 & 1.083 & RG & FRII & HERG &  & e &  & c \\
  125.0 & ? & UNID & - & - &  &  &  & \\
  129.0 & 0.0208 & RG & FRI-NAT &  & yes & k;igm & igm & c-x \\
  129.1 & 0.0222 & RG & FRI &  & yes & igm & igm & c-x \\
  130.0 & 0.109 & RG & FRI-WAT &  &  & igm &  & c \\
  131.0 & ? & UNID & - & - &  &  &  & \\
  132.0 & 0.214 & RG & FRII & LERG & yes &  &  & c-x \\
  133.0 & 0.2775 & RG & FRII & HERG &  &  &  & c \\
  134.0 & ? & UNID & - & - &  &  &  & \\
  135.0 & 0.12738 & RG & FRII & HERG & yes &  &  & c \\
  136.1 & 0.064 & RG & FRII-XS & HERG &  & e &  & c \\
  137.0 & ? & UNID & - & - &  &  &  & \\
  138.0 & 0.759 & QSO & CSS &  &  &  &  & c \\
  139.2 & ? & UNID & - & - &  &  &  & \\
  141.0 & ? & UNID & - & - &  &  &  & \\
  142.1 & 0.4061 & RG & FRII &  &  &  &  & c \\
  147.0 & 0.545 & QSO & CSS &  &  &  &  & c \\
  152.0 & ? & UNID & - & - &  &  &  & \\
  153.0 & 0.2769 & RG & FRII & LERG & yes &  &  & c-x \\
  154.0 & 0.58 & QSO &  &  &  & e;k &  & c \\
  158.0 & ? & UNID & - & - &  &  &  & \\
  165.0 & 0.2957 & RG & FRII & LERG &  & e &  & c \\
  166.0 & 0.2449 & RG & FRII & LERG &  &  &  & c \\
  169.1 & 0.633 & RG & FRII & HERG &  &  &  & c \\
  171.0 & 0.2384 & RG & FRII & HERG &  & e &  & c-x \\
  172.0 & 0.5191 & RG & FRII & HERG &  &  &  & c \\
  173.0 & 1.035 & QSO & CSS & HERG &  &  &  & c \\
  173.1 & 0.2921 & RG & FRII & LERG & yes & h;l &  & c \\
  175.0 & 0.77 & QSO &  &  &  &  &  & c \\
  175.1 & 0.92 & RG & FRII & HERG &  &  &  & c \\
  180.0 & 0.22 & RG & FRII & HERG &  &  &  & c \\
\noalign{\smallskip}
\hline
\end{tabular}\\
\end{center}
\end{table}

\begin{table} 
\caption{The current status of the 3CR \chn\ observations.}
\tiny
\begin{center}
\begin{tabular}{|rrrrrrrrr|}
\hline
3CR  & $z$ & class & radio & optical & Cluster & \chn\ & \xmm\ & X-ray \\
name  &  &  & morph. & class & flag & detections & detections & obs. \\
\hline 
\noalign{\smallskip}
  181.0 & 1.382 & QSO &  &  &  & h &  & c \\
  184.0 & 0.994 & RG & FRII & HERG & yes & l & igm & c-x \\
  184.1 & 0.1182 & RG & FRII & HERG & yes &  &  & c \\
  186.0 & 1.06551 & QSO & CSS &  & yes & igm &  & c \\
  187.0 & 0.465 & RG & FRII & LERG &  & e;l &  & c \\
  190.0 & 0.24639 & QSO & CSS &  &  &  &  & c \\
  191.0 & 1.96810 & QSO &  &  &  & k;l &  & c \\
  192.0 & 0.05968 & RG & FRII-XS & HERG & yes & l & igm & c-x \\
  194.0 & 1.184 & RG & FRII & HERG &  &  &  & c \\
  196.0 & 0.87063 & QSO &  &  &  &  &  & c \\
  196.1 & 0.198 & RG & FRII & LERG & yes & igm &  & c \\
  197.1 & 0.12825 & RG & FRII & HERG & yes &  &  & c \\
  198.0 & 0.08145 & RG & FRII & HERG & yes &  &  & c \\
  200.0 & 0.458 & RG & FRII & LERG & yes & k;l &  & c \\
  204.0 & 1.112 & QSO &  &  &  &  &  & c-x \\
  205.0 & 1.53154 & QSO &  &  &  &  &  & c-x \\
  207.0 & 0.68038 & QSO & LDQ &  & yes & k;l & igm & c-x \\
  208.0 & 1.11151 & QSO &  &  &  &  &  & c \\
  208.1 & 1.02 & QSO &  &  &  &  &  & c-x \\
  210.0 & 1.169 & RG & FRII & HERG & yes & e;h & igm & c-x \\
  212.0 & 1.04931 & QSO & LDQ &  &  & e;h &  & c \\
  213.1 & 0.19405 & RG & FRI-CSS & LERG & yes & e;h &  & c \\
  215.0 & 0.41106 & QSO &  &  & yes & k;l &  & c-x \\
  216.0 & 0.67002 & QSO &  &  &  &  &  & c \\
  217.0 & 0.8975 & RG & FRII & HERG &  &  &  & c \\
  219.0 & 0.17456 & RG & FRII & HERG & yes & k;l &  & c \\
  220.1 & 0.61 & RG & FRII & HERG & yes & igm &  & c \\
  220.2 & 1.15610 & QSO &  &  &  & h;l &  & c \\
  220.3 & 0.68 & RG & FRII &  &  &  &  & c \\
  222.0 & 1.339 & RG & FRI &  &  &  &  & c \\
  223.0 & 0.13673 & RG & FRII & HERG & yes &  & igm & c-x \\
  223.1 & 0.1075 & RG & FRII-XS & HERG &  &  &  & c \\
  225.0A & 1.565 & RG & FRII &  &  &  &  & c \\
  225.0B & 0.58 & RG & FRII & HERG &  & h &  & c \\
  226.0 & 0.8177 & RG & FRII &  &  &  &  & c \\
  227.0 & 0.086272 & RG & FRII & HERG &  & h &  & c \\
  228.0 & 0.5524 & RG & FRII & HERG &  & e;h &  & c \\
  230.0 & 1.487 & RG & FRII & HERG &  &  &  & c \\
  231.0 & 0.000677 & RG & FRI & LERG &  &  &  & c-x \\
  234.0 & 0.184925 & RG & FRII & HERG &  & h &  & c-x \\
  236.0 & 0.1005 & RG & FRII & LERG &  &  &  & c \\
  237.0 & 0.877 & RG & FRII-CSS &  &  &  &  & c \\
  238.0 & 1.405 & RG & FRII & HERG &  &  &  & c \\
  239.0 & 1.781 & RG & FRII &  &  &  &  & x \\
  241.0 & 1.617 & RG & FRII &  &  &  &  & c-x \\
  244.1 & 0.428 & RG & FRII & HERG & yes & e &  & c-x \\
  245.0 & 1.027872 & QSO &  &  &  & k &  & c \\
  247.0 & 0.7489 & RG & FRII & HERG & yes &  &  & c \\
  249.0 & 1.554 & QSO &  &  &  &  &  & x \\
  249.1 & 0.3115 & QSO &  &  &  &  &  & c-x \\
  250.0 & ? & UNID & - & - &  &  &  & \\
\noalign{\smallskip}
\hline
\end{tabular}\\
\end{center}
\end{table}

\begin{table} 
\caption{The current status of the 3CR \chn\ observations.}
\tiny
\begin{center}
\begin{tabular}{|rrrrrrrrr|}
\hline
3CR  & $z$ & class & radio & optical & Cluster & \chn\ & \xmm\ & X-ray \\
name  &  &  & morph. & class & flag & detections & detections & obs. \\
\hline 
\noalign{\smallskip}
  252.0 & 1.1 & RG & FRII &  &  &  &  & c \\
  254.0 & 0.736619 & QSO & LDQ &  &  & e;h &  & c \\
  255.0 & 1.355 & QSO &  &  &  &  &  & c \\
  256.0 & 1.819 & RG & FRII &  &  &  &  & c \\
  257.0 & 2.474 & QSO &  &  &  &  &  & x \\
  258.0 & 0.165 & RG & FRI-CSS & LERG & yes &  &  & c \\
  263.0 & 0.646 & QSO & LDQ &  &  & h &  & c \\
  263.1 & 0.824 & RG & FRII &  &  &  &  & c \\
  264.0 & 0.021718 & RG & FRI & LERG & yes & k & igm & c-x \\
  265.0 & 0.811 & RG & FRII &  &  & h;l &  & c \\
  266.0 & 1.275 & RG & FRII &  &  &  &  & c-x \\
  267.0 & 1.14 & RG & FRII &  &  &  &  & c \\
  268.1 & 0.97 & RG & FRII &  &  & h &  & c \\
  268.2 & 0.362 & RG & FRII &  & yes & e;h &  & c-x \\
  268.3 & 0.37171 & RG & FRII &  &  &  &  & c \\
  268.4 & 1.402200 & QSO &  &  &  &  &  & c-x \\
  270.0 & 0.007378 & RG & FRI & LERG & yes & k & igm & c-x \\
  270.1 & 1.528432 & QSO &  &  &  &  &  & c \\
  272.0 & 0.944 & RG & FRII &  &  &  &  & c \\
  272.1 & 0.003392 & RG & FRI & LERG & yes & k &  & c-x \\
  273.0 & 0.158339 & QSO & CDQ &  &  & k &  & c-x \\
  274.0 & 0.0043 & RG & FRI & LERG & yes & k;igm & igm & c-x \\
  274.1 & 0.422 & RG & FRII & HERG &  & e & l & c-x \\
  275.0 & 0.48 & RG & FRII & LERG & yes &  &  & c \\
  275.1 & 0.5551 & QSO & LDQ &  &  & k;h;l &  & c \\
  277.0 & 0.414 & RG & FRII &  &  &  &  & c \\
  277.1 & 0.31978 & QSO &  &  &  &  &  & c \\
  277.2 & 0.766 & RG & FRII & HERG &  &  &  & c-x \\
  277.3 & 0.085336 & RG & FRII & HERG &  &  &  & c \\
  280.0 & 0.996 & RG & FRII &  & yes & k;h;l &  & c-x \\
  280.1 & 1.667065 & QSO &  &  &  & l &  & \\
  284.0 & 0.239754 & RG & FRII & HERG & yes &  & igm & c-x \\
  285.0 & 0.0794 & RG & FRII & HERG &  & l &  & c \\
  286.0 & 0.849934 & QSO &  &  &  &  &  & c \\
  287.0 & 1.055 & QSO &  &  &  &  &  & c-x \\
  287.1 & 0.2156 & RG & FRII & HERG &  & h &  & c \\
  288.0 & 0.246 & RG & FRI & LERG & yes & igm &  & c \\
  288.1 & 0.96296 & QSO &  &  &  &  &  & c \\
  289.0 & 0.9674 & RG & FRII &  &  &  &  & c \\
  292.0 & 0.71 & RG & FRII & HERG & yes & e & igm & c-x \\
  293.0 & 0.045034 & RG & FRI & LERG &  & e &  & c \\
  293.1 & 0.709 & RG & FRII &  &  &  &  & c \\
  294.0 & 1.779 & RG & FRII &  & yes & h;igm &  & c \\
  295.0 & 0.4641 & RG & FRII & LERG & yes & h;igm &  & c \\
  296.0 & 0.024704 & RG & FRI & LERG & yes & k;igm & igm & c-x \\
  297.0 & 1.4061 & QSO &  &  &  &  &  & c \\
  298.0 & 1.438120 & QSO &  &  & yes &  & igm & c-x \\
  299.0 & 0.367 & RG & FRII &  & yes & h &  & c \\
  300.0 & 0.27 & RG & FRII & HERG &  &  &  & c-x \\
\noalign{\smallskip}
\hline
\end{tabular}\\
\end{center}
\end{table}

\begin{table} 
\caption{The current status of the 3CR \chn\ observations.}
\tiny
\begin{center}
\begin{tabular}{|rrrrrrrrr|}
\hline
3CR  & $z$ & class & radio & optical & Cluster & \chn\ & \xmm\ & X-ray \\
name  &  &  & morph. & class & flag & detections & detections & obs. \\
\hline 
\noalign{\smallskip}
  300.1 & 1.15885 & RG & FRII & HERG &  &  &  & c \\
  303.0 & 0.141186 & RG & FRII & HERG & yes & k;l &  & c \\
  303.1 & 0.2704 & RG & FRII-CSS & HERG &  & e &  & c-x \\
  305.0 & 0.041639 & RG & FRII-CSS & HERG &  & e &  & c-x \\
  305.1 & 1.132 & RG & FRII-CSS & LERG &  &  &  & c \\
  306.1 & 0.441 & RG & FRII & HERG & yes & e &  & c \\
  309.1 & 0.905 & QSO &  &  &  & e &  & c \\
  310.0 & 0.0538 & RG & FRI & LERG & yes & igm &  & c \\
  314.1 & 0.1197 & RG & FRI & LERG & yes &  &  & c-x \\
  313.0 & 0.461 & RG & FRII & HERG & yes & h;igm &  & c \\
  315.0 & 0.1083 & RG & FRI-XS & LERG & yes & e &  & c \\
  317.0 & 0.034457 & RG & FRI & LERG & yes & igm & igm & c-x \\
  318.0 & 1.574 & RG & FRII &  & yes &  &  & c-x \\
  318.1 & 0.045311 & RG & FRI &  & yes & igm & igm & c-x \\
  319.0 & 0.192 & RG & FRII & LERG & yes &  &  & c-x \\
  320.0 & 0.342 & RG & FRII &  & yes & igm &  & c \\
  321.0 & 0.0961 & RG & FRII & HERG &  & h;l &  & c-x \\
  322.0 & 1.681 & RG & FRII &  & yes &  & igm & x \\
  323.0 & 0.679 & RG & FRII &  &  & e &  & c \\
  323.1 & 0.2643 & RG & FRII & HERG & yes &  &  & c \\
  324.0 & 1.2063 & RG & FRII &  & yes & e;h &  & c-x \\
  325.0 & 1.135 & RG & FRII &  &  & h &  & c \\
  326.0 & 0.0895 & RG & FRII & LERG &  & l &  & c \\
  326.1 & 1.825 & RG & FRII &  &  &  &  &  \\
  327.0 & 0.1048 & RG & FRII & HERG & yes & h &  & c \\
  327.1 & 0.462 & RG & FRI & HERG &  & k &  & c-x \\
  330.0 & 0.55 & RG & FRII & HERG & yes & h;l &  & c \\
  332.0 & 0.151019 & RG & FRII & HERG & yes &  &  & c \\
  334.0 & 0.5551 & QSO & LDQ &  &  & k;l &  & c \\
  336.0 & 0.926542 & QSO &  &  &  &  &  & c \\
  337.0 & 0.635 & RG & FRII &  & yes & e &  & c-x \\
  338.0 & 0.030354 & RG & FRI & LERG & yes & igm & igm & c-x \\
  340.0 & 0.7754 & RG & FRII &  &  &  &  & c \\
  341.0 & 0.448 & RG & FRII & HERG & yes & e;k & igm & c-x \\
  343.0 & 0.988 & QSO &  &  &  &  &  & c \\
  343.1 & 0.75 & RG & FRII &  &  &  &  & c \\
  345.0 & 0.5928 & QSO & CDQ &  & yes & k & igm & c-x \\
  346.0 & 0.162012 & RG & FRI & HERG & yes & k &  & c \\
  348.0 & 0.155 & RG & FRI & LERG & yes & igm & igm & c-x \\
  349.0 & 0.205 & RG & FRII & HERG &  & h &  & c-x \\
  351.0 & 0.37194 & RG & FRII &  &  & h &  & c \\
  352.0 & 0.8067 & RG & FRII &  &  &  &  & c \\
  353.0 & 0.030421 & RG & FRII & LERG & yes & k:l:igm & igm & c-x \\
  356.0 & 1.079 & RG & FRII &  &  & e &  & c \\
  357.0 & 0.166148 & RG & FRII & LERG & yes &  &  & c \\
  368.0 & 1.131 & RG & FRII &  &  &  &  & c \\
  371.0 & 0.051 & BLL & BL & - &  & k &  & c \\
  379.1 & 0.256 & RG & FRII & HERG &  &  &  & c \\
  380.0 & 0.692 & QSO & CDQ &  &  & k &  & c \\
  381.0 & 0.1605 & RG & FRII & HERG &  &  &  & c \\
  382.0 & 0.05787 & RG & FRII & HERG &  &  &  & c-x \\
\noalign{\smallskip}
\hline
\end{tabular}\\
\end{center}
\end{table}

\begin{table} 
\caption{The current status of the 3CR \chn\ observations.}
\tiny
\begin{center}
\begin{tabular}{|rrrrrrrrr|}
\hline
3CR  & $z$ & class & radio & optical & Cluster & \chn\ & \xmm\ & X-ray \\
name  &  &  & morph. & class & flag & detections & detections & obs. \\
\hline 
\noalign{\smallskip}
  386.0 & 0.016885 & RG & FRI & LERG & yes &  & igm & c-x \\
  388.0 & 0.0917 & RG & FRII & LERG & yes & e &  & c \\
  389.0 & ? & UNID & - & - &  &  &  & x \\
  390.0 & ? & UNID & - & - &  &  &  & \\
  390.3 & 0.0561 & RG & FRII & HERG &  & k;h &  & c-x \\
  394.0 & ? & UNID & - & - &  &  &  & \\
  399.1 & ? & UNID & - & - &  &  &  & \\
  401.0 & 0.2011 & RG & FRII & LERG & yes & igm &  & c \\
  402.0 & 0.025948 & RG & FRI &  & yes & k &  & c-x \\
  403.0 & 0.059 & RG & FRII & HERG &  & k;h &  & c \\
  403.1 & 0.0554 & RG & FRII & LERG & yes &  &  & c \\
  405.0 & 0.056075 & RG & FRII &  & yes & h;igm & igm & c-x \\
  409.0 & ? & UNID & - & - &  &  &  & \\
  410.0 & 0.2485 & RG & FRII &  &  &  &  & c \\
  411.0 & 0.467 & RG & FRII & HERG &  &  &  & c-x \\
  415.2 & ? & UNID & - & - &  &  &  & \\
  418.0 & 1.686 & QSO &  &  &  &  &  & \\
  424.0 & 0.126988 & RG & FRI & LERG & yes & e &  & c \\
  427.1 & 0.572 & RG & FRII & LERG & yes & l;igm &  & c \\
  428.0 & ? & UNID & - & - &  &  &  & \\
  430.0 & 0.055545 & RG & FRII & LERG & yes & e &  & c \\
  431.0 & ? & UNID & - & - &  &  &  & \\
  432.0 & 1.785 & QSO &  &  &  &  &  & c-x \\
  434.0 & 0.322 & RG & FRII &  & yes &  &  & c \\
  433.0 & 0.1016 & RG & FRII & HERG &  & l &  & c \\
  435.0 & 0.471 & RG & FRII &  &  &  &  & c \\
  436.0 & 0.2145 & RG & FRII & HERG &  & e;h &  & c-x \\
  437.0 & 1.48 & RG & FRII &  &  & e;h &  & c \\
  438.0 & 0.29 & RG & FRII & HERG & yes & igm &  & c \\
  441.0 & 0.708 & RG & FRII &  &  &  &  & c \\
  442.0 & 0.0263 & RG & FRI & LERG & yes & igm &  & c \\
  445.0 & 0.055879 & RG & FRII &  & yes & h &  & c-x \\
  449.0 & 0.017085 & RG & FRI & LERG & yes & igm & igm & c-x \\
  452.0 & 0.0811 & RG & FRII & HERG & yes & h;l & igm & c-x \\
  454.0 & 1.757 & QSO &  &  &  &  &  & x \\
  454.1 & 1.841 & RG & FRII &  & yes &  &  & \\
  454.2 & ? & UNID & - & - &  &  &  & \\
  454.3 & 0.859 & QSO & CDQ &  &  & k &  & c-x \\
  455.0 & 0.543 & QSO &  &  &  &  &  & c \\
  456.0 & 0.233 & RG & FRII & HERG &  &  &  & c \\
  458.0 & 0.289 & RG & FRII & HERG & yes & h &  & c \\
  459.0 & 0.22012 & RG & FRII &  &  & l &  & c-x \\
  460.0 & 0.268 & RG & FRII & LERG & yes &  &  & c-x \\
  465.0 & 0.030221 & RG & FRI-WAT & LERG & yes & k;igm & igm & c-x \\
  468.1 & ? & UNID & - & - &  &  &  & \\
  469.1 & 1.336 & RG & FRII &  &  &  & l & c-x \\
  470.0 & 1.653 & RG & FRII &  &  & h &  & c \\
\noalign{\smallskip}
\hline
\end{tabular}\\
\end{center}
\end{table}

\end{document}